\documentclass[prx,twocolumn,superscriptaddress,floatfix,longbibliography]{revtex4-2}
\usepackage{amssymb}
\usepackage{physics}
\usepackage{graphicx}
\usepackage{siunitx}
\usepackage{times}
\usepackage{amsmath}
\usepackage{amsthm}
\usepackage{amsfonts}
\usepackage[T1]{fontenc}
\usepackage[latin9]{inputenc}
\usepackage{array}
\usepackage{multirow}
\usepackage{color}
\usepackage{esint}
\usepackage{bm}
\usepackage{xcolor}
\usepackage{bbm}
\usepackage{babel}

\usepackage{float}

\usepackage{hyperref}
\hypersetup
{	colorlinks,%
	citecolor=green,%
	linkcolor=red,%
	urlcolor=blue%
}
\allowdisplaybreaks[4]

\newcommand{\moire}{moir\'{e}}

\begin{document}

\title{Self-consistent theory of fractional quantum anomalous Hall states in rhombohedral graphene}

\author{Ke Huang}
\affiliation{Department of Physics, City University of Hong Kong, Kowloon, Hong Kong SAR, China}

\author{Xiao Li}
\email{xiao.li@cityu.edu.hk}
\affiliation{Department of Physics, City University of Hong Kong, Kowloon, Hong Kong SAR, China}

\author{Sankar Das Sarma}
\affiliation{Condensed Matter Theory Center and Joint Quantum Institute,
University of Maryland, College Park, Maryland 20742, USA}

\author{Fan Zhang}
\affiliation{Department of Physics, The University of Texas at Dallas, Richardson, Texas  75080, USA}

\date{\today}
\begin{abstract}
The fractional quantum anomalous Hall (FQAH) effect in rhombohedral pentalayer graphene (PLG) has attracted significant attention due to its potential for observing exotic quantum states. 
In this work, we present a self-consistent Hartree-Fock theory for the FQAH effect in rhombohedral PLG. 
In particular, we focus on the convergence of the Hartree-Fock calculation with various reference fields and discuss the stability of the FQAH states in PLG. 
We show that the so-called charge neutrality scheme provides an unambiguous result for the Hartree-Fock calculation, as it ensures a convergence with respect to the momentum cutoff.
Based on the Hartree-Fock band structure, we further carry out exact diagonalization calculations to explore the stability of the FQAH states in PLG. 
Our work provides an improved and unified (minimal) theoretical framework to understand the FQAH effect in rhombohedral PLG and paves the way for future studies.
\end{abstract}
\maketitle

\section{Introduction}
The interplay between band topology, band geometry, and electron interactions has led to the discovery of a plethora of exotic quantum states in condensed matter systems. 
Among them, the fractional quantum anomalous Hall (FQAH) states, first proposed a decade ago in various toy models~\cite{Tang2011,Sun2011,Neupert2011,Regnault2011,Sheng2011}, have attracted significant attention due to their potential for realizing {topologically ordered states} that otherwise require a strong magnetic field. 
Meanwhile, recent experimental progress in realizing such states in twisted bilayer graphene~\cite{Spanton2018,Xie2021}, twisted bilayer MoTe$_2$~\cite{Cai2023,Zeng2023,Park2023,Xu2023}, and {multilayer rhombohedral} graphene~\cite{Lu2024,XieJian2024} has further stimulated interest in the FQAH states. 
There is enormous current activity on this topic, and the situation is in a state of flux, but it is already clear that the established theoretical framework for Chern insulators~\cite{Tang2011,Sun2011,Neupert2011,Regnault2011,Sheng2011} does not directly apply to the observed QAH and FQAH phenomenologies in graphene multilayers~\cite{Lu2024,XieJian2024}.

A comprehensive theoretical description of these observed FQAH states demands a fundamental understanding of the interplay among band structure, band topology, electron interactions, and the quantum geometry of the Bloch wavefunctions. 
In this regard, the mapping of Chern bands to generalized Landau levels has provided a valuable framework for interpreting the emergence of FQAH states~\cite{Qi2011,Wu2012a,Parameswaran2013,Jackson2015,Claassen2015,Tarnopolsky2019,Ledwith2020,Wang2021,Wang2021a,Ozawa2021,Mera2021,Wang2022,Ledwith2022,Ledwith2023,Dong2023b,Wang2023,Fujimoto2024,Wang2024,MoralesDuran2024,Wu2024}. 
Such an approach highlights the crucial role of quantum geometry---specifically, the uniformity and fluctuations of Berry curvature and quantum metric across the {\moire} Brillouin zone. 
It works particularly well for twisted bilayer MoTe$_2$ because, in that system, the first \moire\ valence band is relatively flat and well isolated {in energy}, has a nonzero Chern number $\abs{C} = 1$, and also nearly saturates the trace inequality~\cite{Wang2024,MoralesDuran2024}. 
The uniformity of the Berry curvature and quantum metric across the \moire\ Brillouin zone can be viewed as a key ingredient for stabilizing the FQAH states in twisted bilayer MoTe$_2$. 
In other words, such FQAH states can be viewed as adiabatically connected to the fractional quantum Hall states {at a strong magnetic field} and to the physics discussed in Refs.~\cite{Tang2011,Sun2011,Neupert2011,Regnault2011,Sheng2011}.

However, the {physics} in rhombohedral pentalayer graphene (PLG) is more complex because, in the noninteracting picture, the lowest \moire\ conduction band is not well isolated from the other bands, especially in a large displacement field. 
Moreover, the noninteracting band structure of PLG is far from saturating the trace inequality. 
In particular, a naive direct band structure theory leads to the system being a trivial metal rather than an insulator, let alone a Chern insulator. 
To resolve this conundrum, most of the existing theories on the FQAH {states} in PLG~\cite{Dong2023,Dong2023a,Zhou2023} adopt a Hartree-Fock (HF) approximation at {integer filling} $\nu=1$ to isolate the lowest {quasiparticle} conduction band from the other bands and then project the Hamiltonian to this band for performing the exact diagonalization (ED) calculation at fractional band fillings $\nu= 1/3, 2/3$, etc, to identify the FQAH states. 
These studies find that the correlated Chern insulator and FQAH states could exist even without the \moire\ potential, suggesting that the \moire\ potential may not be essential for the realization of QAH and FQAH states. 
Soon after, the effect of the remote bands, including both the conduction and valence bands, was incorporated through a renormalization scheme in the HF calculation, generating both FQAH and charge density wave phases~\cite{Guo2023}. 
Further studies on the HF calculation show that the {so-called reference field} and valence bands can significantly affect the phase diagram. 
In particular, the \moire\ potential could be crucial for {the choice of reference field}~\cite{Kwan2023}. 
We note that our terminology ``reference field'' has earlier been alluded to in the literature as ``reference density''~\cite{Kwan2023}, ``energy reference point''~\cite{Christos2022}, ``reference state''~\cite{Parker2021}, or ``subtraction scheme''~\cite{Bultinck2020a}. 
We use the terminology ``reference field'' throughout this paper. 

Nonetheless, the existing approach in the literature is not without its challenges. 
In fact, the whole theoretical scheme of using the HF theory to generate the time-reversal symmetry breaking (and hence a nontrivial Chern number) leading to FQAH and QAH effects in PLG  is thus somewhat controversial since it appears to depend on the specific approximation schemes. 
For example, the convergence of the HF calculation is notably influenced by the selection of the {aforementioned} reference field, making it difficult to determine the optimal choice for the FQAH states in PLG. 
Different choices of the reference field give very different results, including situations with no FQAH states. 
To address this issue, we systematically compare the convergence of the HF calculation with different reference fields in PLG. 
Specifically, we considered three different reference fields: the charge neutrality scheme (reference field A), the non-\moire\ scheme for the \moire\ case (reference field B), and the infinite-temperature scheme (reference field C). 
We show that the charge neutrality scheme (reference field A) provides an unambiguous result for the HF calculation, as it ensures a convergence with respect to the large-momentum cutoff, which we will use in the subsequent calculations. 
We believe that combining the reference field choice and the momentum cutoff provides a unique (and non-arbitrary) theoretical framework for studying QAH and FQAH physics in PLG-based Chern insulators. 

In addition, the impact of including valence bands in the HF calculation on the FQAH effect in PLG is still a matter of debate.
In this work, we manage to include all bands within the momentum cutoff by performing the HF calculation in the plane wave basis~\cite{Xie2020}. 
This allows us to systematically study the effects of including valence bands in the HF calculation. 
We show that the FQAH states in PLG are robust against the inclusion of valence bands in the HF calculation.

Finally, the self-consistency of the theory is another crucial issue. 
Given that the HF calculation may produce different band structures at different fillings, {it may be instructive to} carry out the HF calculation directly at the desired fractional filling to ensure self-consistency. 
This is an important new feature of the current work. 
Furthermore, it is essential to consider whether the HF calculation and the subsequent ED calculation should be approached as distinct steps or integrated into a single self-consistent procedure. 
Specifically, it is unclear whether identifying the global minimum in the HF calculation will invariably result in the most energetically favorable FQAH state in the ED calculation.
To solve this problem, we propose to view the HF calculation as providing a set of basis states for the ED calculation, and we compare three different methods to combine the HF and ED calculations. 
We find that, in our examples, avoiding the global minimum in the HF calculation can lead to a more energetically favorable FQAH state in PLG in the ED calculation. 
We use this new approach to predict the phase diagram for the FQAH states in PLG. 
By addressing these three challenges, we provide an improved and unified framework to understand the FQAH states in rhombohedral PLG and {potentially in other {\moire} flat-band systems}, paving the way for future studies.

The structure of the paper is as follows. 
In Section~\ref{Sec:TheModel}, we introduce the model Hamiltonian for PLG and discuss the HF approximation. 
In Section~\ref{Sec:HF}, we present our theory for the HF approximation. 
We discuss several important issues, including how to include the valence bands in the calculation, the choice of different reference fields, and the dependence of the calculation on the large-momentum cutoff. 
In particular, we discuss the effects of three different reference fields used in the literature: the charge neutrality scheme (reference field A), the non-\moire\ scheme for the \moire\ case (reference field B), and the infinite-temperature scheme (reference field C), and how they affect the convergence of the HF calculation. 
In Section~\ref{Sec:FQAH}, we present our theory for the search of FQAH states through ED calculations. 
In particular, we propose three different methods to combine the HF and ED calculations. 
We compare their results and ultimately predict the phase diagram for the FQAH states in PLG. 
In Section~\ref{Sec:Conclusion}, we provide further discussions and conclude the paper.

\section{The Model \label{Sec:TheModel}}

\begin{figure}[t]
\center
\includegraphics[width=\columnwidth]{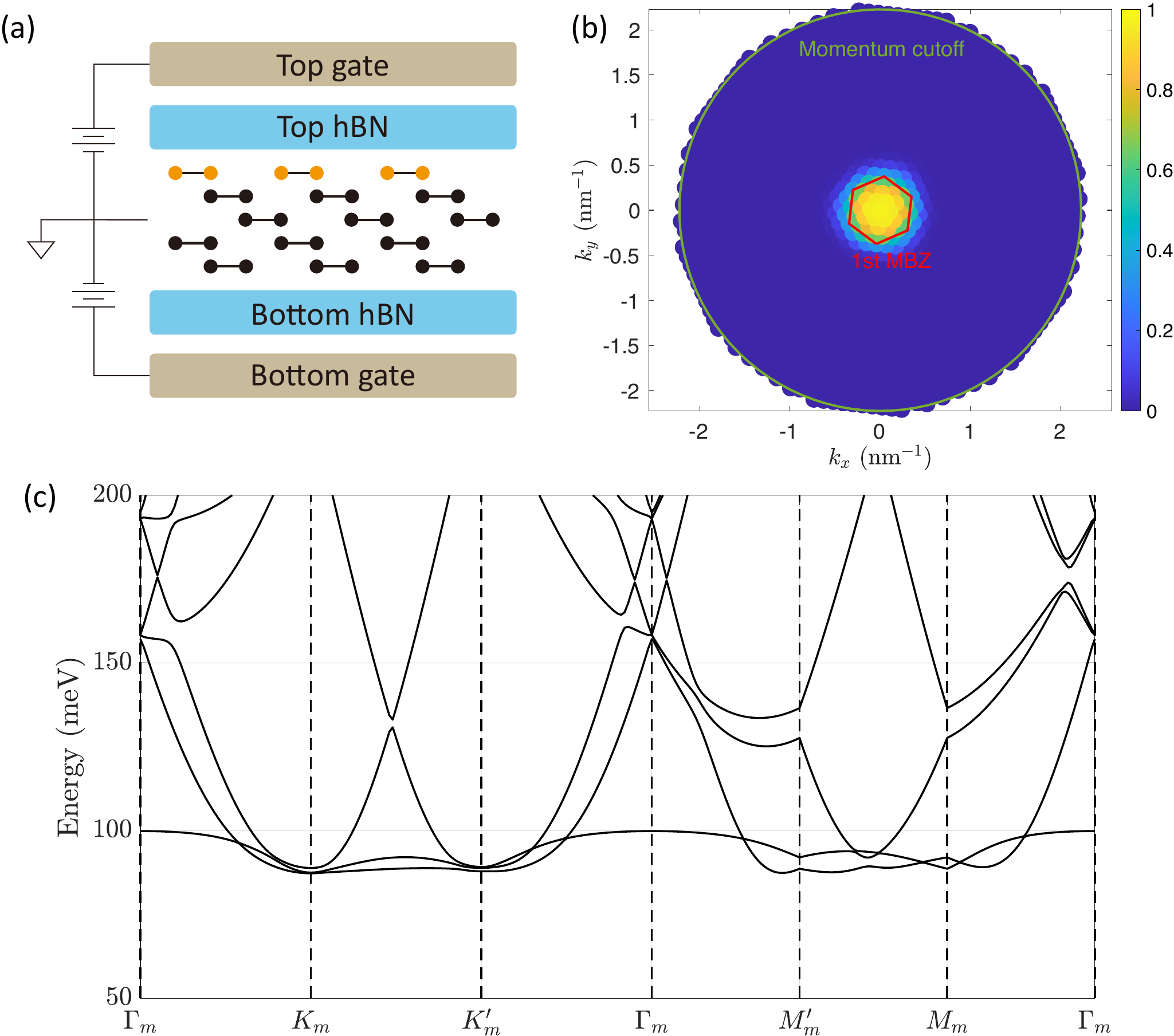}
\caption{\label{Fig:Device}
(a) An illustration of the device configuration used in the experiment~\cite{Lu2024}, where the top hBN is aligned with the top layer of PLG (highlighted in yellow), while the bottom hBN is misaligned with the sample. 
(b) The first MBZ of the \moire\ superlattice and momentum cutoff $\Lambda$ introduced in the calculation. 
The color represents the density in the momentum space (i.e., $\expval*{\Psi_{\vb k}^\dag \Psi_{\vb k}}$) of the HF conduction band of the HF self-consistent solution at filling factor $\nu=1$. 
(c) The noninteracting band structure of PLG with all electrons in the conduction band polarized towards the bottom hBN by an applied electric field. 
Here, the twist angle is $\theta=0.77^\circ$, the applied bias is $u_d=\SI{50}{meV}$, and the momentum cutoff is $\Lambda=3.41 b_m$. 
}
\end{figure}

In this section, we present the model for the PLG we adopt in this work. 
The structure of the PLG is shown in Fig.~\ref{Fig:Device}(a), in accordance with the experiment~\cite{Lu2024}. 
The top hexagonal boron nitride (hBN) is aligned with the top layer of the PLG, creating a \moire\ pattern, while the bottom hBN layer is misaligned with the PLG. 
As a result, such a system can be modeled as~\cite{Moon2014}
\begin{align}
	H_{\text{s}}&=\sum_{\sigma}\sum_{\vb k\in\text{BZ}}\Psi_{\vb k,\sigma}^\dag [H_{\text{PLG}}(\vb k)+\kappa V_{\text{\moire}}(0)] \Psi_{\vb k,\sigma}\nonumber\\
	&\quad+\kappa\sum_{\sigma}\sum_{j}\sum_{\vb k\in\text{BZ}}
	\Psi_{\vb k+\vb g_j,\sigma}^\dag V_{\text{\moire}}(\vb g_j)\Psi_{\vb k,\sigma},
\end{align}
where BZ denotes the Brillouin zone of the pristine (i.e., non-\moire) PLG, $\sigma$ denotes the spin, and $H_{\text{PLG}}(\vb k)$ is the Hamiltonian of the pristine PLG under a perpendicular electric field.  
Moreover, the momentum $\vb k$ is defined in the Brillouin zone of the pristine PLG, and $V_{\text{\moire}}$ is the \moire\ potential generated by the coupling between PLG and hBN. 
Here, $\Psi_{\vb k,\sigma}$ is a column vector composed of $c_{\vb k,\sigma,\alpha}$, with $\alpha$ representing the collection of layer, sublattice, and valley degrees. 
Finally, $V_{\text{\moire}}(0)$ and $V_{\text{\moire}}(\vb g_j)$ are the Fourier coefficients of the \moire\ potential on the top layer of the PLG, 
 \begin{align}
	V_{\text{\moire}}(\vb r)=V_{\text{\moire}}(0)+\sum_j V_{\text{\moire}}(\vb g_j)e^{i\vb r\cdot\vb g_j}.
\end{align}
More details of the model are given in Appendix~\ref{App:BM model}.

The lattice mismatch and the twist angle between PLG and hBN cause a nonzero $\vb g_j$ and lead to a \moire\ Brillouin zone (MBZ), as shown in Fig.~\ref{Fig:Device}(b). 
We choose $\vb g_1,\vb g_2$ as the two \moire\ basis vectors in the momentum space, the length of which is denoted by $b_m$. 
For convenience, we can always decompose the momentum into two parts, $\vb k=\{\vb k\}+[\vb k]$, where $\{\vb k\}$ denotes the part in the first MBZ, and $[\vb k]$ denotes the corresponding \moire\ reciprocal lattice vector. 
We also introduce a parameter $\kappa$ to adjust the {\moire\ potential} manually. 
Specifically, we designate the $\kappa=1$ case as the \moire\ case and the $\kappa=0$ case as the pristine (or non-\moire) case. 
Note that in this work, we will only study the non-\moire\ case using the same set of mean-field order parameters as those in the \moire\ case. 
Other symmetry-breaking phases of the non-\moire\ case are irrelevant to this work.

Finally, we introduce the Coulomb interaction to the system, which is given by 
\begin{align}
	V_{\text{Coulomb}}=\frac{1}{2A}\sum_{\sigma,\sigma'}\sum_{\vb q\in\text{BZ}}
	V(q):\rho_{\vb q,\sigma}^\dag\rho_{\vb q,\sigma'}:,
\end{align}
where $A$ is the area of the system, $:\mathcal O:$ denotes the normal order of an operator, and $\rho_{\vb q,\sigma}=\sum_{\vb k\in\text{BZ}}\Psi_{\vb k+\vb q,\sigma}^\dag\Psi_{\vb k,\sigma}$. 
Here, we use the appropriate gate-screened Coulomb interaction $V(q)=\tanh(dq)/(2\epsilon\epsilon_0 q)$ for $q\neq 0$, where $d = \SI{30}{nm}$ is the distance between the sample and the gate. 
In all calculations, we take $\epsilon=5$ to compare with other works. 
In principle, a smaller (larger) $\epsilon$ will further enhance (suppress) the FQAH states. 
Furthermore, the $V(q=0)$ contribution is proportional to $N^2-N$, where $N$ is the total particle number, so we set $V(q=0)=0$ to remove the $N^2$-dependence of the energy. 
This choice is for convenience only and does not affect our results or conclusions.

\section{The Hartree-Fock approximation \label{Sec:HF}}

As the calculation of FQAH states is strongly limited by the capability of the ED calculation (because of the exponential growth of the interacting Hilbert space), one can typically include only one or two bands in the calculation at most due to computational limitations~\footnote{This is akin to ED calculations for continuum Landau level based strong field FQH physics where the ED is necessarily limited to one or two Landau levels.}. 
If a band in a multi-band system is well isolated from all the other bands,  one can then perform the ED calculation to the Hamiltonian projected to that isolated band, similar to the case of twisted bilayer MoTe$_2$. 
Unfortunately, PLG with a large displacement field does not possess such an isolated band. 
Nonetheless, several recent works~\cite{Dong2023,Dong2023a,Zhou2023} propose that the HF approximation can isolate an HF flat band from the other conduction bands. 
Then, the Hamiltonian is projected onto this HF band for the ED calculation. 

In the HF approximation, the interaction can be regarded as a linear functional of the one-body density matrix, that is, 
\begin{align}
	V_{\text{HF}}(P)=\sum_{\sigma}\sum_{\vb k,\vb q\in\text{BZ}} \Psi_{\vb k+\vb q,\sigma}^\dag [V_{\text{HF}}(P)]_{\sigma,\vb k+\vb q,\vb k} \Psi_{\vb k,\sigma}, 
\end{align}
where $[V_{\text{HF}}(P)]_{\sigma,\vb k+\vb q,\vb k}$ is given by
\begin{align}
	[V_{\text{HF}}(P)]_{\sigma,\vb k+\vb q,\vb k}&=\frac{V(q)}{A}
	\sum_{\vb k',\sigma'}\trace[P_{\sigma',\vb k',\vb k'+\vb q}]\nonumber\\
	&\quad -\frac{1}{A}\sum_{\vb k'}V(\abs*{\vb k-\vb k'})P_{\sigma,\vb k',\vb k'+\vb q}^\intercal. 
\end{align}
Here $P$ is the one-body density matrix of a Gaussian state, 
\begin{align}
P_{\sigma,\vb k,\vb k'}
= \delta_{\vb k,\vb k'}\mathbb{I}
- \delta_{\{\vb k-\vb k'\},0} 
\qty[\expval{\Psi_{\vb k',\sigma}\Psi_{\vb k,\sigma}^\dag}]^{^\intercal},
\end{align}
where $\mathbb{I}$ is the identity matrix. 
For both \moire\ and non-\moire\ cases, the only nonvanishing order parameters allowed are the correlations between $\vb k$ and $\vb k'$ that differ by a \moire\ reciprocal vector.

\subsection{Choice of the reference field}

Apart from the noninteracting continuum model and the Coulomb interaction introduced above, an additional noninteracting term must be included in the Hamiltonian for completeness. 
It originates from the fact that it is impossible to carry out the self-consistent calculation in the whole graphene Brillouin zone (BZ) due to the large number of momentum points involved. 
As a result, a finite momentum cutoff $\Lambda$ is introduced in the calculation, as shown in Fig.~\ref{Fig:Device}(b). 
Specifically, the interaction of a reference field $P^{\text{ref}}$ at the HF level is subtracted from the continuum model and leads to
\begin{align}\label{Eq:Htotal}
	H=H_s-V_{\text{HF}}(P^{\text{ref}})+V_{\text{Coulomb}}.
\end{align}
Note that $P^{\text{ref}}$ is defined within the momentum cutoff $\Lambda$, and thus the subtraction depends on $\Lambda$. 
At the HF level, the Hamiltonian for a density matrix $P$ then becomes
\begin{align}
	H_{\text{HF}}(P)=H_s+V_{\text{HF}}(\delta P),
\end{align}
where we define $\delta P=P-P^{\text{ref}}$, and thus, $H_s$ is exactly the Hamiltonian for $P=P_{\text{ref}}$ at the HF level. 
The choice of the reference field would obviously affect the results.

Depending on the nature of the noninteracting model, there are two ways to argue how a finite momentum cutoff results in an additional term. 
The physics is related to the short-range or the large-momentum part of the interaction. 
The first scenario arises if the continuum model describes the actual noninteracting Hamiltonian of the system, then the momentum cutoff implies that the short-range part of the interaction ($V(q)$ at large $q$) is not included in the calculation. 
Consequently, one should incorporate the short-range effects beyond the cutoff, which can be considered as a noninteracting term at the HF level. 
However, we emphasize that as the short-range (i.e., atomistic) part of the interaction sensitively depends on the lattice structure and the screening from the ionic cores, it will be drastically different from the gate-screened Coulomb interaction and can only be obtained through detailed ab initio calculations (and is generally unknown).  
The second scenario is that the continuum model is not the pristine noninteracting Hamiltonian but an effective model that has already included the interaction. 
This situation may occur in the density functional theory calculations or when fitting an effective model to experimental data.
To avoid double counting the interaction within the cutoff, one must subtract a reference field in the HF calculation as we do.  

{Having concluded that a proper reference of density matrix is necessary} in the HF calculation, we now discuss the choice of the reference field $P^{\text{ref}}$. 
Unfortunately, {there has yet a} consensus on the choice of the reference field in the literature, and some empirical choices of the reference field are 
\begin{itemize}
	\item \textbf{The charge neutrality scheme (A)}: $P^{\text{ref}}=P^{\kappa=1}$ for the \moire\ case and $P^{\text{ref}}=P^{\kappa=0}$ for the non-\moire\ case, where $P^{\kappa=1}$ and $P^{\kappa=0}$ are respectively the noninteracting ground state of the \moire\ case and the non-\moire\ case at the charge neutrality point (CNP);
	\item \textbf{The non-\moire\ scheme for the \moire\ case (B)}: In this case one chooses $P^{\text{ref}}=P^{\kappa=0}$;
	\item \textbf{The infinite-temperature scheme (C)}: $P^{\text{ref}}=\mathbb{I}/2$, the infinite-temperature reference field.
\end{itemize}
The large-momentum cutoff is valid only if the density matrix at large momentum has only a small correction at low energies. 
Some works~\cite{Dong2023,Dong2023a,Zhou2023,Kwan2023} introduced a band cutoff by projecting the Hamiltonian to a few low-energy bands;
one should also check the convergence with respect to the band cutoff. 
In the following, we will perform the HF calculations using the plane wave basis and include all the \moire\ states within the momentum cutoff~\cite{Xie2020}.

\subsection{Divergence for reference fields B and C}

The divergence with respect to the momentum cutoff can be more readily understood at the CNP. 
The reason is that a large displacement-field-induced band gap of PLG at the CNP results in an unambiguous HF ground state so that interaction and \moire\ potentials can be considered as perturbations. 
In what follows, we will treat the true HF ground state as a perturbation around the noninteracting ground state. 
Specifically, we investigate two blocks of the HF interaction $V_{\text{HF}}(\delta P)$ matrix: the $\vb k=0$ diagonal block and the off-diagonal block between $\vb k=\vb g_1$ and $\vb k'=0$.
Note that the HF interaction is exactly zero for reference field A at the CNP because $P = P^\text{ref}$ in this particular case, ensuring trivial convergence. 
Therefore, this section only discusses the \moire/non-\moire\ cases with reference field C and the \moire\ cases with reference field B.

\begin{figure}[t]
\center
\includegraphics[width=0.9\columnwidth]{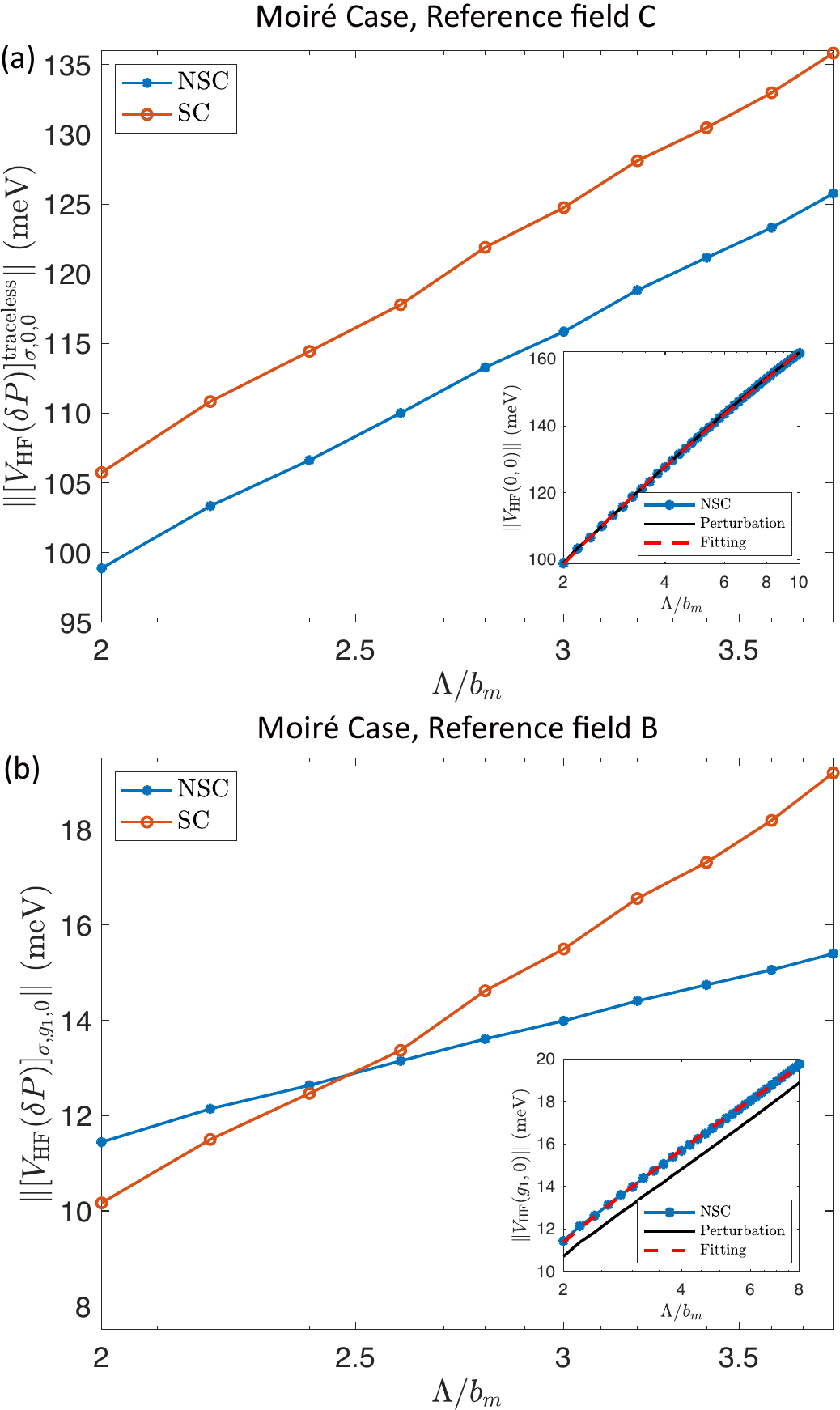}
\caption{\label{Fig:Converge}
A plot of how the HF interaction $V_{\text{HF}}(\delta P)$ depends on the momentum cut-off $\Lambda$.
(a) The traceless part of the $\vb k=0$ diagonal block of the HF interaction for reference field C. 
(b) The off-diagonal block of the HF interaction between $k=\vb g_1$ and $k'=0$ for reference field B. 
Note that the horizontal axis is in logarithmic scale in all panels.
Note that the operator norm is precisely zero for reference field A. 
In the main panels, the blue lines represent the non-self-consistent (NSC) result using the one-body density matrix of the noninteracting ground state. 
The orange lines represent the self-consistent (SC) result using the self-consistent HF ground state. 
The insets show the NSC (blue lines) and perturbation (black lines) results for much larger cutoffs, and the red dashed lines represent a fitting of the NSC results to $a+b[\ln (\Lambda/b_m)]^c$ with $c=0.8$. 
Here, all calculations are performed in the \moire\ case on a $6\times 6$ mesh in the MBZ at the CNP, and we take the twist angle to be $0.77^\circ$ and $u_d=\SI{30}{meV}$.
}
\end{figure}

\subsubsection{Divergence for reference field C}
We first theoretically and numerically analyze the \moire\ case with reference field C and argue that the same phenomenon also happens in the non-\moire\ case.
As the reference field cannot cancel the diagonal part of the density matrix, the leading order appears on the $\vb k=0$ diagonal block of the HF interaction. 
The trace of the diagonal blocks introduces a slowly varying momentum-space potential, which shifts the energy of the state but does not significantly alter the wave function. 
By contrast, the traceless part of the diagonal block $[V_{\text{HF}}(\delta P)]_{\sigma,0,0}^{\text{traceless}}$ can profoundly affect the wave function, and thus we only focus on the traceless part hereafter.
Because the ground state at CNP is gapped, the \moire\ potential can also be regarded as a perturbation, and the effect of the \moire\ potential on the diagonal block appears as a second-order perturbation. 
Hence, we have
\begin{align}\label{Eq:diagonal}
	&[V_{\text{HF}}(\delta P^{\text{SC}})]_{\sigma,0,0}^{\text{traceless}}
	\approx [V_{\text{HF}}(\delta P^{\text{NSC}})]_{\sigma,0,0}^{\text{traceless}}\nonumber\\
	&\approx [V_{\text{HF}}(\delta P^{\kappa=0})]_{\sigma,0,0}^{\text{traceless}}\nonumber\\
	&=-\int_{0}^{\Lambda}\frac{\dd k}{(2\pi)^2}kV(k)\int_{0}^{2\pi}(P^{{\kappa=0}}_{\sigma,\vb k,\vb k}-\mathbb{I}/2)\dd \theta,
\end{align}
where the non-self-consistent (NSC) result is the noninteracting ground state of the \moire\ case, $P^{\text{NSC}}=P^{\kappa=1}$.
A naive estimation would be $\norm{[V_{\text{HF}}(\delta P^{\text{SC}})]_{\sigma,0,0}^{\text{traceless}}}\sim O(\Lambda)$ considering that $\norm{P^{\kappa=0}_{\sigma,\vb k,\vb k}-\mathbb{I}/2}=1/2$. 
However, $P^{\kappa=0}_{\sigma,\vb k,\vb k}$ cancels each other in the first integration because of the $C_3$ rotational symmetry, and we actually have 
\begin{align}
	\int_{0}^{2\pi} (P^{{\kappa=0}}_{\sigma,\vb k,\vb k}-\mathbb{I}/2)\dd \theta\sim O(1/k). 
\end{align}
Hence, the leading-order perturbation suggests a logarithmic divergence with the momentum cutoff $\Lambda$, as shown in the inset of Fig.~\ref{Fig:Converge}(a). 
The argument is further validated by the excellent agreement between the NSC result and the leading-order perturbation result in the inset of Fig.~\ref{Fig:Converge}(a).  
For the self-consistent (SC) solution, the difference between the SC and NSC results is on the order of $O(1)$ for all numerically accessible cutoffs ($\Lambda/b_m\leq 3.8$), and we anticipate that the diagonal logarithmic divergence also happens in the SC result. 
We emphasize that this argument applies to both \moire\ and non-\moire\ cases and is robust against changes in the details of the \moire\ potential. 
This robustness arises because the leading-order perturbation on the diagonal block is given by the pristine PLG and independent of the \moire\ potential. 
Therefore, a more accurate high-energy correction to the continuum model will not remedy the divergence. 
However, the short-range part of the interaction can modify $V(q)$ at large $q$ and potentially remove the divergence. 

\subsubsection{Divergence for reference field B}
For the \moire\ case with reference field B, the reference field cancels the leading-order contributions from the diagonal part of the density matrix. 
Thus, the dominant contribution comes from the off-diagonal part. 
Although an accurate estimation of the correction from the interaction is difficult, the leading-order correction from the \moire\ potential can be calculated analytically from the first-order perturbation theory. 
Specifically, up to the leading order in the \moire\ potential, the ground state is approximated by
\begin{align}
	\ket{\psi^{\kappa=1}}&\approx\ket{\psi^{\kappa=0}}\\
	+&\sum_{\vb k,j}\sum_{\alpha,\beta} \frac{[V_{\text{\moire}}]_{c,\vb k+\vb g_j, \beta; v,\vb k, \alpha}   }{E_{v,\vb k, \alpha}-E_{c,\vb k+\vb g_j,\beta}}
	 c^\dag_{c,\vb k+\vb g_j,\beta}c_{v,\vb k,\alpha}\ket{\psi^{\kappa=0}},\nonumber
\end{align}
where $c_{c,\vb k, \alpha}$ and $c_{v,\vb k, \alpha}$ are respectively the fermion operators of the conduction and valence band of the non-\moire\ Hamiltonian with energies $E_{c,\vb k, \alpha}, E_{v,\vb k, \alpha}$, and $\ket{\psi^{\kappa=1}},\ket{\psi^{\kappa=0}}$ are the noninteracting ground state of the \moire\ and non-\moire\ cases, respectively.
In addition, $[V_{\text{\moire}}]_{v,\vb k+\vb g_j, \beta; c,\vb k, \alpha}$ denotes the matrix element of the \moire\ potential in the eigenstate basis of the non-\moire\ case. 
Then, the first-order correction to the one-body density matrix can be written as 
\begin{align*}
	&\mel{\psi^{\kappa=1}}{c^\dag_{c,\vb k+\vb g_1,\beta}c_{v,\vb k,\alpha}}{\psi^{\kappa=1}}\approx\frac{[V_{\text{\moire}}]_{c,\vb k+\vb g_1, \beta; v,\vb k, \alpha}^*}{E_{v,\vb k, \alpha}-E_{c,\vb k+\vb g_1,\beta}},\\
	&\mel{\psi^{\kappa=1}}{c^\dag_{v,\vb k+\vb g_1,\alpha}c_{c,\vb k,\beta}}{\psi^{\kappa=1}}\approx\frac{[V_{\text{\moire}}]_{c,\vb k, \beta; v,\vb k+\vb g_1, \alpha}   }{E_{v,\vb k+\vb g_1, \alpha}-E_{c,\vb k,\beta}}.
\end{align*}
Consequently, we estimate that
\begin{align}
	\delta P^{\text{NSC}}_{\sigma,\vb k +\vb g_1, \vb k}\sim V_{\text{\moire}}(\vb g_1)/\delta E_{cv}(\vb k)\sim O(1/k),
\end{align}
where $\delta E_{cv}(\vb k)$ is the energy gap between the valence and conduction bands in the non-\moire\ case.  
Moreover, because $\vb g_1$ becomes $\vb g_2$ under a $C_3$ rotation, the off-diagonal blocks of the density matrix cannot cancel one another as the diagonal blocks do in Eq.~\eqref{Eq:diagonal}. 
Hence, the off-diagonal parts of the HF interaction possess a logarithmic divergence, as numerically verified in the inset of Fig.~\ref{Fig:Converge}(b). 
Also, we numerically show that the difference between the NSC result and the leading-order perturbation is just on the order of $O(1)$, verifying the legitimacy of the first-order perturbation. 
However, the agreement between the SC result and the NSC result for reference field B is not as good as that for reference field C. 
In fact, we find that the SC result is even more divergent than the NSC result for $\Lambda/b_m\leq 3.8$, implying that the interaction may also contribute to a logarithmic divergence. 
Note that, unlike reference C, the divergence here is sensitive to the details of the \moire\ potential because it comes from the first-order perturbation. 
Therefore, both corrections to the high-energy part of the continuum model and the short-range part of the interaction can affect the result.

\subsection{Convergence for reference field A}

\begin{figure}[t]
\center
\includegraphics[width=0.9\columnwidth]{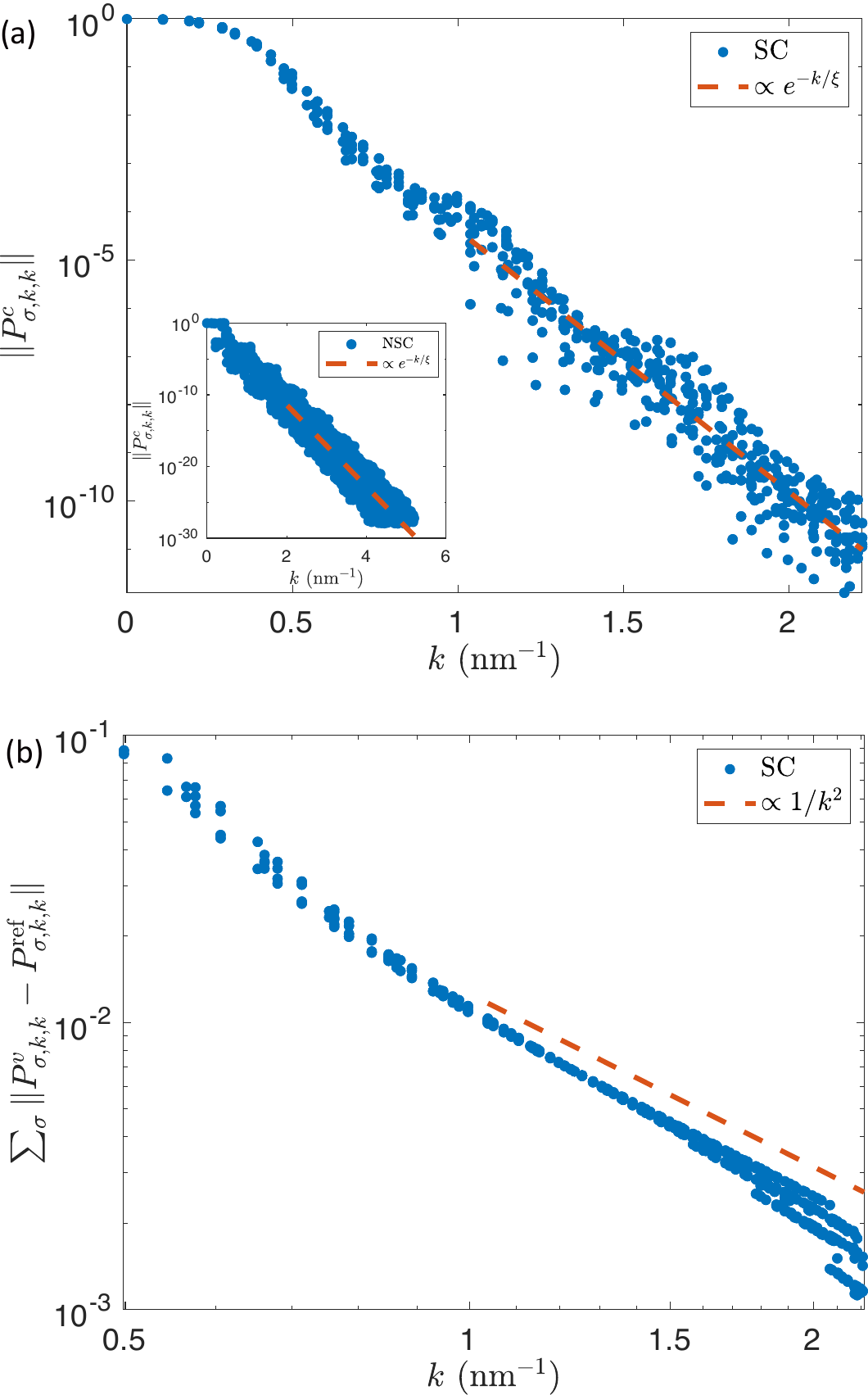}
\caption{\label{Fig:Density}
(a) Operator norm of the one-body density matrix of the lowest conduction $\norm{P_{\sigma,\vb k,\vb k}^c}$. 
The main panel shows the result of the SC solution, while the inset shows the NSC result. The red dashed lines represent the filling to $\propto e^{-k/\xi}$ with $\xi=0.076\,\text{nm}^{-1}$ for the NSC result and $\xi=0.080\,\text{nm}^{-1}$ for the SC result. (b) Operator norm of the difference between the SC valence bands and the reference field $\sum_\sigma\norm{P_{\sigma,\vb k,\vb k}^v-P_{\sigma,\vb k,\vb k}^{\text{ref}}}$. The red dashed line is proportional to $1/k^2$. 
In both panels, we take $u_d=\SI{50}{meV}$ and $\theta=0.77^\circ$, and all calculations are performed in the \moire\ case with reference field A on a $6\times 6$ mesh in the MBZ at filling factor $\nu=1$.
}
\end{figure}

In the previous section, the divergence for reference fields B and C is discussed, while the \moire/non-\moire\ cases with reference field A are trivial because there is a fine-tuned cancellation between $P$ and $P^{\text{ref}}$ at the CNP. 
Hence, one cannot immediately conclude the convergence for the two cases at a finite filling from their trivial convergence at the CNP. 
Nonetheless, arguments similar to those in the previous section can also be applied to finite fillings, and we will show in this section that the HF calculation is indeed convergent at any finite fillings for the two cases. 
In the following, we focus on the \moire\ case with reference field A because the same argument also applies to the non-\moire\ case.

At a finite filling, the one-body density matrix can be separated into the contribution from the filled HF conduction band and that from all HF valence bands, i.e., $P=P^c+P^v$. 
Therefore, the HF interaction becomes
\begin{align}
	V_{\text{HF}}(\delta P)=V_{\text{HF}}(P^c)+V_{\text{HF}}(P^v-P^{\text{ref}}),
\end{align}
where the first term represents the interaction of the conduction band, and the second term is no longer trivial because the interaction of the conduction band will excite the valence bands. 
Though the reference field does not cancel the interaction of the conduction band, the conduction band actually leads to an exponential convergence, in sharp contrast to the filled valence bands. 
The essential difference is that all valence bands are filled, so the wave function is spread over the whole BZ. 
In contrast, only a finite number of conduction bands are filled at a finite filling, resulting in a highly localized wave function concentrated around the low-energy region of the BZ.  
In Fig.~\ref{Fig:Device}(b), we calculate the density in the momentum space of the HF conduction band of the self-consistent solution at the filling factor $\nu=1$. 
The density forms a plateau with $\expval{\Psi_{\vb k}^\dag\Psi_{\vb k}}\approx 1$ in the first MBZ, but decays rapidly outside the first MBZ, as expected. 
Additionally, we note that the noninteracting conduction bands are exponentially localized on rings centered at $\vb k=0$ because of the confinement of the PLG kinetic energy. 
In the inset of Fig.~\ref{Fig:Density}(a), we numerically calculate the operator norm of $P_{\sigma,\vb k,\vb k}$ of the lowest conduction band, and extract its localization ``length'' in the momentum space $\xi=\SI{0.076}{nm^{-1}}$ by fitting the momentum-space density to $\propto e^{-k/\xi}$. 
Thus, the wave function is rather localized, considering that the typical ``length'' scale is $b_m=0.65\,\text{nm}^{-1}$. 
As the kinetic energy dominates at large $k$, we anticipate that the spreading of the HF conduction bands is also exponentially hindered by the kinetic energy. 
In the main panel of Fig.~\ref{Fig:Density}(a), we numerically verify the exponential localization of the lowest HF conduction band at filling factor $\nu=1$. 
Moreover, the localization ``length'' of the HF band is $\xi=0.080\,\text{nm}^{-1}$, very close to the noninteracting result, which further confirms the dominance of the kinetic energy at large $k$. Hence, we conclude that $V_{\text{HF}}(P^c)$ is always exponentially convergent.

As for the excitation of the valence bands $P^v-P^{\text{ref}}$, we again adopt the first-order perturbation theory by regarding the interaction between the filled conduction band and the valence bands at large $k$ as a perturbation. As the conduction band is localized around $k=0$, we estimate that
\begin{align}
	P^v_{\sigma,\vb k,\vb k}-P^{\text{ref}}_{\sigma,\vb k,\vb k}\sim V(k)/\delta E_{cv}(\vb k)\sim O(1/k^2),
\end{align}
as numerically verified in Fig.~\ref{Fig:Density}(b). 
Consequently, we conclude that $V_{\text{HF}}(P^v-P^{\text{ref}})$ in the low-energy region should converge in the same way as $O(1/\Lambda)$.

To conclude, we have shown in this section that the HF calculation is convergent at any finite filling for the \moire\ case with reference field A, and the same argument also applies to the non-\moire\ case. 
Because of the divergence for reference fields B and C, all of our following calculations will only be based on reference field A.
We rule out B and C in favor of A for reasons described in depth above.

\section{Search for the FQAH states in PLG \label{Sec:FQAH}}

Having established the convergence of the HF calculation with reference field A, we now search for the FQAH states in the PLG. 
We begin by noting that the calculation of the FQAH ground state is much more convoluted in the PLG than that in the toy models~\cite{Tang2011,Sun2011,Neupert2011,Regnault2011,Sheng2011, Wu2012,Liu2012}, twisted bilayer graphene~\cite{Repellin2020}, twisted double bilayer graphene~\cite{Liu2021}, and twisted bilayer MoTe${_2}$~\cite{Li2021,Wang2024}, where an isolated flat band exists at the noninteracting level. 
Nonetheless, the two kinds of calculations can be reconciled from a variational perspective. 
As ED always necessitates a subspace, the results should be considered a variational solution dependent on the subspace.  
A natural ansatz for such a subspace is
\begin{align}\label{Eq:Ansatz}
	\mathcal P=\text{span}\qty{\prod_{i=1}^{N} f_{c,\vb k_i}^\dag\prod_{\alpha,\vb k}{f_{v,\alpha,\vb k}^\dag}\ket{0}: \vb k_i\in\text{MBZ}}.
\end{align}
Here, $f_{c,\vb k_i}$ and $f_{v,\alpha,\vb k}$ can be respectively regarded as the orbital of the partially filled conduction band and the completely filled valence bands. 
The projected Hamiltonian is thus given by
\begin{align}
	\mathcal P H \mathcal P&=E_{v}+\sum_{\vb k}\varepsilon(\vb k) f_{c,\vb k}^\dag f_{c,\vb k}+\sum_{\vb k}\varepsilon_{cv}(\vb k) f_{c,\vb k}^\dag f_{c,\vb k}\nonumber\\
	&\quad +\sum_{\vb k_1, \vb k_2, \vb q}V_{\vb k_1,\vb k_2,\vb q}
	f_{c,\vb k_1 -\vb q}^\dag f_{c,\vb k_2+\vb q}^\dag f_{c,\vb k_2}f_{c,\vb k_1}.
\end{align}
Here, the first term is the total energy of the filled valence bands, given by $E_v=\mel{\psi_v}{H}{\psi_v}$ with $\ket{\psi_v}=\prod_{\alpha,\vb k}{f_{v,\alpha,\vb k}^\dag}\ket{0}$, which can be readily calculated by Wick's theorem.
The second term is the kinetic energy of the conduction band, given by 
\begin{align}
	\varepsilon(\vb k)=\mel{0}{f_{c,\vb k}[H_s-V_{\text{HF}}(P^{\text{ref}})]f_{c,\vb k}^\dag}{0}. 
\end{align} 
The third term represents the inter-band interaction between the valence bands and the conduction band, which becomes a noninteracting term in this subspace. 
Such an energy is explicitly given by
\begin{align}
	\varepsilon_{cv}(\vb k)=\sum_{\alpha,\vb p}\mel{0}{f_{v,\alpha,\vb p}f_{c,\vb k}V_{\text{Coulomb}}f^\dag_{c,\vb k}f^\dag_{v,\alpha,\vb p}}{0}.
\end{align}
The last term is the intra-band interaction of conduction-band electrons, given by
\begin{align}
	V_{\vb k_1,\vb k_2,\vb q}=\frac14 \mel{0}{f_{c,\vb k_2+\vb q}f_{c,\vb k_1-\vb q}V_{\text{Coulomb}}f^\dag_{c,\vb k_1}f^\dag_{c,\vb k_2}}{0}.
\end{align}
Note that $\varepsilon(\vb k)+\varepsilon_{cv}(\vb k)$ is different from the HF band energy because the HF band energy also includes the intra-band interaction within the conduction band. Hence, one should not use the HF band energy to avoid double counting the intra-band interaction.

\subsection{Three methods to search for the FQAH states in PLG} 
Having discussed the framework to search for the FQAH state in PLG, we discuss an important issue in the calculation.
In particular, finding the global energy minimum through a complete variational calculation is numerically inaccessible. 
Instead, one can only compare the ground state energies in select subspaces and choose the best candidate. 
In the presence of an isolated noninteracting conduction band, choosing the orbital of the noninteracting conduction and valance bands is reasonable. 
However, ambiguity appears in the HF calculation of the PLG. 
Therefore, we propose three methods to conduct the HF calculation and search for the FQAH states in PLG.

The first method (Method~I) uses the HF band at filling factor $\nu=1$ as the subspace to carry out the ED calculations at fractional fillings, as was done in several recent works~\cite{Dong2023,Dong2023a,Zhou2023}. 
However, the HF band energy and wave functions at a fractional filling (i.e., $\nu = 2/3$) may deviate from those at an integer filling because a fractional filling is usually not a few-particle excitation away from an integer filling.  

The second method (Method~II) uses the HF bands from the unconstrained HF calculation directly at the desired fractional filling $\nu$. 
Here, ``unconstrained HF'' refers to the fact that the HF calculation finds the Slater determinant that minimizes the HF energy. 
There is also a downside to this method, however. 
An unconstrained HF calculation will always result in a ground state resembling a Fermi liquid (FL). 
In other words, the ground state will be characterized by a Fermi surface in the lowest conduction band, where orbitals below the Fermi surface are fully occupied, and those above it are completely empty (i.e., a discontinuity at the Fermi surface).
In contrast, the FQAH ground state is non-Fermi-liquid (NFL) like and generally tends to occupy each noninteracting orbital evenly because the interaction dominates the kinetic energy and has no preference over any orbitals~\cite{Abouelkomsan2020}. 
Therefore, it is not clear whether the unconstrained HF calculation can capture the NFL FQAH ground state.

Inspired by the uniform occupation number of the FQAH states, we propose a third choice for the HF calculation (Method~III). 
Instead of using the unconstrained HF, we perform a constrained HF in which all orbitals in the lowest HF conduction band are occupied but only at the desired fractional filling $\nu$. 
Meanwhile, all HF valence bands are still fully occupied. 
One might worry that this constrained HF calculation treats the lowest conduction band differently from the other bands. 
However, in Appendix~\ref{App:HF}, we prove that this constraint is equivalent to requiring the Gaussian state to be spin-polarized with a uniform density in the MBZ relative to the CNP. 
Therefore, our constraint does not give preferential treatment to any specific band. 
In addition, the constraint guarantees that the self-consistent solution produces an exactly uniform occupation number for all noninteracting orbitals, as applied to FQAH states. 
This method is equivalent to treating the inter-band interaction between the valence and conduction bands at the HF level but treating the intra-band interaction within the lowest conduction band at the ED level.

Here, we summarize the three different methods as follows: 
\begin{itemize}
	\item \textbf{Method~I}: To conduct the unconstrained HF calculation at filling factor $1$, and use the obtained first HF quasiparticle conduction band to perform the ED at a fractional filling $\nu$;
	\item \textbf{Method~II}: To conduct the unconstrained HF calculation at fractional filling $\nu$ and use this partially filled conduction band to perform the ED at the same filling;
	\item \textbf{Method~III}: To conduct the constrained HF calculation at fractional filling $\nu$ and use this partially but uniformly filled conduction band to perform the ED at the same filling. 
\end{itemize}
The rationale for the HF calculation here is to generate the basis that should be used to construct the subspace for the ED calculation. 
This subspace is composed of completely filled valence bands, and only one partially filled conduction band, and the HF bands need to be identified with the valence bands and the unique conduction band used in Eq.~\eqref{Eq:Ansatz}. 
Hence, the three methods are viable only if there is only one partially filled band in the HF calculation, while all the other bands are either completely filled or completely empty.
Method~III is always viable because of this constraint. 
However, Methods~I and~II may not be viable if there are more than one partially filled bands. 
In what follows, we will produce some numerical results to compare the three methods and then use them to search for the FQAH states in the PLG. 
Note that because we only use reference field A, the divergence issue is not a concern in the following calculations. 
In addition, all of our results will be labeled as AII, AIII, etc., to indicate the reference field and the method used.

\begin{figure*}[!]
	\center
	\includegraphics[width=0.9\textwidth]{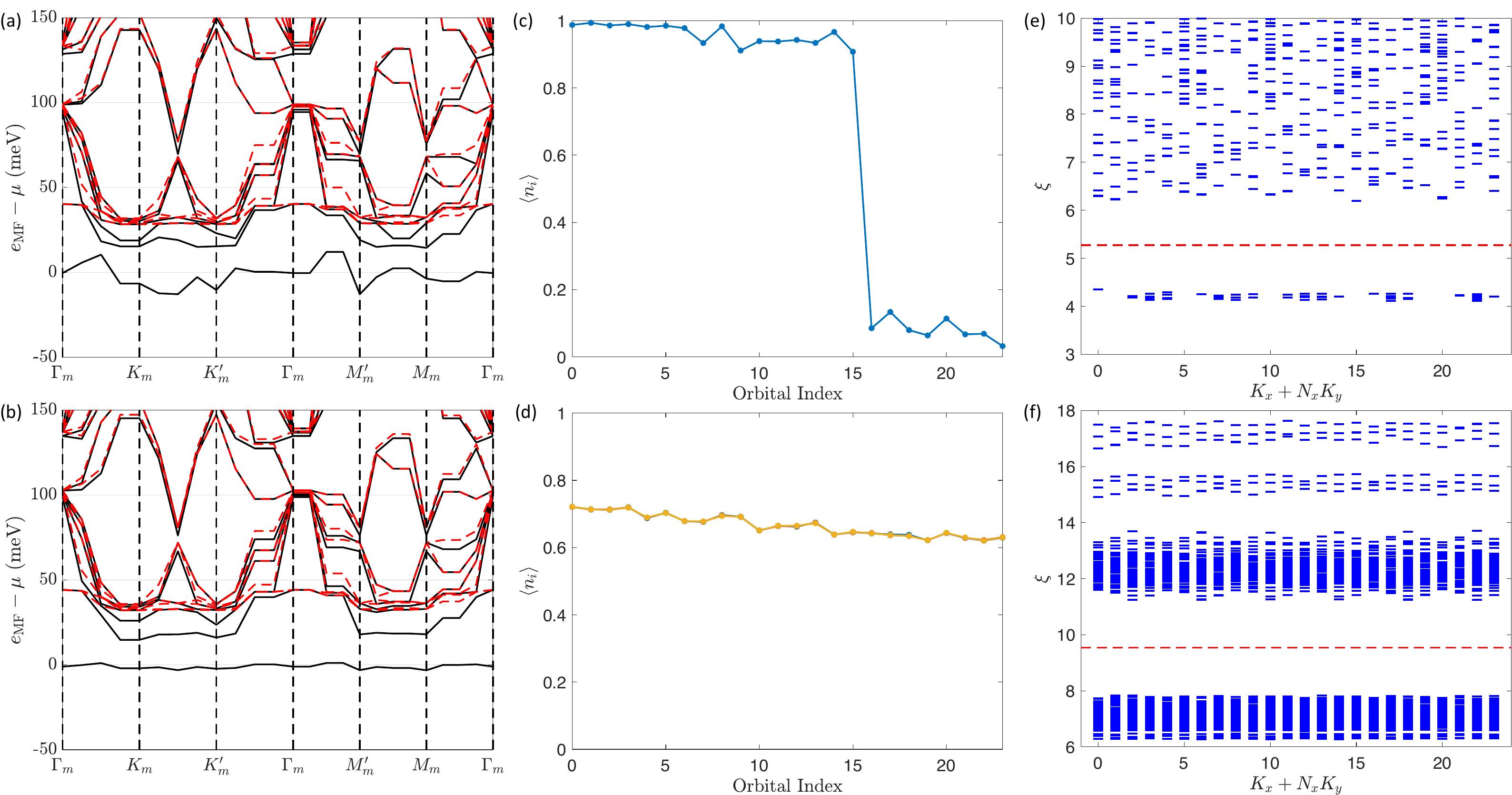}
	\caption{\label{Fig:2/3_spec}
	Comparison between AII (top row) and AIII (bottom row) in the \moire\ case on a $4\times 6$ mesh in the MBZ at $2/3$ filling. 
	(a,b) HF band structure obtained using methods AII and AIII. 
	The black and red lines represent the spin-up and spin-down sectors, respectively. 
	(c,d) Occupation number of the ground states using methods AII and AIII. The indices of the orbitals are ordered according to their HF energies. 
	Note that we plot the occupation number of all three-fold degenerate ground states for AIII in (d). (e) Particle entanglement spectrum with a subsystem of three holes for the particle-hole conjugate of AII. There are $56$ states below the entanglement gap, corresponding to the quasihole excitations (8 choose 3) of Fermi Liquid (FL). (f) Particle entanglement spectrum with a subsystem of three holes for the particle-hole conjugate of AIII. 
	There are $1088$ states below the entanglement gap, corresponding to the (1,3)-permissible excitations of FQAH states. 
Here, the twist angle is $0.77^\circ$, and we take $u_d=\SI{50}{meV}$.
	The ground state energy of AIII is lower than that of AII, as shown in Fig.~\ref{Fig:2/3_e}.
	}
\end{figure*}

\subsection{Comparison between Methods II and III}

In this section, we first compare Methods II and III. 
In Fig.~\ref{Fig:2/3_spec}, we numerically use Method II and III in the \moire\ case on a $N_{x}\times N_y=4\times6$ mesh in the MBZ to search for the ground state at $\nu=2/3$ filling. 
We find that the unconstrained HF produces a much more dispersive conduction band than the constrained HF, as shown in Fig.~\ref{Fig:2/3_spec}(a) and~\ref{Fig:2/3_spec}(b). 
This difference arises because the unconstrained HF can lower the occupied orbitals' energies by sacrificing the unoccupied orbitals' energies. 
In contrast, the constraint HF forces all orbitals of the lowest HF conduction band to be uniformly occupied. 
This more dispersive band, in turn, favors an FL ground state in the ED calculation, while the flat band from the constrained HF makes an NFL FQAH ground state possible. 

The expectation is numerically verified in Fig.~\ref{Fig:2/3_spec}(c-f). 
Indeed, Method II leads to an apparent discontinuity of the density at the Fermi surface (16 particles out of 24 orbitals). 
In contrast, Method III leads to an almost uniform occupation of all HF orbitals. 
These results also confirm that the HF calculation and the ED calculation in Method III are, to some extent, consistent. 
We can further use the particle entanglement spectrum (PES) to distinguish the FL state from the FQAH state. 
Specifically, we partition $N$ particles into two sets, A and B, with $N_A$ and $N_B$ particles in each set. 
The reduced density matrix of a state $\rho$ is defined by $\rho_\text{A}=\trace_{\text{B}}{\rho}$, where $\trace_{\text{B}}$ is the partial trace over set B. 
If we denote the eigenvalues of $\rho_{\text{A}}$ as $\lambda_j$, the resulting PES is defined as $\xi_j=-\ln \lambda_j$. 
Note that as the ground state respects the translational symmetry, $\rho_{\text{A}}$ also respects the translational symmetry. 
Thus, the PES can be evaluated in each momentum sector. 
We take $N_A=3$ and calculate the PES of the particle-hole conjugate of the Method II and III results at the $\nu = 2/3$ filling. 
For Method II, we take $\rho=\dyad{\psi_{\text{FL}}}$, where $\ket{\psi_{\text{FL}}}$ is the unique ground state, which is a FL. 
For Method III, we take 
\begin{align}
	\rho=\dfrac{1}{3}\sum_{i=1}^3\dyad{\psi_{\text{FQAH}},i},
\end{align}
where $\ket{\psi_{\text{FQAH}},i}$ are the three-fold degenerate ground states. 
As the PES calculates the quasihole excitations, the PES of an FL state should have $\binom{N}{N_A}$ states below the entanglement gap, as verified in Fig.~\ref{Fig:2/3_spec}(e). 
Note that we take $N=8$ here because of the particle-hole conjugate. 
By contrast, the PES of an FQAH state should follow the $(1,3)$-permissible excitations of FQAH states~\cite{Regnault2011}, as verified in Fig.~\ref{Fig:2/3_spec}(f). 
Similar results are also obtained at the $\nu=3/5$ filling (see Appendix~\ref{App:3/5}).

\subsection{Comparison between Methods I and III}\label{Sec:Compare13}

In this section, we discuss the difference and connection between Methods I and III, both of which can produce FQAH ground states. 
The essential difference lies in their different HF methods, and we will demonstrate that {for certain parameters,} the unconstrained and constrained HF solutions can be considered two solutions in the same phase that are connected adiabatically.

If the unconstrained HF has an insulating ground state at filling factor $1$, then the constrained HF at the same filling is equivalent to the unconstrained one. 
However, at a smaller filling $0<\nu<1$, the constrained HF has an additional factor $\nu$ for the lowest conduction band. 
Effectively, only the lowest band is occupied for reference field A because the reference field approximately cancels the contributions from the valence bands.
The additional factor $\nu$ is approximately an overall factor in front of $V_{\text{HF}}(\delta P)$ because $V_{\text{HF}}$ is a linear functional. 
We anticipate that the constrained HF at filling factor $0< \nu <1$ with a full interaction strength $V(k)$ should be similar to the unconstrained HF at filling factor $1$ with interaction strength reduced to $\nu V(k)$. 
Consequently, the unconstrained and constrained HF results should belong to the same phase if the reduction of the interaction strength does not result in a phase transition. To validate the argument, we propose a fourth method:
\begin{itemize}
	\item \textbf{Method~IV}: To conduct the unconstrained HF calculation at filling factor 1 but with the full interaction strength $V(k)$ replaced by $\nu V(k)$ and use the lowest conduction band to do the ED at filling factor $\nu$ with the full interaction strength $V(k)$. 
\end{itemize}

\begin{figure*}[t]
	\center
	\includegraphics[width=1\textwidth]{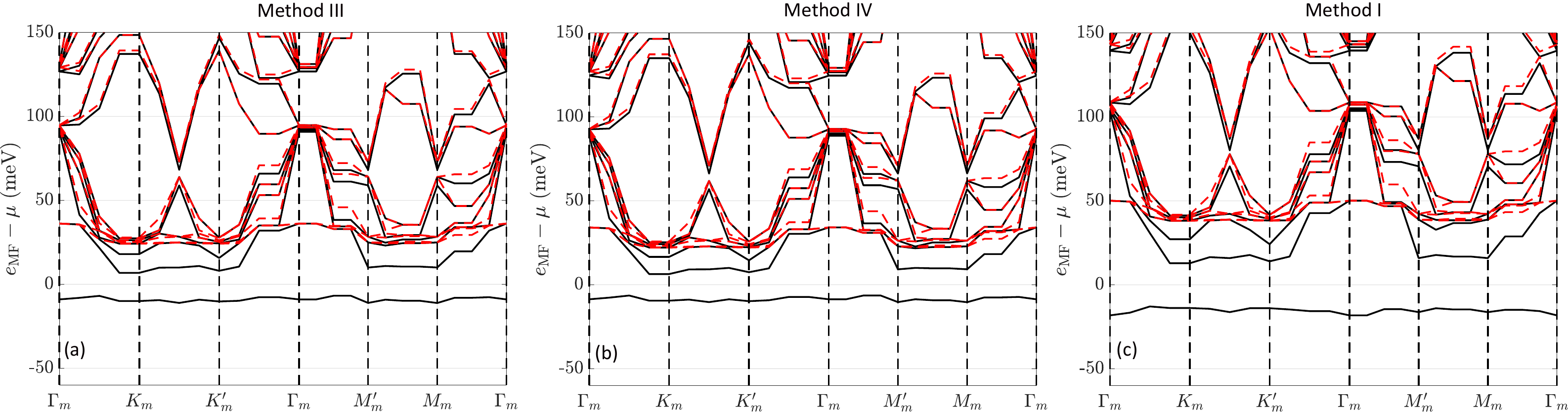}
	\caption{\label{Fig:compare}
	HF band structure of (a) Method III, (b) Method IV, and (c) Method I. Here, we calculate the \moire\ case with reference field A at $2/3$ filling on a $N_x\times N_y=4\times6$ mesh in the MBZ.  In addition, we take $u_d=\SI{50}{meV}$ and $\theta=0.77^\circ$.
	}
\end{figure*}

First, we calculate the HF band structure of Methods III, IV, and I with $\theta=0.77^\circ$ and $u_d=\SI{50}{meV}$ in Fig.~\ref{Fig:compare}, in search for an FQAH state at $\nu = 2/3$. 
The HF bands from Methods III and IV are quite similar, while the bands from Method I show noticeable differences, including the band gap and the dispersion of the lowest conduction band. 
However, all three solutions exhibit a conduction band with Chern number $C=1$, indicating they belong to the same phase. 
Second, all three methods produce FQAH states but with different energies. 
Specifically, if we set the ground-state energy of Method III as $E_{\text{Method III}}= 0$, then the ground-state energies of the other two methods are
\begin{align}
	E_{\text{Method IV}}= 1.30\,\text{meV}, \quad 
	E_{\text{Method I}}= 13.99\,\text{meV}. 
\end{align}
Therefore, while Method III yields the lowest energy, the result from Method IV is only slightly higher.
In contrast, the energy produced by Method I is significantly higher than the other two.  
We can also examine the many-body wave function overlap among the three solutions, which is given by the following matrix: 
\begin{align}
	\Xi= 
	\begin{pmatrix}
		1 & 0.9957 & 0.9418 \\
		0.9957 & 1 & 0.9118 \\
		0.9418 & 0.9118 & 1
	\end{pmatrix}, 
\end{align}
where the matrix element $\Xi_{ij}$ is the overlap between Method $i$ and Method $j$ with $i,j=1,2,3$.
Note that we use the numerical label $i = 1, 2, 3$ to represent Method III, IV, and I, respectively. 
Here, we show the overlap in the zero-momentum sector, and the overlaps in the three momentum sectors are the same up to the third decimal place.
We thus find that the HF wave functions produced by  Methods III and IV are very close to each other and have a high overlap. 
Meanwhile, the HF wave functions produced by Methods III and I are also quite similar, although the overlap is smaller. 
However, the overlap is still above $0.9$, suggesting that they are still adiabatically connected. 

We also note that the unconstrained and constrained HF calculations lead to the same phase for the parameters and filling factors used in this section. 
However, in principle, they can produce ground states in different phases for other parameters or filling factors.
As the PLG has a gapped interacting ground state at filling factor $1$ but a gapless noninteracting ground state, there must be a phase transition for a specific interaction strength. 
Consequently, the constrained HF should experience a phase transition at some filling factor. 
There could be two kinds of phase transitions, one of which is a transition from one insulating phase to another insulating phase. 
In this scenario, Method III opens up the opportunity for gapped states other than an FQAH state. 
The other kind of transition is from an insulating phase to a metallic phase, and the global gap between the occupied bands and the vacant bands vanishes. 
The absence of a sizable global band gap renders the projection ansatz questionable.
Hence, the result of Method III becomes unreliable, although Method III is still viable. 
We note that Method I is also viable but unreliable at small fillings because the HF conduction band is expected to be drastically renormalized.
The PLG is likely to belong to the second scenario, which will be elaborated in Section~\ref{Sec:Fillings}.

Finally, we emphasize that the discussion in this section is limited to reference field A and filling factor $0\leq \nu\leq 1$. 
For other reference fields or higher fillings, the additional factor for the partially filled conduction band cannot be simplified as an overall renormalization of the interaction strength.

\subsection{Phase diagram for the FQAH states in PLG}

In this section, we will compare the three methods to derive a phase diagram of the PLG. 
In particular, we fix the twist angle to be $\theta=0.77^\circ$ to emulate the experiment~\cite{Lu2024} and use the displacement field $u_d$ as a tuning parameter. 

\begin{figure*}[t]
\center
\includegraphics[width=0.8\textwidth]{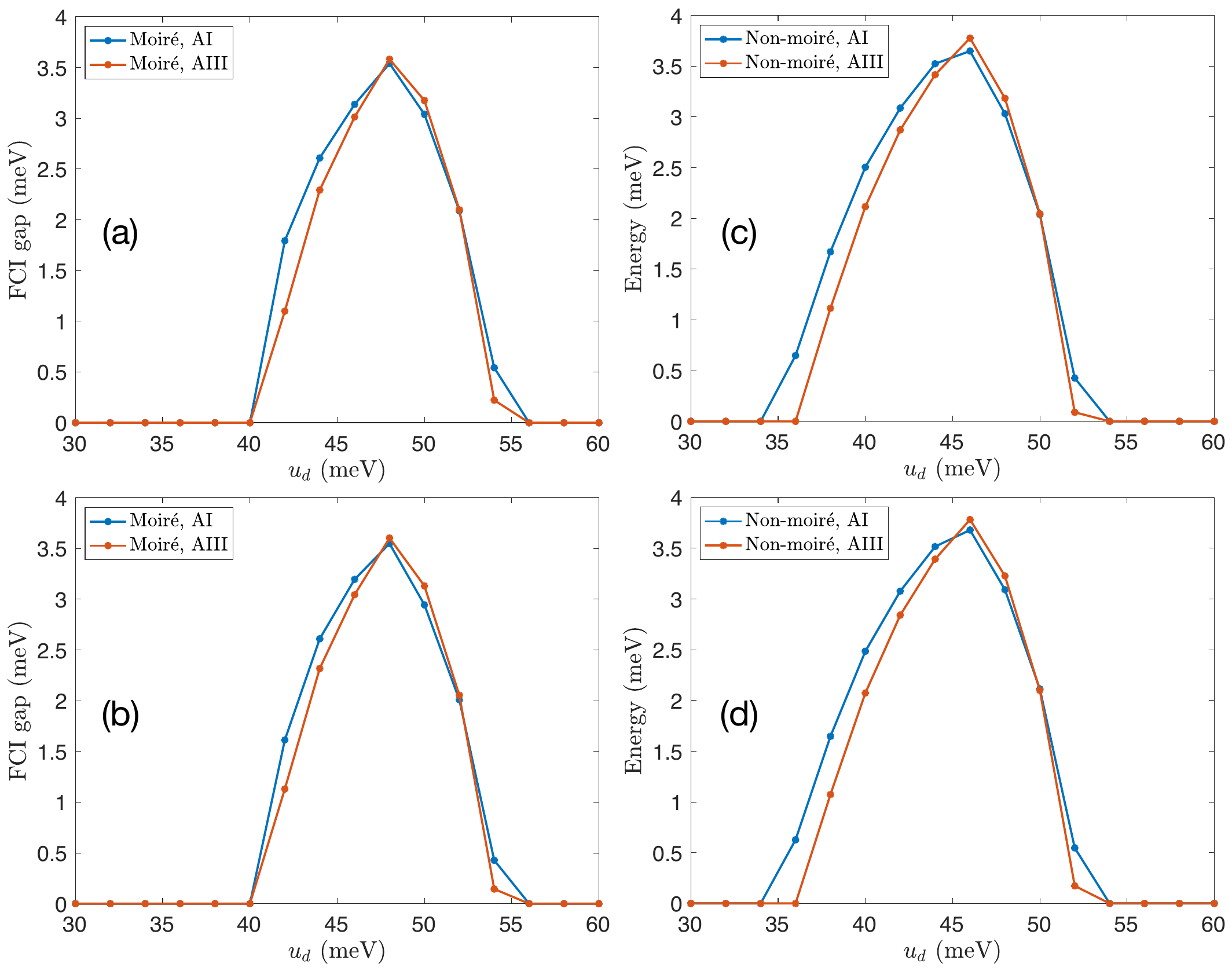}
\caption{\label{Fig:2/3_fci}
FQAH energy gap of the three methods. 
In the upper panels, we take $\Lambda=2.34 b_m$, equivalent to $20$ MBZ on average. 
In the lower panels, we take $\Lambda=3.41 b_m$, equivalent to $42$ MBZ on average. 
We study (a)-(b) the \moire\ case and (c)-(d) the non-\moire\ case with reference field A. 
Here, the twist angle is $0.77^\circ$, and the calculation is performed on a $4\times6$ mesh in the MBZ.
}
\end{figure*}

First, we calculate the FQAH energy gap to characterize the FQAH ground states of Methods I and III. 
Dictated by the (1,3)-permissible rule, the FQAH ground states only appear in specific momentum sectors~\cite{Regnault2011}. 
Consequently, we define the FQAH energy gap as the gap between the highest ground energy in the dictated momentum sectors and the lowest energy of the rest of the spectrum. 
Further, we set the FQAH gap to zero if the ground states do not appear in the dictated sectors.
We perform the calculation with two momentum cutoffs: (1) $\Lambda\leq 2.34 b_m$ (corresponding to 20 MBZ on average) in Fig.~\ref{Fig:2/3_fci}(a) and~\ref{Fig:2/3_fci}(c); 
(2) $\Lambda\leq 3.41 b_m$ (corresponding to 42 MBZ on average) in Fig.~\ref{Fig:2/3_fci}(b) and~\ref{Fig:2/3_fci}(d). 
By comparing the results obtained with these two different momentum cutoffs, we find that the convergence (with respect to the cutoff) is rather good for both \moire\ and non-\moire\ cases.
We also find that the FQAH gaps for both cases and both methods are qualitatively similar. 
The similarity between \moire\ and non-\moire\ cases suggests that the formation of an FQAH state in PLG may not be sensitive to the details of the \moire\ potential. 

\begin{figure*}[t]
\center
\includegraphics[width=0.8\textwidth]{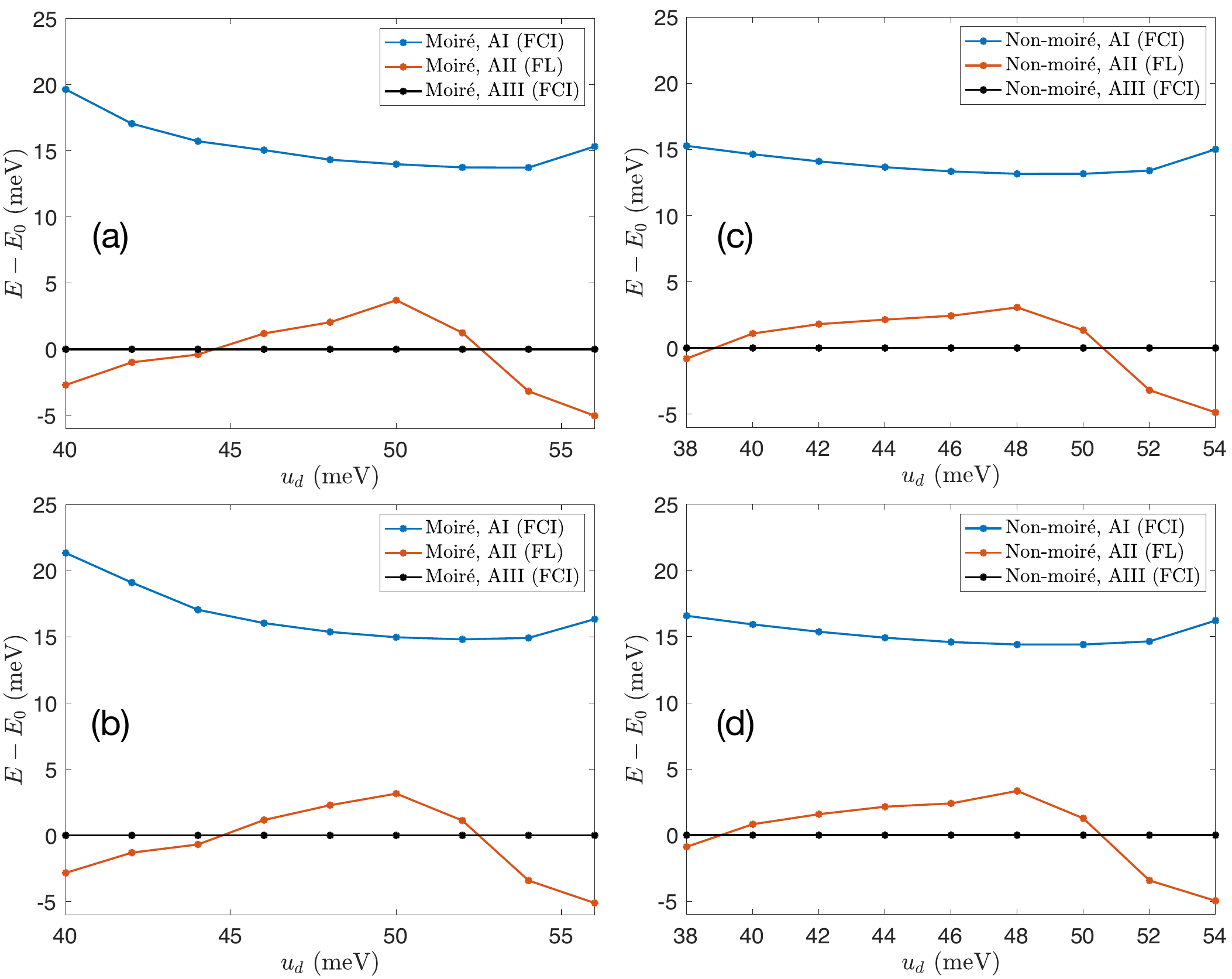}
\caption{\label{Fig:2/3_e}
Energies of the ground states produced by the three methods AI, AII, and AIII. 
We denote the energy of Method III as $E_0$. 
In the upper panels, we take $\Lambda=2.34 b_m$, equivalent to $20$ MBZ on average. 
In the lower panels, we take $\Lambda=3.41 b_m$, equivalent to $42$ MBZ on average. 
We study (a)-(b) the \moire\ case and (c)-(d) the non-\moire\ case with reference field A. 
Here, the twist angle is $0.77^\circ$, and the calculation is performed on a $4\times6$ mesh in the MBZ. 
}
\end{figure*}

Second, we compare the many-body energies of the three methods in the regime where Methods I and III have a nonvanishing FQAH gap in Fig.~\ref{Fig:2/3_e}.
Note that we again calculate the energies for the two cutoffs, which further verifies the convergence with respect to the cutoff for both cases. 
For Method I, we observe that it produces a much higher energy than the other two methods for all calculated parameters. 
This is because Method I overestimates the interaction by adding $1/3$ more filling. 
Additionally, we indeed find that Method III can have a lower energy than Method II for $\SI{45}{meV}\lesssim u_d \lesssim \SI{53}{meV}$ {in the \moire\ case and for $\SI{39}{meV}\lesssim u_d \lesssim \SI{51}{meV}$ in the non-\moire\ case}, suggesting a possible FQAH phase.  
However, {the FQAH phase} in this figure is smaller than the region where a finite FQAH gap exists because of the competition between the FQAH and the FL phase.

\subsection{Method III and its FQAH gaps at different fillings}\label{Sec:Fillings}

\begin{figure}[t]
\center
\includegraphics[width=0.9\columnwidth]{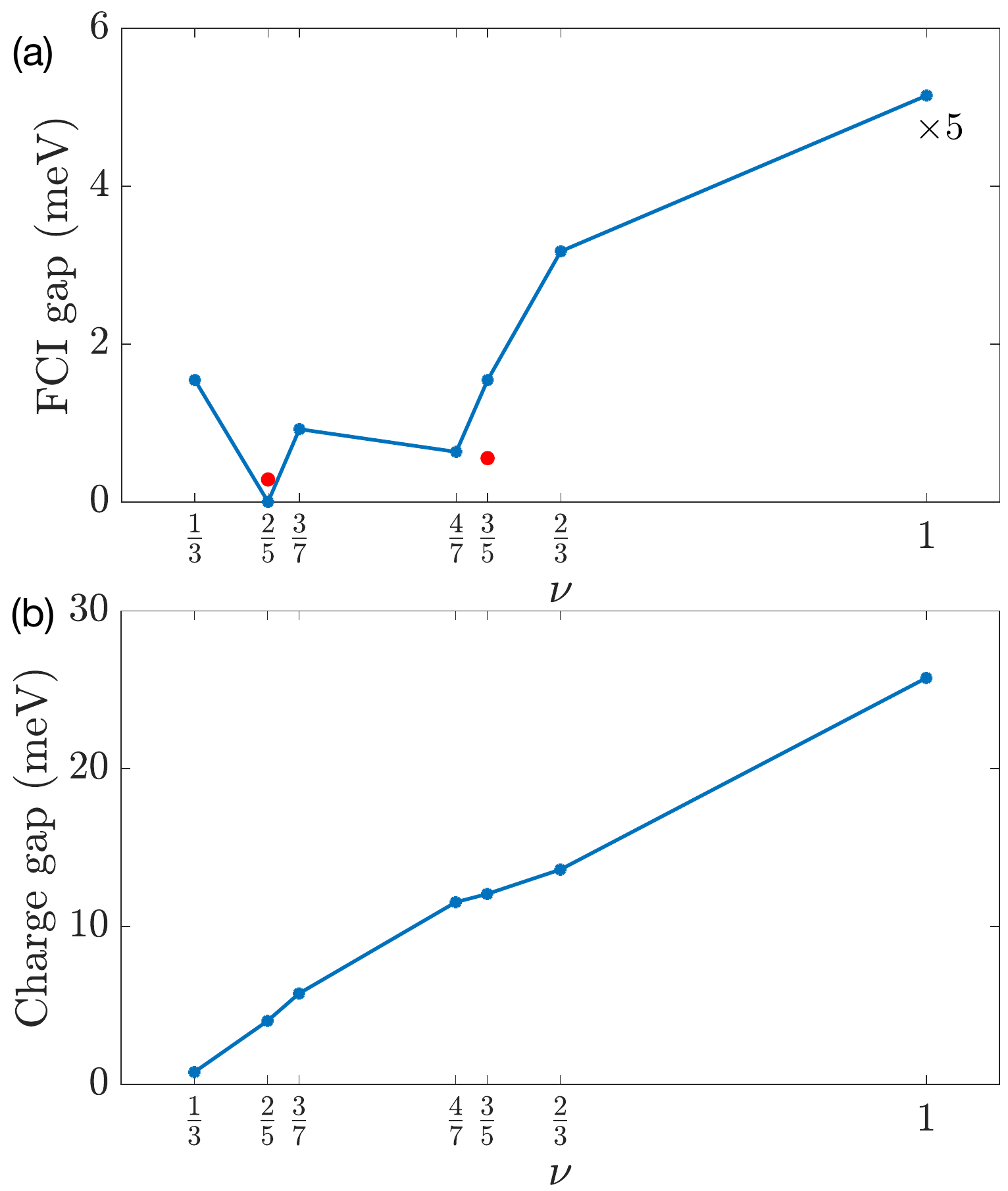}
\caption{\label{Fig:Gap}
(a) FQAH gap obtained using Method III. The gap at $\nu=1$ is about $\SI{25}{meV}$, so we write $\times 5$ for that data point. 
(b) Charge gap between occupied and unoccupied bands in the constrained HF calculations. 
For the blue dots, the calculation is performed using a momentum cutoff $\Lambda=2.34b_m$ on a $4\times 6$ mesh in the MBZ for $\nu=1/3,2/3,1$, a $5\times 5$ mesh for $\nu=2/5,3/5$, and a $4\times 7$ mesh for $\nu=3/7,4/7$. 
For the two red dots in (a), the calculation is performed on a $4\times 5$ mesh in the MBZ. 
}
\end{figure}

In this section, we explore how the results produced by Method III depend on the filling factors. 
In particular, we calculate the \moire\ case with reference field A, and take $\theta=0.77^\circ$ and $u_d=\SI{50}{meV}$. 
In Fig.~\ref{Fig:Gap}(a), we show the FQAH gaps at different fillings. 
For $\nu=1$ in Fig.~\ref{Fig:Gap}(a), we show the charge gap between the occupied bands and the unoccupied bands as a benchmark. 
For $\nu=1/3,2/3,2/5,3/5,3/7,4/7$, the FQAH gap is defined as the gap between the ground states in the dictated momentum sectors and the other states, and the FQAH gap is set to zero if there is no such gap. 
Our results show that $\nu=2/3$ has the most prominent FQAH gap, whereas $\nu=2/5$ does not (accidentally) have an FQAH gap in the $N_x\times N_y=5\times5$ system (but does for a different system size). 
However, we note that the finite-size effects are inevitable in the ED calculation, so the exact values of the gap should not be taken too seriously since there are likely to be severe finite-size corrections arising from the necessarily small ED system sizes feasible for the computation.
For example, in the same figure, we show that if we decrease the system size to $4\times 5$, the FQAH gap at $\nu=2/5$ becomes finite, while that at $\nu = 3/5$ becomes smaller (see the red dots). 
These differences provide a crude measure for the finite size corrections involved in the ED gap estimates, but an accurate finite size scaling is impossible at this stage because of the current computational constraints on ED.

For the results presented in Fig.~\ref{Fig:Gap}(a), we ignore the premise of Method III and the projection ansatz Eq.~\eqref{Eq:Ansatz} that a large charge gap must exist between the occupied bands and unoccupied bands. 
According to Section~\ref{Sec:Compare13}, reducing the filling factor is similar to reducing the interaction strength. 
As the system is gapless without interaction, there must be a filling factor $0<\nu<1$ at which the global gap in the constrained HF vanishes. 
In Fig.~\ref{Fig:Gap}(b), we calculate the charge gap in the constrained HF for various filling factors. 
The results show that the charge gap almost has a linear dependence on the filling factor and vanishes around $\nu=1/3$. 
We further verify that the Chern number of the lowest HF conduction band is always 1 for $\nu>1/3$, indicating no further transition in the gapped phase. 
Hence, the PLG belongs to the second scenario discussed in Section~\ref{Sec:Compare13}. 
For large filling factors like $\nu=2/3$, the charge gap is much larger than the FQAH gap, and we deem the projection to the lowest band valid. 
However, for smaller filling factors, particularly $\nu=1/3$, the charge gap is of the same order as the FQAH gap. 
Therefore, Method III or even the projection ansatz Eq.~\eqref{Eq:Ansatz} fails in this case, suggesting that this is not a single-band physics.  
If this is the case, the physics in the PLG at small filling factors is not adiabatically connected to the physics of continuum Landau level FQH in strong field 2D systems.

\section{Summary and outlook \label{Sec:Conclusion}}

We note that the FQAH states we find in PLG are all spin-and-valley polarized. 
In this context, it is instructive to compare our results with the quantum Hall ferromagnets where an integer quantum Hall effect (IQHE) can occur (in the lowest spinful Landau level) despite the Land\'{e} $g$-factor being zero since electron-electron interactions even at the HF level will spontaneously open a spin gap even if there is no Zeeman gap~\cite{Nomura2006,Young2012,Li2016}. 
In systems where the lowest conduction band is well isolated from the other bands~\cite{Zou2018,Zhang2019,Wu2019,Bultinck2020,Xie2020}, the spin-valley polarization can be understood using the Stoner model. 
In such cases, the interaction only lifts the degeneracy but does not alter the wave function. As a result, the Chern number can be determined at the noninteracting level.
For the PLG, the spin-valley polarized ground state is still favored in both the unconstrained and constrained HF calculations at integer and fractional fillings despite the absence of an isolated band at the noninteracting level. 
However, unlike the Stoner model, the wave function for the HF ground state in the PLG is drastically different from its noninteracting counterpart due to the mixture between different noninteracting bands. 
Therefore, the Chern number may also change. 
In particular, both the wave function and the Chern number of the HF ground state depend on the filling factor in the unconstrained HF calculation. 
This implies that the QAH problem in PLG is likely to be qualitatively different from the corresponding quantum Hall ferromagnetism in the continuum Landau level problem in spite of some superficial similarity.

It is important to address the use of mean-field theory at fractional fillings in this work. 
While mean-field theory is generally unreliable for directly determining many-body ground states at fractional fillings, our approach extends beyond conventional mean-field methods. 
Our methodology employs an expanded variational ansatz, as described in Eq.~\eqref{Eq:Ansatz}, which surpasses the standard Slater determinant ansatz used in conventional mean-field theory. 
This ansatz assumes that FQAH physics in PLG is confined to a single quasiparticle band, with interband interactions treated at the HF level. 
In our framework, the HF solution serves primarily to provide quasiparticle band wavefunctions, which are the variables in our ansatz. The accuracy of the HF solution itself is less critical; what matters is the ED result using these quasiparticle bands. 
As our method is variational, its efficacy is ultimately judged by its ability to yield a lower-energy ground state in ED calculations.
Importantly, we do not need to pursue the lowest-energy HF solution, as the HF solution's energy is not our primary concern. 
This principle underpins Method III described in the main text. 
As shown in Fig.~\ref{Fig:2/3_e}, we can achieve a more energetically favorable FQAH state using quasiparticle bands from a constrained HF solution, even when the energy of the constrained HF solution exceeds that of the unconstrained HF solution. 
This result demonstrates the superiority and justification of the theoretical framework we established in exploring FQAH states.

We mention that although our work establishes a unified theoretical framework for studying FQAH in PLG, any quantitative comparison to the experimental data~\cite{Lu2024} is not feasible at this stage since the experiment at this point appears to be seriously affected by unknown background disorder effects, leading to very large resistivity instead of the zero resistivity expected for all quantum Hall effects on general grounds~\cite{XieMing2024}.
In addition, the nominally extracted experimental  FQAH activation gaps manifest a constancy for all fractions, which is also a mystery, most likely also connected to the unknown background disorder problem~\cite{XieMing2024}. 
An additional mystery is the absence of the $\nu=1/3$ FQAH state in the experiment, although FQAH is observed for fractions like $\nu=2/5$, $3/7$, $4/7$, $3/5$, $2/3$, etc. 
The absence of the $\nu=1/3$ FQAH may be related to our finding of a very small HF charge gap at $\nu=1/3$ [Fig.~\ref{Fig:Gap}(b)], making the $\nu=1/3$ FQAH particularly fragile and unstable to other competing non-FQAH phases such as a pinned Wigner crystal or a disorder-induced Anderson localized insulator. 
It may be important to emphasize that the corresponding extensively studied (1982-2024) Landau level-based high-field continuum FQH also always disappears at some small fraction $\nu \sim 1/5$-$1/7$ because of competing compressible Anderson localized or pinned Wigner crystal phases which manifest no gaps.



To summarize, we have studied the FQAH states in the PLG using the HF and ED methods.
After setting up the continuum model for the PLG and the interaction Hamiltonian, we first discussed the HF calculation in the PLG.
In particular, we carefully compared the different choices of the reference field and studied the divergence issue associated with momentum cutoff in the HF calculation. 
We concluded that reference field A is the most suitable choice for the HF calculation in the PLG. 
We then discussed the ED calculation in the PLG. 
In particular, we proposed three methods to search for the FQAH states in the PLG and compared their results. 
We found that the constrained HF calculation can produce a nearly uniform occupation number for all HF orbitals, which is essential for the FQAH states. 
Our work thus provides a self-consistent framework for understanding the FQAH states in the PLG and resolves the current controversy on the optimal theoretical techniques to study the FQAH in PLG.

The current theory of FQAH is primarily an analog of the FQHE in the lowest Landau level (LLL) in the extensively studied 2D high-field systems (e.g. GaAs), where the physics merely happens within one energetically and geometrically uniform band defined essentially by a single LL. 
However, there is no \emph{a priori} reason that the FQAH states in PLG must be adiabatically connected to the FQHE in the LLL, and it is premature to compare the theoretical results with the experiment. 
Particularly, the longitudinal resistance in the PLG experiment, probably caused by unknown disorder effects, is much larger than that in the fractional quantum Hall effect, and the FQAH gaps fitted to the experimental data are also much smaller than the theoretical gaps~\cite{XieMing2024}. 
The possibility of FQAH physics beyond the FQHE in the LLL is also hinted at by our calculation of Method III at small filling factors or weak interaction, where the charge gap is very small or even zero, making the projection to a single band illegitimate. 
More experimental results with better data are necessary for the next step in our understanding of the FQAH physics in PLG.

\section*{Acknowledgement}
We acknowledge helpful discussions with 
A. Bernevig, Z.~Dong, L.~Ju, Z.~ Lu, 
J.~Pixley, T.~Senthil, I.~Sodemann, M.~Xie, X.~Xu, J.~Yu, and Y.~Zhang. 
K.H. and X.L. are supported by the Research Grants Council of Hong Kong (Grants No.~CityU~11300421, CityU~11304823, and C7012-21G), and City University of Hong Kong (Projects No.~9610428 and 7005938). 
K.H. is also supported by the Hong Kong PhD Fellowship Scheme. 
S.D.S. is supported by the Laboratory for Physical
Sciences through the Condensed Matter Theory Center (CMTC) at the University of Maryland. 
This work was performed in part at the Aspen Center for Physics, which is supported by National Science Foundation grant PHY-2210452.
\appendix

\section{The continuum model for rhombodedral multilayer graphene}\label{App:BM model}

In this appendix, we will briefly review the continuum model for the rhombohedral n-layer graphene (RnG)~\cite{Zhang2010,Jung2013,Moon2014}. 
In monolayer graphene, the three nearest neighbor vectors are given by $\vb*\delta_n =R_{2\pi n/3}[0, a_{\text{G}}/\sqrt{3}]^\intercal$ for $n=0,1,2$, where $a_\text{G}\approx 2.46\,\text{\AA}$ is the lattice constant of graphene, and $R_\varphi$ is counterclockwise rotation by angle $\varphi$. 
Correspondingly, the basis vectors in the real space are
\begin{align}
	\vb R_1=a_\text{G}[1,0]^\intercal,\quad \vb R_2=a_\text{G}[1/2,\sqrt{3}/2]^\intercal,
\end{align}
and the reciprocal basis vectors of graphene are
\begin{align}
	\vb G_1=\frac{2\pi}{a_{\text{G}}}\qty[1,-\frac{1}{\sqrt{3}}]^\intercal,\quad \vb G_2=\frac{2\pi}{a_{\text{G}}}\qty[0,\frac{2}{\sqrt{3}}]^\intercal.
\end{align}
The two corners of the Brillouin zone are defined as 
\[
\vb K=-\vb K'=\dfrac{2}{3} \vb G_1+ \dfrac{1}{3} \vb G_2.
\]

The Hamiltonian of the pristine RnG in a perpendicular displacement field is modeled as
\begin{align}
H_{\text{RnG}}(\vb k)=H_0(\vb k)+V_{d},
\end{align}
where $H_0(\vb k)$ represents the kinetic energy, and the displacement field is $[V_{d}]_{ll'}=\delta_{ll'}[l-(n_l-1)/2]u_d$, where $n_l$ is the number of layers.
Here, $l=0$ represents the top layer, and $l=n_l-1$ represents the bottom layer.
The kinetic term $H_{0}(\vb k)$ is given by
\begin{align}
H_{0}(\vb k)
 =\left[\begin{array}{ccccc}
 h^{(0)} & h^{(1)} & h^{(2)} &  & \\
 h^{(1)\dag} & h^{(0)} & \ddots & \ddots &  \\
 h^{(2)\dag} & \ddots & \ddots & h^{(1)} & h^{(2)} \\
  & \ddots & h^{(1)\dag} & h^{(0)} & h^{(1)} \\
  &  & h^{(2)\dag} & h^{(1)\dag} & h^{(0)}\end{array}\right],
\end{align}
where
\begin{align}
&h^{(0)}=-t_0\left[\begin{array}{cc}0 & f_{\vb k} \\ f_{\vb k}^* & 0\end{array}\right],\quad
h^{(1)}=\left[\begin{array}{cc}t_4f_{\vb k} & t_3 f_{\vb k}^* \\t_1 & t_4f_{\vb k} \end{array}\right],\nonumber\\
&h^{(2)}=\left[\begin{array}{cc}0 & \frac{t_2}{2} \\0 & 0\end{array}\right],\quad f_{\vb k}=\sum_{j}\exp(i\vb k\cdot \delta_j).
\end{align}
Following Ref.~\cite{Wang2024a}, we take 
\begin{align*}
	(t_0,t_1,t_2,t_3,t_4)=(3100,380,-21,290,141)\, \text{meV}. 
\end{align*}
In the main text, we relegate the valley degree of freedom to the same level as the sublattice or layer degrees of freedom to allow the inter-valley correlation in the HF calculation. Moreover, the momentum is measured from $\vb K$ or $\vb K'$ for the two valleys, respectively. Hence, the model used in this work is given by
\begin{align}
H_{\text{PLG}}(\vb k)=\left[\begin{array}{cc} H_{\text{R5L}(\vb K+\vb k)} &  0_{10\times10}\\ 0_{10\times10} & H_{\text{R5L}(\vb K'+\vb k)}\end{array}\right].
\end{align}

The reciprocal basis vectors of the aligned hBN is
\begin{align}
	\vb G_{i}'=\frac{a_\text{G}}{a_\text{hBN}}R_{\theta}\vb G_i
\end{align}
with a lattice mismatch $a_\text{hBN}/a_\text{G}\approx 1.018$. The mismatch leads to the following \moire\ reciprocal vectors
$\vb g_1=\vb G_1-\vb G_1'$, $\vb g_n=R_{2\pi (n-1)/3}\vb g_1$ for $n=2,3$, and $\vb g_n=-\vb g_{n-3}$ for $n=4,5,6$. Thus, the \moire\ potential on the top layer for the $\vb K$ valley takes the following form,
 \begin{align}
	V_{\text{\moire}}^{\text{top}}(\vb r)=V_{\text{\moire}}^{\text{top}}(0)+\sum_{j=1}^6 V_{\text{\moire}}^{\text{top}}(\vb g_j)e^{i\vb r\cdot\vb g_j}.
\end{align}
Because of the hermiticity and the $C_3$ rotational symmetry, the \moire\ potential satisfies
 \begin{align}
	&V_{\text{\moire}}^{\text{top}}(0)=V_{\text{\moire}}^{\text{top}}(0)^\dag,\quad V_{\text{\moire}}^{\text{top}}(\vb g_j)=V_{\text{\moire}}^{\text{top}}(-\vb g_{j})^\dag, \nonumber\\
	&V_{\text{\moire}}^{\text{top}}(0)=U_{3}V_{\text{\moire}}^{\text{top}}(0)U_{3}^\dag, \nonumber\\
	&V_{\text{\moire}}^{\text{top}}(R_{2\pi/3}\vb g_j)=U_{3}V_{\text{\moire}}^{\text{top}}(\vb g_{j})U_{3}^\dag,
\end{align}
where $U_3=\text{diag}(1,\omega)$ with $\omega=\exp(2\pi i/3)$ is the representation of $C_3$ in RnG. Following Ref.~\cite{Moon2014}, we take
 \begin{align}
	V_{\text{\moire}}^{\text{top}}(0)=V_0\mathbb{I},\quad
	V_{\text{\moire}}^{\text{top}}(\vb g_1)=V_1e^{-i\psi}\left[\begin{array}{cc} 1 & 1 \\ \omega & \omega\end{array}\right],
\end{align}
where $(V_0,V_1,\psi)=(28.9\,\text{meV},21\,\text{meV},-0.29)$. The $\vb K'$ valley is generated via time-reversal symmetry. 
Therefore, the complete form of the \moire\ potential used in this work is 
 \begin{align}
	V_{\text{\moire}}(\vb g)=
	\left[\begin{array}{cccc} 
	V_{\text{\moire}}^{\text{top}}(\vb g) & 0_{2\times 8} & 0_{2\times 2} &0_{2\times 8} \\ 
	0_{8\times 2} & 0_{8\times 8} & 0_{8\times 2} & 0_{8\times 8} \\
	 0_{2\times 2} & 0_{2\times 8} & V_{\text{\moire}}^{\text{top}}(\vb g) &0_{2\times 8} \\
	  0_{8\times 2} & 0_{8\times 8} & 0_{8\times 2} & 0_{8\times 8}
	 \end{array}\right],
\end{align}
where $\vb g$ takes value from $0$ and $\vb g_i$.


\section{Hartree-Fock approximation\label{App:HF}}




The Hartree-Fock approximation is a variational method based on the ansatz that the ground state is a Gaussian state. 
Such states satisfy Wick's theorem, and thus can be identified with their one-body density matrix $P$ satisfying: (1) $P$ is a positive-semidefinite hermitian matrix; (2) the operator norm of $P$ is less than or equal to 1. 
The two conditions guarantee that the occupation number must be nonnegative and less than or equal to 1. 
Throughout this work, we only allow correlations in the same spin sector and between two momentums connected by a \moire\ reciprocal vector. 
Thus, both the one-body density matrix and the HF Hamiltonian are block diagonal, and each block $P(\sigma,\vb k)$ or $[H_{\text{HF}}(P)](\sigma,\vb k)$ is labeled by its spin $\sigma$ and momentum $\vb k$ in the first MBZ. 
The energy of a Gaussian state is given by
\begin{align}
	E(P)&=\trace[H_sP^\intercal]+\frac12\trace[P^\intercal V_{\text{HF}}(P)]\nonumber\\
	&=\sum_{\sigma,\vb k}\trace[H_s(\sigma,\vb k)P(\sigma,\vb k)^\intercal]\nonumber\\
	&\quad+\sum_{\sigma,\vb k}\trace[P(\sigma,\vb k)^\intercal [V_{\text{HF}}(P)](\sigma,\vb k)]/2.
\end{align}
Because of the circular cutoff in the momentum space, the dimensions of each block are different. For a $\vb k$ in the first MBZ, we denote the number of $\vb p$ that $\{\vb p\}=\vb k$ by $N(\vb k)$, and the dimension of $P(\sigma,\vb k)$ is $D(\vb k)=gN(\vb k)$ with $g=g_vn_ln_{\text{sub}}$, where $g_v$ is the valley degree of freedom, $n_l$ is the number of the layers, and $n_{\text{sub}}$ the number of the sublattices. 
At the CNP, the total particle number is given by 
\begin{align}
	N_{\text{CNP}}= \dfrac{g_s}{2}\sum_{\vb k\in\text{MBZ}}D(\vb k), 
\end{align}
where $g_s$ is the spin degree of freedom. 

\begin{figure}[t]
	\center
	\includegraphics[width=0.9\columnwidth]{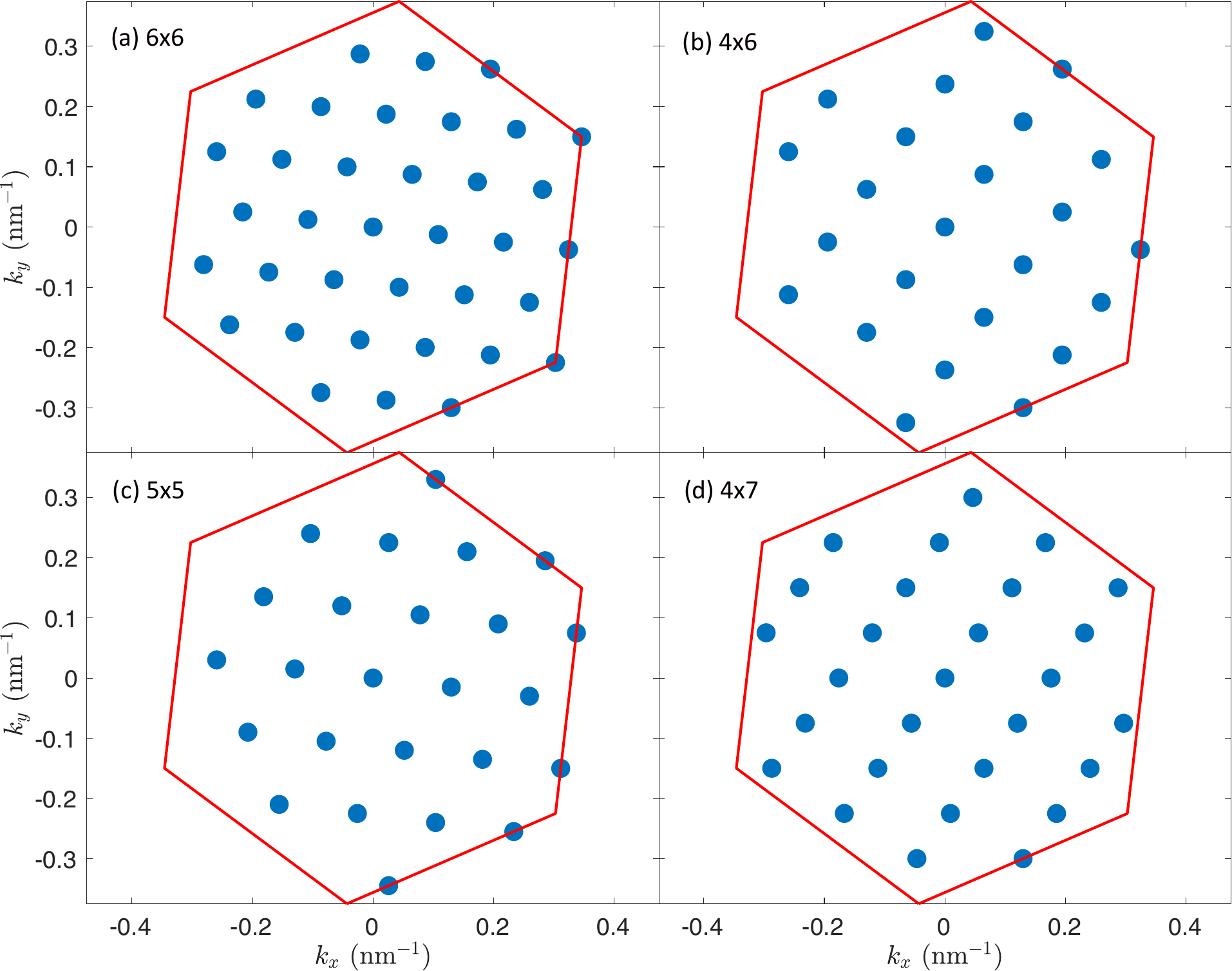}
	\caption{\label{Fig:Mesh}
	The mesh in the MBZ we used in the calculations. (a) $N_x\times N_y=6\times6$, (b) $N_x\times N_y=4\times6$, (c) $N_x\times N_y=5\times5$, and (d) $N_x\times N_y=4\times7$. Here, the twist angle is $0.77^\circ$.
	}
\end{figure}

Finally, our calculations are carried out on a $N_x\times N_y$ mesh in the MBZ defined by 
\begin{align}
	\vb k=\qty{\dfrac{i}{N_x}\vb g_1+ \dfrac{j}{N_y} \vb g_2},
\end{align}
with $i=0,\cdots,N_x-1$ and $j=0,\cdots,N_y-1$. 
In the main text, we used four types of meshes, as illustrated in Fig.~\ref{Fig:Mesh}.
Therefore, at a finite filling $\nu$, the total particle number is $N=N_{\text{CNP}}+\nu N_xN_y$. 

\subsection{Unconstrained Hartree-Fock calculations}

In the unconstrained Hartree-Fock calculation, we aim to minimize $E(P)$ without any constraint on $P$, which can be achieved by the algorithm in~\cite{Kudin2002}. 
The procedure is summarized as follows. At each iteration, we start with $P_i$ and diagonalize $H_{\text{HF}}(P_i)$.  Then we choose the lowest $N$ orbitals of $H_{\text{HF}}(P_i)$, which constitute a one-body density matrix $\tilde P$, given by 
\begin{align}
	\tilde P=\sum_{\alpha=1}^{N}\psi_{\alpha}^*\psi_{\alpha}^\intercal,
\end{align}
where the column vectors $\psi_{\alpha}$ are the eigenstates of $H_{\text{HF}}(P_i)$.
Then, $P_\lambda:=(1-\lambda)P_i+\lambda \tilde P$ also represents a Gaussian state for $0\leq\lambda\leq 1$ because $P_\lambda$ is positive-semidefinite and 
\begin{align}
	\norm{P_\lambda}\leq \lambda\norm{P_i}+(1-\lambda)\norm*{\tilde P}\leq 1. 
\end{align}
By minimizing $E(P_\lambda)$, we obtain $P_{i+1}$ for the next round of iteration.

The convergence of the above algorithm is guaranteed by the monotone convergence theorem. 
Moreover, the converged result is a local minimum. 
To see this, we note that 
\begin{align}
	E(P_\lambda)&=E(P_i)+\lambda^2\trace[(\tilde P-P_i) ^\intercal V_{\text{HF}}(\tilde P-P_i)]/2\nonumber\\
	&\quad+\lambda \trace[(\tilde P-P_i) ^\intercal H_{\text{HF}}(P_i)].
\end{align}
Note that the third term is always nonpositive because $\tilde P$ is the ground state of $H_{\text{HF}}(P_i)$. Hence, we have $P_i=\tilde P$ if the convergence is achieved. 
Because $\trace[(\tilde P-P) ^\intercal H_{\text{HF}}(\tilde P)]\geq 0$ for all $P$, we know that $\tilde P$ is a local minimum.
As a corollary, we also know that convergent result is always a Slater determinant, though the intervening states are generally not Slater determinant.
Because the algorithm conserves the total particle number but not the particle number in each block, the unconstrained HF can lead to metallic states or insulating states.


\subsection{Constrained Hartree-Fock calculations}

In the constrained Hartree-Fock calculation, we aim to minimize $E(P)$ for $P$ satisfying the following conditions: 
\begin{align*}
	\trace[P(\uparrow,\vb k)]=\dfrac{g N(\vb k)}{2}+\nu, \qq{and}
	\trace[P(\downarrow,\vb k)]=\dfrac{g N(\vb k)}{2}. 
\end{align*}
The constraint makes $P$ spin-polarized, and the density in the MBZ is uniform relative to the CNP. 
The constrained HF equation can be solved by slightly modifying the algorithm in the previous section. 
Specifically, at each round of iteration, we start with $P_i$ and diagonalize $H_{\text{HF}}(P_i)$. 
In sector $(\uparrow,\vb k)$, the orbitals are $\psi_{\uparrow,\vb k,\alpha}$ for $\alpha=1,2,\cdots,D(\vb k)$. 
Therefore, $\tilde P(\sigma,\vb k)$ is defined as
\begin{align}
	\tilde P(\sigma,\vb k)&=\{\nu\}\,\psi_{\uparrow,\vb k, N(\vb k)/2+[\nu]+1}^*\psi_{\uparrow,\vb k,N(\vb k)/2+[\nu]+1}^\intercal \nonumber\\
	&\quad+\sum_{\alpha=1}^{g N(\vb k)/2+[\nu]}\psi_{\uparrow,\vb k,\alpha}^*\psi_{\uparrow,\vb k,\alpha}^\intercal,
\end{align}
where $[\nu]$ and $\{\nu\}$ denote the integer and decimal parts of $\nu$, respectively. 
The construction guarantees that $\tilde P$ satisfies the constraint, and so does $P_{\lambda}:=(1-\lambda)P_i+\lambda \tilde P$. 
By minimizing $E(P_\lambda)$, we obtain $P_{i+1}$ for the next round of iteration.

First, the convergence is obviously guaranteed. Second, we need to prove that the convergent result is a local minimum. Compared to the proof in the previous section, we only need to prove that $\trace[(\tilde P-P) ^\intercal H_{\text{HF}}(\tilde P)]\geq 0$ for all $P$ with the constraint. Further, we only need to prove the statement in each sector, 
\begin{align}
	\trace[\tilde P(\sigma,\vb k) ^\intercal [H_{\text{HF}}(\tilde P)](\sigma,\vb k)]\geq \trace[P(\sigma,\vb k) ^\intercal [H_{\text{HF}}(\tilde P)](\sigma,\vb k)].
\end{align}
This is equivalent to proving the following inequality,
\begin{align}
	\sum_{\alpha=1}^{d} e_\alpha n_\alpha\geq \sum_{\alpha=1}^{[n]} e_\alpha+\{n\}e_{[n]+1},
\end{align}
where $e_\alpha\leq e_{\alpha+1}$, $0\leq n\leq d$, $0\leq n_\alpha\leq 1$, and we have $\sum_{\alpha=1}^dn_\alpha=n$. 
The inequality can be proved as follows: 
\begin{align}
	&\sum_{\alpha=1}^{[n]} e_\alpha(1-n_\alpha)+\{n\}e_{[n]+1}-\sum_{\alpha=[n]+1}^de_\alpha n_\alpha \notag\\
	&\leq e_{[n]+1}\qty([n]- \sum_{\alpha=1}^{[n]}n_\alpha)+\{n\}e_{[n]+1}-e_{[n]+1}\sum_{\alpha=[n]+1}^d n_\alpha \notag\\
	&= 0.
\end{align}
As a result, we know that the constraint here is equivalent to the constraint in the main text.

\section{Effect of Hartree-Fock bandwidth on FCI}

\begin{figure}[t]
\center
\includegraphics[width=\columnwidth]{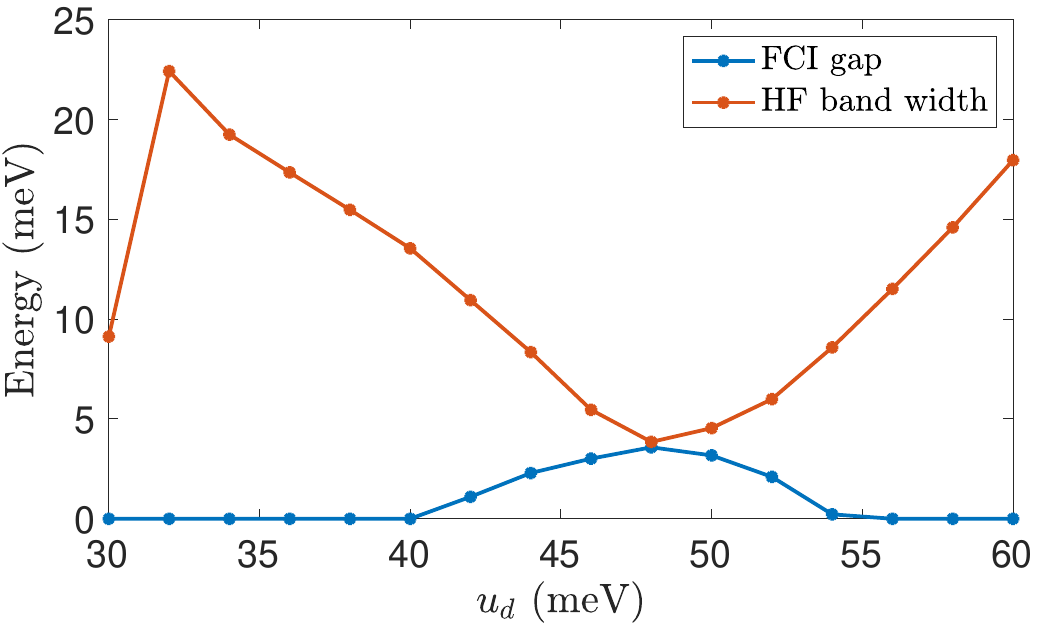}
\caption{\label{Fig:Width}
FCI gap and HF bandwidth derived from Method III at $\nu=2/3$, performed on a $4\times 6$ mesh in the MBZ.
Here, the twist angle is $0.77^\circ$, and the displacement field is $u_d=\SI{50}{meV}$.
}
\end{figure}

The band dispersion is known to obstruct the formation of FCI states, and causes transition to FL states. In the PLG, the bandwidth cannot be varied independently, but it is effectively tuned by the displacement field.  In Fig.~\ref{Fig:Width}, we calculate the HF bandwidth and the FCI gap at $\nu=2/3$ using Method III. 
As shown in the figure, the HF bandwidth is minimized at $u_d=\SI{48}{meV}$, at which point the FCI gap is maximized. 
As the displacement field deviates from $u_d=\SI{48}{meV}$, the bandwidth significantly increases and culminates in varnishing FCI gaps. 
We also note that the discontinuity of the bandwidth at $u_d=\SI{30}{meV}$ is caused by a phase transition in the HF calculation. 
The Chern number of the HF band at $u_d=\SI{30}{meV}$ is $0$, while the Chern number is $1$ for all the other parameters shown in Fig.~\ref{Fig:Width}.

\begin{figure*}[!]
\center
\includegraphics[width=0.9\textwidth]{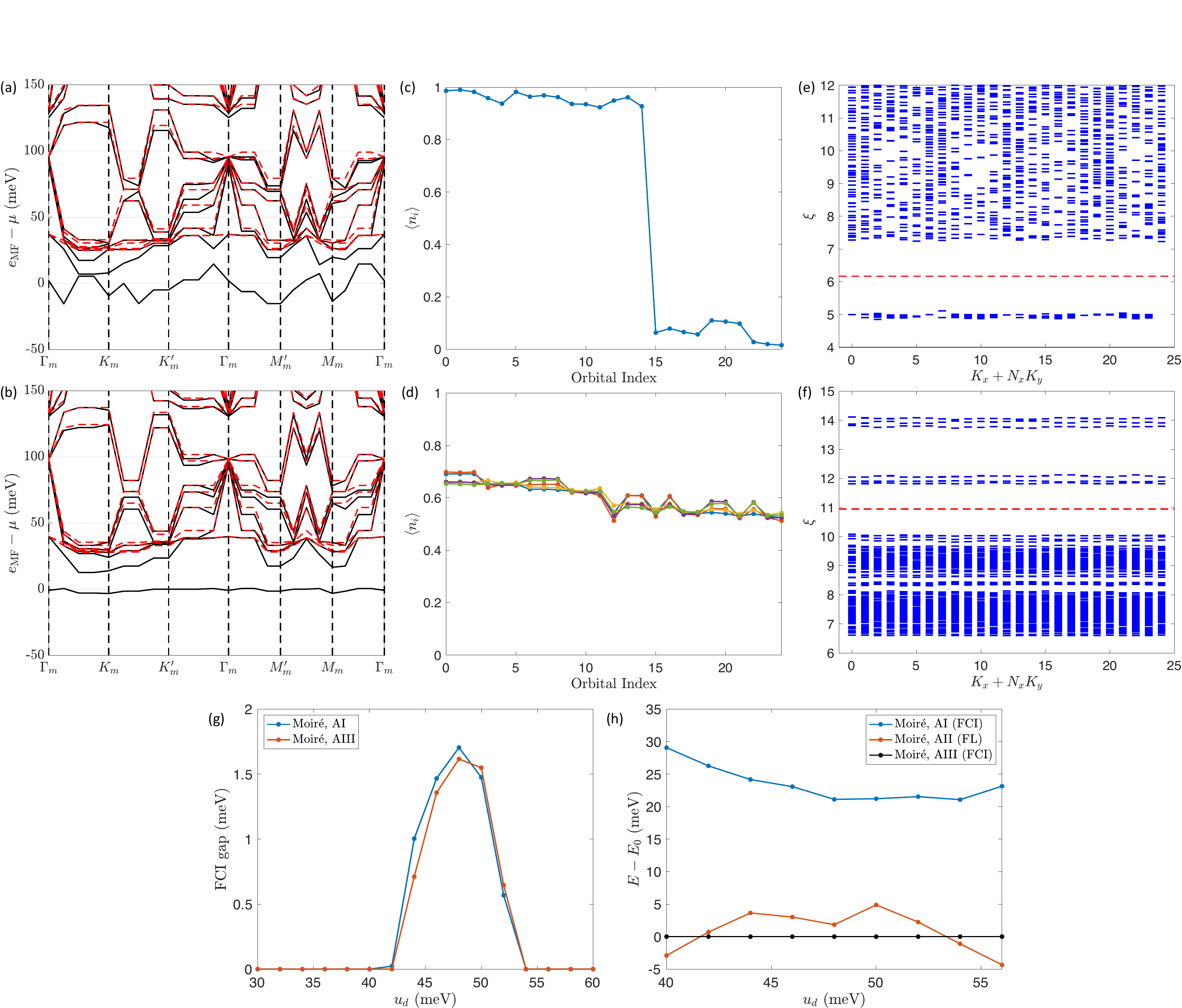}
\caption{\label{Fig:3/5_spec}
Comparison between AII (top row) and AIII (bottom row) in the \moire\ case on a $5\times 5$ mesh in the MBZ at $3/5$ filling. 
(a,b) HF band structure of AII and AIII. The black and red lines represent the spin-up and spin-down sectors, respectively. 
(c,d) Occupation number of the ground states of AII and AIII. 
The indices of the orbitals are ordered according to their HF energies. Note that we plot the occupation number of all five-fold degenerate ground states for AIII in (d). 
(e) Particle entanglement spectrum with a subsystem of three holes for the particle-hole conjugate of AII. There are $120$ states below the entanglement gap, corresponding to the quasihole excitations (10 choose 3) of FL. 
(f) Particle entanglement spectrum with a subsystem of three holes for the particle-hole conjugate of AIII. 
There are $2150$ states below the entanglement gap, corresponding to the (2,5)-permissible excitations of FQAH. In (a-f), we take $u_d=\SI{50}{meV}$.
(g) Energies of the three methods at $3/5$ filling for various $u_d$. We set $E_0$ to be the energy of Method III. (h) FQAH energy gap at $3/5$ filling for various $u_d$. In all panels, we take $\Lambda=2.34b_m$ and $\theta=0.77^\circ$. 
}
\end{figure*}

\section{FQAH results at the $3/5$ filling}\label{App:3/5}

In this appendix, we provide the additional data for the $3/5$ filling for the \moire\ case with reference field A. In Fig.~\ref{Fig:3/5_spec}(a-f), we numerically verify that Methods II and III also generate FL states and FQAH states, respectively. 
The results at $3/5$ filling manifest similar features to the results at $2/3$ filling, including the HF band dispersion, particle density, and entanglement gap. 
We further compare the three methods for various $u_d$. 
First, we calculate the FQAH energy gap Fig.~\ref{Fig:3/5_spec}(g), indicating that both Methods~I and III produce FQAH states. 
Second, we show in Fig.~\ref{Fig:3/5_spec}(h) that there is a region where Method~III gives lower energy than Method~II. 
We also emphasize that this region completely resides in the region where the FQAH gap does not vanish.

\section{FQAH states in rhombohedral hexalayer graphene}

\begin{figure*}[!]
\center
\includegraphics[width=\textwidth]{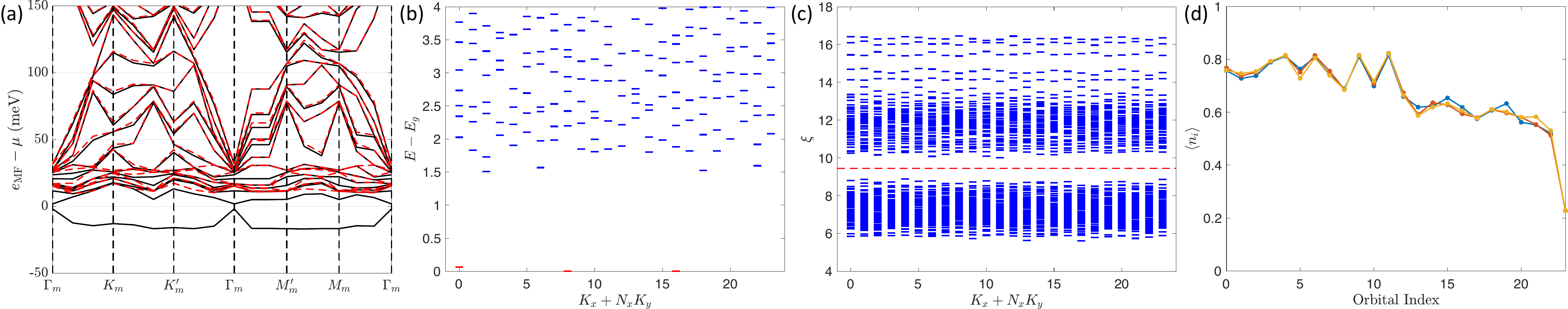}
\caption{\label{Fig:HLG}
Calculation of Method III in the HLG with reference field A performed on a $4\times 6$ mesh in the MBZ at $2/3$ filling. (a) HF band structure. 
The black and red lines represent the spin-up and spin-down sectors, respectively. 
(b) Many-body energy spectrum. The three-fold degenerate states appear in the momentum sector dictated by the (1,3)-permissible rule, and we confirm that their energies are lower than that of the FL ground state obtained by Method II. 
(c) Particle entanglement spectrum with a subsystem of three holes for the particle-hole conjugate of the ground state. There are $1088$ states below the entanglement gap, corresponding to the (1,3)-permissible excitations of FQAH. 
(d) Occupation number of the ground states. The indices of the orbitals are ordered according to their HF energy, and the color of the lines represents the three-fold degenerate ground states. 
Here, we take $\theta=0.2^\circ$ and $u_d=\SI{46}{meV}$. 
}
\end{figure*}

Beyond PLG, HF calculation also suggested a topologically nontrivial band in rhombohedral hexalayer graphene (HLG)~\cite{Dong2023,Dong2023a,Zhou2023,Kwan2023}, alluding to a possible FQAH phase, which has been experimentally observed~\cite{XieJian2024}. 
In this appendix, we calculate the possible FQAH states in HLG at $2/3$ filling using Method III in the main text. 
The continuum model of HLG is given in Appendix~\ref{App:BM model}, and the interaction is the screened Coulomb interaction introduced in the main text with a relative permittivity $\epsilon=5$. 
Further, we take the twist angle and the displacement field close to the experimental conditions with $\theta=0.2^\circ$ and $u_d=46\,\text{meV}$.

In Fig.~\ref{Fig:HLG}(a), we perform the constrained HF calculation on a $N_x\times N_y=4\times 6$ mesh in the MBZ. 
Note that the lowest HF band in the HLG with $\theta=0.2^\circ$ is not as ideal as that in the PLG with $\theta=0.77^\circ$. 
Specifically, although the band is almost flat, there is a peak at $\Gamma_m$, which could potentially hinder the formation of an FQAH phase. 
Nonetheless, the ED results in a finite $4\times 6$ system indeed suggest the existence of the FQAH phase, featured by both the degeneracy in the energy spectrum in Fig.~\ref{Fig:HLG}(b) and the gap in the PES in Fig.~\ref{Fig:HLG}(c). 
We further verify that the energy of the FQAH ground state is lower than that of the FL ground state obtained by Method II. 

However, we also observe that the FQAH phase here may not be as stable as that in the PLG. 
First, we note the energy difference between the FQAH and FL state is just $\SI{0.4}{meV}$, much smaller than that in the PLG (which is about $\SI{4}{meV}$). 
Second, the energy peak causes an obvious particle number deficiency at $\Gamma_m$, as shown in Fig.~\ref{Fig:HLG}(d). 
Therefore, the occupation number is much less uniform than that in the PLG.

\bibliography{PLG_bib_v6.bib}

\begin{thebibliography}{62}%
\makeatletter
\providecommand \@ifxundefined [1]{%
 \@ifx{#1\undefined}
}%
\providecommand \@ifnum [1]{%
 \ifnum #1\expandafter \@firstoftwo
 \else \expandafter \@secondoftwo
 \fi
}%
\providecommand \@ifx [1]{%
 \ifx #1\expandafter \@firstoftwo
 \else \expandafter \@secondoftwo
 \fi
}%
\providecommand \natexlab [1]{#1}%
\providecommand \enquote  [1]{``#1''}%
\providecommand \bibnamefont  [1]{#1}%
\providecommand \bibfnamefont [1]{#1}%
\providecommand \citenamefont [1]{#1}%
\providecommand \href@noop [0]{\@secondoftwo}%
\providecommand \href [0]{\begingroup \@sanitize@url \@href}%
\providecommand \@href[1]{\@@startlink{#1}\@@href}%
\providecommand \@@href[1]{\endgroup#1\@@endlink}%
\providecommand \@sanitize@url [0]{\catcode `\\12\catcode `\$12\catcode
  `\&12\catcode `\#12\catcode `\^12\catcode `\_12\catcode `\%12\relax}%
\providecommand \@@startlink[1]{}%
\providecommand \@@endlink[0]{}%
\providecommand \url  [0]{\begingroup\@sanitize@url \@url }%
\providecommand \@url [1]{\endgroup\@href {#1}{\urlprefix }}%
\providecommand \urlprefix  [0]{URL }%
\providecommand \Eprint [0]{\href }%
\providecommand \doibase [0]{https://doi.org/}%
\providecommand \selectlanguage [0]{\@gobble}%
\providecommand \bibinfo  [0]{\@secondoftwo}%
\providecommand \bibfield  [0]{\@secondoftwo}%
\providecommand \translation [1]{[#1]}%
\providecommand \BibitemOpen [0]{}%
\providecommand \bibitemStop [0]{}%
\providecommand \bibitemNoStop [0]{.\EOS\space}%
\providecommand \EOS [0]{\spacefactor3000\relax}%
\providecommand \BibitemShut  [1]{\csname bibitem#1\endcsname}%
\let\auto@bib@innerbib\@empty
\bibitem [{\citenamefont {Tang}\ \emph {et~al.}(2011)\citenamefont {Tang},
  \citenamefont {Mei},\ and\ \citenamefont {Wen}}]{Tang2011}%
  \BibitemOpen
  \bibfield  {author} {\bibinfo {author} {\bibfnamefont {E.}~\bibnamefont
  {Tang}}, \bibinfo {author} {\bibfnamefont {J.-W.}\ \bibnamefont {Mei}},\ and\
  \bibinfo {author} {\bibfnamefont {X.-G.}\ \bibnamefont {Wen}},\ }\bibfield
  {title} {\bibinfo {title} {{High-Temperature Fractional Quantum Hall
  States}},\ }\href {https://doi.org/10.1103/physrevlett.106.236802} {\bibfield
   {journal} {\bibinfo  {journal} {Phys. Rev. Lett.}\ }\textbf {\bibinfo
  {volume} {106}},\ \bibinfo {pages} {236802} (\bibinfo {year}
  {2011})}\BibitemShut {NoStop}%
\bibitem [{\citenamefont {Sun}\ \emph {et~al.}(2011)\citenamefont {Sun},
  \citenamefont {Gu}, \citenamefont {Katsura},\ and\ \citenamefont
  {Das~Sarma}}]{Sun2011}%
  \BibitemOpen
  \bibfield  {author} {\bibinfo {author} {\bibfnamefont {K.}~\bibnamefont
  {Sun}}, \bibinfo {author} {\bibfnamefont {Z.}~\bibnamefont {Gu}}, \bibinfo
  {author} {\bibfnamefont {H.}~\bibnamefont {Katsura}},\ and\ \bibinfo {author}
  {\bibfnamefont {S.}~\bibnamefont {Das~Sarma}},\ }\bibfield  {title} {\bibinfo
  {title} {{Nearly Flatbands with Nontrivial Topology}},\ }\href
  {https://doi.org/10.1103/physrevlett.106.236803} {\bibfield  {journal}
  {\bibinfo  {journal} {Phys. Rev. Lett.}\ }\textbf {\bibinfo {volume} {106}},\
  \bibinfo {pages} {236803} (\bibinfo {year} {2011})}\BibitemShut {NoStop}%
\bibitem [{\citenamefont {Neupert}\ \emph {et~al.}(2011)\citenamefont
  {Neupert}, \citenamefont {Santos}, \citenamefont {Chamon},\ and\
  \citenamefont {Mudry}}]{Neupert2011}%
  \BibitemOpen
  \bibfield  {author} {\bibinfo {author} {\bibfnamefont {T.}~\bibnamefont
  {Neupert}}, \bibinfo {author} {\bibfnamefont {L.}~\bibnamefont {Santos}},
  \bibinfo {author} {\bibfnamefont {C.}~\bibnamefont {Chamon}},\ and\ \bibinfo
  {author} {\bibfnamefont {C.}~\bibnamefont {Mudry}},\ }\bibfield  {title}
  {\bibinfo {title} {{Fractional Quantum {Hall} States at Zero Magnetic
  Field}},\ }\href {https://doi.org/10.1103/physrevlett.106.236804} {\bibfield
  {journal} {\bibinfo  {journal} {Phys. Rev. Lett.}\ }\textbf {\bibinfo
  {volume} {106}},\ \bibinfo {pages} {236804} (\bibinfo {year}
  {2011})}\BibitemShut {NoStop}%
\bibitem [{\citenamefont {Regnault}\ and\ \citenamefont
  {Bernevig}(2011)}]{Regnault2011}%
  \BibitemOpen
  \bibfield  {author} {\bibinfo {author} {\bibfnamefont {N.}~\bibnamefont
  {Regnault}}\ and\ \bibinfo {author} {\bibfnamefont {B.~A.}\ \bibnamefont
  {Bernevig}},\ }\bibfield  {title} {\bibinfo {title} {{Fractional Chern
  Insulator}},\ }\href {https://doi.org/10.1103/physrevx.1.021014} {\bibfield
  {journal} {\bibinfo  {journal} {Phys. Rev. X}\ }\textbf {\bibinfo {volume}
  {1}},\ \bibinfo {pages} {021014} (\bibinfo {year} {2011})}\BibitemShut
  {NoStop}%
\bibitem [{\citenamefont {Sheng}\ \emph {et~al.}(2011)\citenamefont {Sheng},
  \citenamefont {Gu}, \citenamefont {Sun},\ and\ \citenamefont
  {Sheng}}]{Sheng2011}%
  \BibitemOpen
  \bibfield  {author} {\bibinfo {author} {\bibfnamefont {D.}~\bibnamefont
  {Sheng}}, \bibinfo {author} {\bibfnamefont {Z.-C.}\ \bibnamefont {Gu}},
  \bibinfo {author} {\bibfnamefont {K.}~\bibnamefont {Sun}},\ and\ \bibinfo
  {author} {\bibfnamefont {L.}~\bibnamefont {Sheng}},\ }\bibfield  {title}
  {\bibinfo {title} {{Fractional quantum Hall effect in the absence of Landau
  levels}},\ }\href {https://doi.org/10.1038/ncomms1380} {\bibfield  {journal}
  {\bibinfo  {journal} {Nat. Commun.}\ }\textbf {\bibinfo {volume} {2}},\
  \bibinfo {pages} {389} (\bibinfo {year} {2011})}\BibitemShut {NoStop}%
\bibitem [{\citenamefont {Spanton}\ \emph {et~al.}(2018)\citenamefont
  {Spanton}, \citenamefont {Zibrov}, \citenamefont {Zhou}, \citenamefont
  {Taniguchi}, \citenamefont {Watanabe}, \citenamefont {Zaletel},\ and\
  \citenamefont {Young}}]{Spanton2018}%
  \BibitemOpen
  \bibfield  {author} {\bibinfo {author} {\bibfnamefont {E.~M.}\ \bibnamefont
  {Spanton}}, \bibinfo {author} {\bibfnamefont {A.~A.}\ \bibnamefont {Zibrov}},
  \bibinfo {author} {\bibfnamefont {H.}~\bibnamefont {Zhou}}, \bibinfo {author}
  {\bibfnamefont {T.}~\bibnamefont {Taniguchi}}, \bibinfo {author}
  {\bibfnamefont {K.}~\bibnamefont {Watanabe}}, \bibinfo {author}
  {\bibfnamefont {M.~P.}\ \bibnamefont {Zaletel}},\ and\ \bibinfo {author}
  {\bibfnamefont {A.~F.}\ \bibnamefont {Young}},\ }\bibfield  {title} {\bibinfo
  {title} {{Observation of fractional Chern insulators in a van der Waals
  heterostructure}},\ }\href {https://doi.org/10.1126/science.aan8458}
  {\bibfield  {journal} {\bibinfo  {journal} {Science}\ }\textbf {\bibinfo
  {volume} {360}},\ \bibinfo {pages} {62} (\bibinfo {year} {2018})}\BibitemShut
  {NoStop}%
\bibitem [{\citenamefont {Xie}\ \emph {et~al.}(2021)\citenamefont {Xie},
  \citenamefont {Pierce}, \citenamefont {Park}, \citenamefont {Parker},
  \citenamefont {Khalaf}, \citenamefont {Ledwith}, \citenamefont {Cao},
  \citenamefont {Lee}, \citenamefont {Chen}, \citenamefont {Forrester},
  \citenamefont {Watanabe}, \citenamefont {Taniguchi}, \citenamefont
  {Vishwanath}, \citenamefont {Jarillo-Herrero},\ and\ \citenamefont
  {Yacoby}}]{Xie2021}%
  \BibitemOpen
  \bibfield  {author} {\bibinfo {author} {\bibfnamefont {Y.}~\bibnamefont
  {Xie}}, \bibinfo {author} {\bibfnamefont {A.~T.}\ \bibnamefont {Pierce}},
  \bibinfo {author} {\bibfnamefont {J.~M.}\ \bibnamefont {Park}}, \bibinfo
  {author} {\bibfnamefont {D.~E.}\ \bibnamefont {Parker}}, \bibinfo {author}
  {\bibfnamefont {E.}~\bibnamefont {Khalaf}}, \bibinfo {author} {\bibfnamefont
  {P.}~\bibnamefont {Ledwith}}, \bibinfo {author} {\bibfnamefont
  {Y.}~\bibnamefont {Cao}}, \bibinfo {author} {\bibfnamefont {S.~H.}\
  \bibnamefont {Lee}}, \bibinfo {author} {\bibfnamefont {S.}~\bibnamefont
  {Chen}}, \bibinfo {author} {\bibfnamefont {P.~R.}\ \bibnamefont {Forrester}},
  \bibinfo {author} {\bibfnamefont {K.}~\bibnamefont {Watanabe}}, \bibinfo
  {author} {\bibfnamefont {T.}~\bibnamefont {Taniguchi}}, \bibinfo {author}
  {\bibfnamefont {A.}~\bibnamefont {Vishwanath}}, \bibinfo {author}
  {\bibfnamefont {P.}~\bibnamefont {Jarillo-Herrero}},\ and\ \bibinfo {author}
  {\bibfnamefont {A.}~\bibnamefont {Yacoby}},\ }\bibfield  {title} {\bibinfo
  {title} {{Fractional Chern insulators in magic-angle twisted bilayer
  graphene}},\ }\href {https://doi.org/10.1038/s41586-021-04002-3} {\bibfield
  {journal} {\bibinfo  {journal} {Nature}\ }\textbf {\bibinfo {volume} {600}},\
  \bibinfo {pages} {439} (\bibinfo {year} {2021})}\BibitemShut {NoStop}%
\bibitem [{\citenamefont {Cai}\ \emph {et~al.}(2023)\citenamefont {Cai},
  \citenamefont {Anderson}, \citenamefont {Wang}, \citenamefont {Zhang},
  \citenamefont {Liu}, \citenamefont {Holtzmann}, \citenamefont {Zhang},
  \citenamefont {Fan}, \citenamefont {Taniguchi}, \citenamefont {Watanabe},
  \citenamefont {Ran}, \citenamefont {Cao}, \citenamefont {Fu}, \citenamefont
  {Xiao}, \citenamefont {Yao},\ and\ \citenamefont {Xu}}]{Cai2023}%
  \BibitemOpen
  \bibfield  {author} {\bibinfo {author} {\bibfnamefont {J.}~\bibnamefont
  {Cai}}, \bibinfo {author} {\bibfnamefont {E.}~\bibnamefont {Anderson}},
  \bibinfo {author} {\bibfnamefont {C.}~\bibnamefont {Wang}}, \bibinfo {author}
  {\bibfnamefont {X.}~\bibnamefont {Zhang}}, \bibinfo {author} {\bibfnamefont
  {X.}~\bibnamefont {Liu}}, \bibinfo {author} {\bibfnamefont {W.}~\bibnamefont
  {Holtzmann}}, \bibinfo {author} {\bibfnamefont {Y.}~\bibnamefont {Zhang}},
  \bibinfo {author} {\bibfnamefont {F.}~\bibnamefont {Fan}}, \bibinfo {author}
  {\bibfnamefont {T.}~\bibnamefont {Taniguchi}}, \bibinfo {author}
  {\bibfnamefont {K.}~\bibnamefont {Watanabe}}, \bibinfo {author}
  {\bibfnamefont {Y.}~\bibnamefont {Ran}}, \bibinfo {author} {\bibfnamefont
  {T.}~\bibnamefont {Cao}}, \bibinfo {author} {\bibfnamefont {L.}~\bibnamefont
  {Fu}}, \bibinfo {author} {\bibfnamefont {D.}~\bibnamefont {Xiao}}, \bibinfo
  {author} {\bibfnamefont {W.}~\bibnamefont {Yao}},\ and\ \bibinfo {author}
  {\bibfnamefont {X.}~\bibnamefont {Xu}},\ }\bibfield  {title} {\bibinfo
  {title} {{Signatures of fractional quantum anomalous Hall states in twisted
  MoTe$_2$}},\ }\href {https://doi.org/10.1038/s41586-023-06289-w} {\bibfield
  {journal} {\bibinfo  {journal} {Nature}\ }\textbf {\bibinfo {volume} {622}},\
  \bibinfo {pages} {63} (\bibinfo {year} {2023})}\BibitemShut {NoStop}%
\bibitem [{\citenamefont {Zeng}\ \emph {et~al.}(2023)\citenamefont {Zeng},
  \citenamefont {Xia}, \citenamefont {Kang}, \citenamefont {Zhu}, \citenamefont
  {Knüppel}, \citenamefont {Vaswani}, \citenamefont {Watanabe}, \citenamefont
  {Taniguchi}, \citenamefont {Mak},\ and\ \citenamefont {Shan}}]{Zeng2023}%
  \BibitemOpen
  \bibfield  {author} {\bibinfo {author} {\bibfnamefont {Y.}~\bibnamefont
  {Zeng}}, \bibinfo {author} {\bibfnamefont {Z.}~\bibnamefont {Xia}}, \bibinfo
  {author} {\bibfnamefont {K.}~\bibnamefont {Kang}}, \bibinfo {author}
  {\bibfnamefont {J.}~\bibnamefont {Zhu}}, \bibinfo {author} {\bibfnamefont
  {P.}~\bibnamefont {Knüppel}}, \bibinfo {author} {\bibfnamefont
  {C.}~\bibnamefont {Vaswani}}, \bibinfo {author} {\bibfnamefont
  {K.}~\bibnamefont {Watanabe}}, \bibinfo {author} {\bibfnamefont
  {T.}~\bibnamefont {Taniguchi}}, \bibinfo {author} {\bibfnamefont {K.~F.}\
  \bibnamefont {Mak}},\ and\ \bibinfo {author} {\bibfnamefont {J.}~\bibnamefont
  {Shan}},\ }\bibfield  {title} {\bibinfo {title} {{Thermodynamic evidence of
  fractional Chern insulator in moir\'{e} MoTe$_2$}},\ }\href
  {https://doi.org/10.1038/s41586-023-06452-3} {\bibfield  {journal} {\bibinfo
  {journal} {Nature}\ }\textbf {\bibinfo {volume} {622}},\ \bibinfo {pages}
  {69} (\bibinfo {year} {2023})}\BibitemShut {NoStop}%
\bibitem [{\citenamefont {Park}\ \emph {et~al.}(2023)\citenamefont {Park},
  \citenamefont {Cai}, \citenamefont {Anderson}, \citenamefont {Zhang},
  \citenamefont {Zhu}, \citenamefont {Liu}, \citenamefont {Wang}, \citenamefont
  {Holtzmann}, \citenamefont {Hu}, \citenamefont {Liu}, \citenamefont
  {Taniguchi}, \citenamefont {Watanabe}, \citenamefont {Chu}, \citenamefont
  {Cao}, \citenamefont {Fu}, \citenamefont {Yao}, \citenamefont {Chang},
  \citenamefont {Cobden}, \citenamefont {Xiao},\ and\ \citenamefont
  {Xu}}]{Park2023}%
  \BibitemOpen
  \bibfield  {author} {\bibinfo {author} {\bibfnamefont {H.}~\bibnamefont
  {Park}}, \bibinfo {author} {\bibfnamefont {J.}~\bibnamefont {Cai}}, \bibinfo
  {author} {\bibfnamefont {E.}~\bibnamefont {Anderson}}, \bibinfo {author}
  {\bibfnamefont {Y.}~\bibnamefont {Zhang}}, \bibinfo {author} {\bibfnamefont
  {J.}~\bibnamefont {Zhu}}, \bibinfo {author} {\bibfnamefont {X.}~\bibnamefont
  {Liu}}, \bibinfo {author} {\bibfnamefont {C.}~\bibnamefont {Wang}}, \bibinfo
  {author} {\bibfnamefont {W.}~\bibnamefont {Holtzmann}}, \bibinfo {author}
  {\bibfnamefont {C.}~\bibnamefont {Hu}}, \bibinfo {author} {\bibfnamefont
  {Z.}~\bibnamefont {Liu}}, \bibinfo {author} {\bibfnamefont {T.}~\bibnamefont
  {Taniguchi}}, \bibinfo {author} {\bibfnamefont {K.}~\bibnamefont {Watanabe}},
  \bibinfo {author} {\bibfnamefont {J.-H.}\ \bibnamefont {Chu}}, \bibinfo
  {author} {\bibfnamefont {T.}~\bibnamefont {Cao}}, \bibinfo {author}
  {\bibfnamefont {L.}~\bibnamefont {Fu}}, \bibinfo {author} {\bibfnamefont
  {W.}~\bibnamefont {Yao}}, \bibinfo {author} {\bibfnamefont {C.-Z.}\
  \bibnamefont {Chang}}, \bibinfo {author} {\bibfnamefont {D.}~\bibnamefont
  {Cobden}}, \bibinfo {author} {\bibfnamefont {D.}~\bibnamefont {Xiao}},\ and\
  \bibinfo {author} {\bibfnamefont {X.}~\bibnamefont {Xu}},\ }\bibfield
  {title} {\bibinfo {title} {{Observation of fractionally quantized anomalous
  Hall effect}},\ }\href {https://doi.org/10.1038/s41586-023-06536-0}
  {\bibfield  {journal} {\bibinfo  {journal} {Nature}\ }\textbf {\bibinfo
  {volume} {622}},\ \bibinfo {pages} {74} (\bibinfo {year} {2023})}\BibitemShut
  {NoStop}%
\bibitem [{\citenamefont {Xu}\ \emph {et~al.}(2023)\citenamefont {Xu},
  \citenamefont {Sun}, \citenamefont {Jia}, \citenamefont {Liu}, \citenamefont
  {Xu}, \citenamefont {Li}, \citenamefont {Gu}, \citenamefont {Watanabe},
  \citenamefont {Taniguchi}, \citenamefont {Tong}, \citenamefont {Jia},
  \citenamefont {Shi}, \citenamefont {Jiang}, \citenamefont {Zhang},
  \citenamefont {Liu},\ and\ \citenamefont {Li}}]{Xu2023}%
  \BibitemOpen
  \bibfield  {author} {\bibinfo {author} {\bibfnamefont {F.}~\bibnamefont
  {Xu}}, \bibinfo {author} {\bibfnamefont {Z.}~\bibnamefont {Sun}}, \bibinfo
  {author} {\bibfnamefont {T.}~\bibnamefont {Jia}}, \bibinfo {author}
  {\bibfnamefont {C.}~\bibnamefont {Liu}}, \bibinfo {author} {\bibfnamefont
  {C.}~\bibnamefont {Xu}}, \bibinfo {author} {\bibfnamefont {C.}~\bibnamefont
  {Li}}, \bibinfo {author} {\bibfnamefont {Y.}~\bibnamefont {Gu}}, \bibinfo
  {author} {\bibfnamefont {K.}~\bibnamefont {Watanabe}}, \bibinfo {author}
  {\bibfnamefont {T.}~\bibnamefont {Taniguchi}}, \bibinfo {author}
  {\bibfnamefont {B.}~\bibnamefont {Tong}}, \bibinfo {author} {\bibfnamefont
  {J.}~\bibnamefont {Jia}}, \bibinfo {author} {\bibfnamefont {Z.}~\bibnamefont
  {Shi}}, \bibinfo {author} {\bibfnamefont {S.}~\bibnamefont {Jiang}}, \bibinfo
  {author} {\bibfnamefont {Y.}~\bibnamefont {Zhang}}, \bibinfo {author}
  {\bibfnamefont {X.}~\bibnamefont {Liu}},\ and\ \bibinfo {author}
  {\bibfnamefont {T.}~\bibnamefont {Li}},\ }\bibfield  {title} {\bibinfo
  {title} {{Observation of Integer and Fractional Quantum Anomalous Hall
  Effects in Twisted Bilayer MoTe$_2$}},\ }\href
  {https://doi.org/10.1103/physrevx.13.031037} {\bibfield  {journal} {\bibinfo
  {journal} {Phys. Rev. X}\ }\textbf {\bibinfo {volume} {13}},\ \bibinfo
  {pages} {031037} (\bibinfo {year} {2023})}\BibitemShut {NoStop}%
\bibitem [{\citenamefont {Lu}\ \emph {et~al.}(2024)\citenamefont {Lu},
  \citenamefont {Han}, \citenamefont {Yao}, \citenamefont {Reddy},
  \citenamefont {Yang}, \citenamefont {Seo}, \citenamefont {Watanabe},
  \citenamefont {Taniguchi}, \citenamefont {Fu},\ and\ \citenamefont
  {Ju}}]{Lu2024}%
  \BibitemOpen
  \bibfield  {author} {\bibinfo {author} {\bibfnamefont {Z.}~\bibnamefont
  {Lu}}, \bibinfo {author} {\bibfnamefont {T.}~\bibnamefont {Han}}, \bibinfo
  {author} {\bibfnamefont {Y.}~\bibnamefont {Yao}}, \bibinfo {author}
  {\bibfnamefont {A.~P.}\ \bibnamefont {Reddy}}, \bibinfo {author}
  {\bibfnamefont {J.}~\bibnamefont {Yang}}, \bibinfo {author} {\bibfnamefont
  {J.}~\bibnamefont {Seo}}, \bibinfo {author} {\bibfnamefont {K.}~\bibnamefont
  {Watanabe}}, \bibinfo {author} {\bibfnamefont {T.}~\bibnamefont {Taniguchi}},
  \bibinfo {author} {\bibfnamefont {L.}~\bibnamefont {Fu}},\ and\ \bibinfo
  {author} {\bibfnamefont {L.}~\bibnamefont {Ju}},\ }\bibfield  {title}
  {\bibinfo {title} {{Fractional quantum anomalous Hall effect in multilayer
  graphene}},\ }\href {https://doi.org/10.1038/s41586-023-07010-7} {\bibfield
  {journal} {\bibinfo  {journal} {Nature}\ }\textbf {\bibinfo {volume} {626}},\
  \bibinfo {pages} {759} (\bibinfo {year} {2024})}\BibitemShut {NoStop}%
\bibitem [{\citenamefont {Xie}\ \emph {et~al.}(2024)\citenamefont {Xie},
  \citenamefont {Huo}, \citenamefont {Lu}, \citenamefont {Feng}, \citenamefont
  {Zhang}, \citenamefont {Wang}, \citenamefont {Yang}, \citenamefont
  {Watanabe}, \citenamefont {Taniguchi}, \citenamefont {Liu}, \citenamefont
  {Song}, \citenamefont {Xie}, \citenamefont {Liu},\ and\ \citenamefont
  {Lu}}]{XieJian2024}%
  \BibitemOpen
  \bibfield  {author} {\bibinfo {author} {\bibfnamefont {J.}~\bibnamefont
  {Xie}}, \bibinfo {author} {\bibfnamefont {Z.}~\bibnamefont {Huo}}, \bibinfo
  {author} {\bibfnamefont {X.}~\bibnamefont {Lu}}, \bibinfo {author}
  {\bibfnamefont {Z.}~\bibnamefont {Feng}}, \bibinfo {author} {\bibfnamefont
  {Z.}~\bibnamefont {Zhang}}, \bibinfo {author} {\bibfnamefont
  {W.}~\bibnamefont {Wang}}, \bibinfo {author} {\bibfnamefont {Q.}~\bibnamefont
  {Yang}}, \bibinfo {author} {\bibfnamefont {K.}~\bibnamefont {Watanabe}},
  \bibinfo {author} {\bibfnamefont {T.}~\bibnamefont {Taniguchi}}, \bibinfo
  {author} {\bibfnamefont {K.}~\bibnamefont {Liu}}, \bibinfo {author}
  {\bibfnamefont {Z.}~\bibnamefont {Song}}, \bibinfo {author} {\bibfnamefont
  {X.~C.}\ \bibnamefont {Xie}}, \bibinfo {author} {\bibfnamefont
  {J.}~\bibnamefont {Liu}},\ and\ \bibinfo {author} {\bibfnamefont
  {X.}~\bibnamefont {Lu}},\ }\href {https://doi.org/10.48550/ARXIV.2405.16944}
  {\bibinfo {title} {{Even- and Odd-denominator Fractional Quantum Anomalous
  Hall Effect in Graphene Moire Superlattices}}} (\bibinfo {year} {2024}),\
  \Eprint {https://arxiv.org/abs/2405.16944} {arXiv:2405.16944
  [cond-mat.mes-hall]} \BibitemShut {NoStop}%
\bibitem [{\citenamefont {Qi}(2011)}]{Qi2011}%
  \BibitemOpen
  \bibfield  {author} {\bibinfo {author} {\bibfnamefont {X.-L.}\ \bibnamefont
  {Qi}},\ }\bibfield  {title} {\bibinfo {title} {{Generic Wave-Function
  Description of Fractional Quantum Anomalous Hall States and Fractional
  Topological Insulators}},\ }\href
  {https://doi.org/10.1103/physrevlett.107.126803} {\bibfield  {journal}
  {\bibinfo  {journal} {Phys. Rev. Lett.}\ }\textbf {\bibinfo {volume} {107}},\
  \bibinfo {pages} {126803} (\bibinfo {year} {2011})}\BibitemShut {NoStop}%
\bibitem [{\citenamefont {Wu}\ \emph {et~al.}(2012{\natexlab{a}})\citenamefont
  {Wu}, \citenamefont {Regnault},\ and\ \citenamefont {Bernevig}}]{Wu2012a}%
  \BibitemOpen
  \bibfield  {author} {\bibinfo {author} {\bibfnamefont {Y.-L.}\ \bibnamefont
  {Wu}}, \bibinfo {author} {\bibfnamefont {N.}~\bibnamefont {Regnault}},\ and\
  \bibinfo {author} {\bibfnamefont {B.~A.}\ \bibnamefont {Bernevig}},\
  }\bibfield  {title} {\bibinfo {title} {{Gauge-fixed Wannier wave functions
  for fractional topological insulators}},\ }\href
  {https://doi.org/10.1103/physrevb.86.085129} {\bibfield  {journal} {\bibinfo
  {journal} {Phys. Rev. B}\ }\textbf {\bibinfo {volume} {86}},\ \bibinfo
  {pages} {085129} (\bibinfo {year} {2012}{\natexlab{a}})}\BibitemShut
  {NoStop}%
\bibitem [{\citenamefont {Parameswaran}\ \emph {et~al.}(2013)\citenamefont
  {Parameswaran}, \citenamefont {Roy},\ and\ \citenamefont
  {Sondhi}}]{Parameswaran2013}%
  \BibitemOpen
  \bibfield  {author} {\bibinfo {author} {\bibfnamefont {S.~A.}\ \bibnamefont
  {Parameswaran}}, \bibinfo {author} {\bibfnamefont {R.}~\bibnamefont {Roy}},\
  and\ \bibinfo {author} {\bibfnamefont {S.~L.}\ \bibnamefont {Sondhi}},\
  }\bibfield  {title} {\bibinfo {title} {{Fractional quantum Hall physics in
  topological flat bands}},\ }\href
  {https://doi.org/10.1016/j.crhy.2013.04.003} {\bibfield  {journal} {\bibinfo
  {journal} {Comptes Rendus. Physique}\ }\textbf {\bibinfo {volume} {14}},\
  \bibinfo {pages} {816} (\bibinfo {year} {2013})}\BibitemShut {NoStop}%
\bibitem [{\citenamefont {Jackson}\ \emph {et~al.}(2015)\citenamefont
  {Jackson}, \citenamefont {M\"{o}ller},\ and\ \citenamefont
  {Roy}}]{Jackson2015}%
  \BibitemOpen
  \bibfield  {author} {\bibinfo {author} {\bibfnamefont {T.~S.}\ \bibnamefont
  {Jackson}}, \bibinfo {author} {\bibfnamefont {G.}~\bibnamefont
  {M\"{o}ller}},\ and\ \bibinfo {author} {\bibfnamefont {R.}~\bibnamefont
  {Roy}},\ }\bibfield  {title} {\bibinfo {title} {{Geometric stability of
  topological lattice phases}},\ }\href {https://doi.org/10.1038/ncomms9629}
  {\bibfield  {journal} {\bibinfo  {journal} {Nat. Commun.}\ }\textbf {\bibinfo
  {volume} {6}},\ \bibinfo {pages} {8629} (\bibinfo {year} {2015})}\BibitemShut
  {NoStop}%
\bibitem [{\citenamefont {Claassen}\ \emph {et~al.}(2015)\citenamefont
  {Claassen}, \citenamefont {Lee}, \citenamefont {Thomale}, \citenamefont
  {Qi},\ and\ \citenamefont {Devereaux}}]{Claassen2015}%
  \BibitemOpen
  \bibfield  {author} {\bibinfo {author} {\bibfnamefont {M.}~\bibnamefont
  {Claassen}}, \bibinfo {author} {\bibfnamefont {C.~H.}\ \bibnamefont {Lee}},
  \bibinfo {author} {\bibfnamefont {R.}~\bibnamefont {Thomale}}, \bibinfo
  {author} {\bibfnamefont {X.-L.}\ \bibnamefont {Qi}},\ and\ \bibinfo {author}
  {\bibfnamefont {T.~P.}\ \bibnamefont {Devereaux}},\ }\bibfield  {title}
  {\bibinfo {title} {{Position-Momentum Duality and Fractional Quantum Hall
  Effect in Chern Insulators}},\ }\href
  {https://doi.org/10.1103/physrevlett.114.236802} {\bibfield  {journal}
  {\bibinfo  {journal} {Phys. Rev. Lett.}\ }\textbf {\bibinfo {volume} {114}},\
  \bibinfo {pages} {236802} (\bibinfo {year} {2015})}\BibitemShut {NoStop}%
\bibitem [{\citenamefont {Tarnopolsky}\ \emph {et~al.}(2019)\citenamefont
  {Tarnopolsky}, \citenamefont {Kruchkov},\ and\ \citenamefont
  {Vishwanath}}]{Tarnopolsky2019}%
  \BibitemOpen
  \bibfield  {author} {\bibinfo {author} {\bibfnamefont {G.}~\bibnamefont
  {Tarnopolsky}}, \bibinfo {author} {\bibfnamefont {A.~J.}\ \bibnamefont
  {Kruchkov}},\ and\ \bibinfo {author} {\bibfnamefont {A.}~\bibnamefont
  {Vishwanath}},\ }\bibfield  {title} {\bibinfo {title} {{Origin of Magic
  Angles in Twisted Bilayer Graphene}},\ }\href
  {https://doi.org/10.1103/physrevlett.122.106405} {\bibfield  {journal}
  {\bibinfo  {journal} {Phys. Rev. Lett.}\ }\textbf {\bibinfo {volume} {122}},\
  \bibinfo {pages} {106405} (\bibinfo {year} {2019})}\BibitemShut {NoStop}%
\bibitem [{\citenamefont {Ledwith}\ \emph {et~al.}(2020)\citenamefont
  {Ledwith}, \citenamefont {Tarnopolsky}, \citenamefont {Khalaf},\ and\
  \citenamefont {Vishwanath}}]{Ledwith2020}%
  \BibitemOpen
  \bibfield  {author} {\bibinfo {author} {\bibfnamefont {P.~J.}\ \bibnamefont
  {Ledwith}}, \bibinfo {author} {\bibfnamefont {G.}~\bibnamefont
  {Tarnopolsky}}, \bibinfo {author} {\bibfnamefont {E.}~\bibnamefont
  {Khalaf}},\ and\ \bibinfo {author} {\bibfnamefont {A.}~\bibnamefont
  {Vishwanath}},\ }\bibfield  {title} {\bibinfo {title} {Fractional chern
  insulator states in twisted bilayer graphene: An analytical approach},\
  }\href {https://doi.org/10.1103/physrevresearch.2.023237} {\bibfield
  {journal} {\bibinfo  {journal} {Phys. Rev. Res.}\ }\textbf {\bibinfo {volume}
  {2}},\ \bibinfo {pages} {023237} (\bibinfo {year} {2020})}\BibitemShut
  {NoStop}%
\bibitem [{\citenamefont {Wang}\ \emph
  {et~al.}(2021{\natexlab{a}})\citenamefont {Wang}, \citenamefont {Cano},
  \citenamefont {Millis}, \citenamefont {Liu},\ and\ \citenamefont
  {Yang}}]{Wang2021}%
  \BibitemOpen
  \bibfield  {author} {\bibinfo {author} {\bibfnamefont {J.}~\bibnamefont
  {Wang}}, \bibinfo {author} {\bibfnamefont {J.}~\bibnamefont {Cano}}, \bibinfo
  {author} {\bibfnamefont {A.~J.}\ \bibnamefont {Millis}}, \bibinfo {author}
  {\bibfnamefont {Z.}~\bibnamefont {Liu}},\ and\ \bibinfo {author}
  {\bibfnamefont {B.}~\bibnamefont {Yang}},\ }\bibfield  {title} {\bibinfo
  {title} {{Exact Landau Level Description of Geometry and Interaction in a
  Flatband}},\ }\href {https://doi.org/10.1103/physrevlett.127.246403}
  {\bibfield  {journal} {\bibinfo  {journal} {Phys. Rev. Lett.}\ }\textbf
  {\bibinfo {volume} {127}},\ \bibinfo {pages} {246403} (\bibinfo {year}
  {2021}{\natexlab{a}})}\BibitemShut {NoStop}%
\bibitem [{\citenamefont {Wang}\ \emph
  {et~al.}(2021{\natexlab{b}})\citenamefont {Wang}, \citenamefont {Zheng},
  \citenamefont {Millis},\ and\ \citenamefont {Cano}}]{Wang2021a}%
  \BibitemOpen
  \bibfield  {author} {\bibinfo {author} {\bibfnamefont {J.}~\bibnamefont
  {Wang}}, \bibinfo {author} {\bibfnamefont {Y.}~\bibnamefont {Zheng}},
  \bibinfo {author} {\bibfnamefont {A.~J.}\ \bibnamefont {Millis}},\ and\
  \bibinfo {author} {\bibfnamefont {J.}~\bibnamefont {Cano}},\ }\bibfield
  {title} {\bibinfo {title} {{Chiral approximation to twisted bilayer graphene:
  Exact intravalley inversion symmetry, nodal structure, and implications for
  higher magic angles}},\ }\href
  {https://doi.org/10.1103/physrevresearch.3.023155} {\bibfield  {journal}
  {\bibinfo  {journal} {Phys. Rev. Res.}\ }\textbf {\bibinfo {volume} {3}},\
  \bibinfo {pages} {023155} (\bibinfo {year} {2021}{\natexlab{b}})}\BibitemShut
  {NoStop}%
\bibitem [{\citenamefont {Ozawa}\ and\ \citenamefont {Mera}(2021)}]{Ozawa2021}%
  \BibitemOpen
  \bibfield  {author} {\bibinfo {author} {\bibfnamefont {T.}~\bibnamefont
  {Ozawa}}\ and\ \bibinfo {author} {\bibfnamefont {B.}~\bibnamefont {Mera}},\
  }\bibfield  {title} {\bibinfo {title} {{Relations between topology and the
  quantum metric for Chern insulators}},\ }\href
  {https://doi.org/10.1103/physrevb.104.045103} {\bibfield  {journal} {\bibinfo
   {journal} {Phys. Rev. B}\ }\textbf {\bibinfo {volume} {104}},\ \bibinfo
  {pages} {045103} (\bibinfo {year} {2021})}\BibitemShut {NoStop}%
\bibitem [{\citenamefont {Mera}\ and\ \citenamefont {Ozawa}(2021)}]{Mera2021}%
  \BibitemOpen
  \bibfield  {author} {\bibinfo {author} {\bibfnamefont {B.}~\bibnamefont
  {Mera}}\ and\ \bibinfo {author} {\bibfnamefont {T.}~\bibnamefont {Ozawa}},\
  }\bibfield  {title} {\bibinfo {title} {{K\"{a}hler geometry and Chern
  insulators: Relations between topology and the quantum metric}},\ }\href
  {https://doi.org/10.1103/physrevb.104.045104} {\bibfield  {journal} {\bibinfo
   {journal} {Phys. Rev. B}\ }\textbf {\bibinfo {volume} {104}},\ \bibinfo
  {pages} {045104} (\bibinfo {year} {2021})}\BibitemShut {NoStop}%
\bibitem [{\citenamefont {Wang}\ and\ \citenamefont {Liu}(2022)}]{Wang2022}%
  \BibitemOpen
  \bibfield  {author} {\bibinfo {author} {\bibfnamefont {J.}~\bibnamefont
  {Wang}}\ and\ \bibinfo {author} {\bibfnamefont {Z.}~\bibnamefont {Liu}},\
  }\bibfield  {title} {\bibinfo {title} {{Hierarchy of Ideal Flatbands in
  Chiral Twisted Multilayer Graphene Models}},\ }\href
  {https://doi.org/10.1103/physrevlett.128.176403} {\bibfield  {journal}
  {\bibinfo  {journal} {Phys. Rev. Lett.}\ }\textbf {\bibinfo {volume} {128}},\
  \bibinfo {pages} {176403} (\bibinfo {year} {2022})}\BibitemShut {NoStop}%
\bibitem [{\citenamefont {Ledwith}\ \emph {et~al.}(2022)\citenamefont
  {Ledwith}, \citenamefont {Vishwanath},\ and\ \citenamefont
  {Khalaf}}]{Ledwith2022}%
  \BibitemOpen
  \bibfield  {author} {\bibinfo {author} {\bibfnamefont {P.~J.}\ \bibnamefont
  {Ledwith}}, \bibinfo {author} {\bibfnamefont {A.}~\bibnamefont
  {Vishwanath}},\ and\ \bibinfo {author} {\bibfnamefont {E.}~\bibnamefont
  {Khalaf}},\ }\bibfield  {title} {\bibinfo {title} {{Family of Ideal Chern
  Flatbands with Arbitrary Chern Number in Chiral Twisted Graphene
  Multilayers}},\ }\href {https://doi.org/10.1103/physrevlett.128.176404}
  {\bibfield  {journal} {\bibinfo  {journal} {Phys. Rev. Lett.}\ }\textbf
  {\bibinfo {volume} {128}},\ \bibinfo {pages} {176404} (\bibinfo {year}
  {2022})}\BibitemShut {NoStop}%
\bibitem [{\citenamefont {Ledwith}\ \emph {et~al.}(2023)\citenamefont
  {Ledwith}, \citenamefont {Vishwanath},\ and\ \citenamefont
  {Parker}}]{Ledwith2023}%
  \BibitemOpen
  \bibfield  {author} {\bibinfo {author} {\bibfnamefont {P.~J.}\ \bibnamefont
  {Ledwith}}, \bibinfo {author} {\bibfnamefont {A.}~\bibnamefont
  {Vishwanath}},\ and\ \bibinfo {author} {\bibfnamefont {D.~E.}\ \bibnamefont
  {Parker}},\ }\bibfield  {title} {\bibinfo {title} {{Vortexability: A unifying
  criterion for ideal fractional Chern insulators}},\ }\href
  {https://doi.org/10.1103/physrevb.108.205144} {\bibfield  {journal} {\bibinfo
   {journal} {Phys. Rev. B}\ }\textbf {\bibinfo {volume} {108}},\ \bibinfo
  {pages} {205144} (\bibinfo {year} {2023})}\BibitemShut {NoStop}%
\bibitem [{\citenamefont {Dong}\ \emph
  {et~al.}(2023{\natexlab{a}})\citenamefont {Dong}, \citenamefont {Ledwith},
  \citenamefont {Khalaf}, \citenamefont {Lee},\ and\ \citenamefont
  {Vishwanath}}]{Dong2023b}%
  \BibitemOpen
  \bibfield  {author} {\bibinfo {author} {\bibfnamefont {J.}~\bibnamefont
  {Dong}}, \bibinfo {author} {\bibfnamefont {P.~J.}\ \bibnamefont {Ledwith}},
  \bibinfo {author} {\bibfnamefont {E.}~\bibnamefont {Khalaf}}, \bibinfo
  {author} {\bibfnamefont {J.~Y.}\ \bibnamefont {Lee}},\ and\ \bibinfo {author}
  {\bibfnamefont {A.}~\bibnamefont {Vishwanath}},\ }\bibfield  {title}
  {\bibinfo {title} {{Many-body ground states from decomposition of ideal
  higher Chern bands: Applications to chirally twisted graphene multilayers}},\
  }\href {https://doi.org/10.1103/physrevresearch.5.023166} {\bibfield
  {journal} {\bibinfo  {journal} {Phys. Rev. Res.}\ }\textbf {\bibinfo {volume}
  {5}},\ \bibinfo {pages} {023166} (\bibinfo {year}
  {2023}{\natexlab{a}})}\BibitemShut {NoStop}%
\bibitem [{\citenamefont {Wang}\ \emph {et~al.}(2023)\citenamefont {Wang},
  \citenamefont {Klevtsov},\ and\ \citenamefont {Liu}}]{Wang2023}%
  \BibitemOpen
  \bibfield  {author} {\bibinfo {author} {\bibfnamefont {J.}~\bibnamefont
  {Wang}}, \bibinfo {author} {\bibfnamefont {S.}~\bibnamefont {Klevtsov}},\
  and\ \bibinfo {author} {\bibfnamefont {Z.}~\bibnamefont {Liu}},\ }\bibfield
  {title} {\bibinfo {title} {{Origin of model fractional Chern insulators in
  all topological ideal flatbands: Explicit color-entangled wave function and
  exact density algebra}},\ }\href
  {https://doi.org/10.1103/physrevresearch.5.023167} {\bibfield  {journal}
  {\bibinfo  {journal} {Phys. Rev. Res.}\ }\textbf {\bibinfo {volume} {5}},\
  \bibinfo {pages} {023167} (\bibinfo {year} {2023})}\BibitemShut {NoStop}%
\bibitem [{\citenamefont {Fujimoto}\ \emph {et~al.}(2024)\citenamefont
  {Fujimoto}, \citenamefont {Parker}, \citenamefont {Dong}, \citenamefont
  {Khalaf}, \citenamefont {Vishwanath},\ and\ \citenamefont
  {Ledwith}}]{Fujimoto2024}%
  \BibitemOpen
  \bibfield  {author} {\bibinfo {author} {\bibfnamefont {M.}~\bibnamefont
  {Fujimoto}}, \bibinfo {author} {\bibfnamefont {D.~E.}\ \bibnamefont
  {Parker}}, \bibinfo {author} {\bibfnamefont {J.}~\bibnamefont {Dong}},
  \bibinfo {author} {\bibfnamefont {E.}~\bibnamefont {Khalaf}}, \bibinfo
  {author} {\bibfnamefont {A.}~\bibnamefont {Vishwanath}},\ and\ \bibinfo
  {author} {\bibfnamefont {P.}~\bibnamefont {Ledwith}},\ }\href@noop {}
  {\bibinfo {title} {{Higher vortexability: zero field realization of higher
  Landau levels}}} (\bibinfo {year} {2024}),\ \Eprint
  {https://arxiv.org/abs/2403.00856} {arXiv:2403.00856 [cond-mat.mes-hall]}
  \BibitemShut {NoStop}%
\bibitem [{\citenamefont {Wang}\ \emph
  {et~al.}(2024{\natexlab{a}})\citenamefont {Wang}, \citenamefont {Zhang},
  \citenamefont {Liu}, \citenamefont {He}, \citenamefont {Xu}, \citenamefont
  {Ran}, \citenamefont {Cao},\ and\ \citenamefont {Xiao}}]{Wang2024}%
  \BibitemOpen
  \bibfield  {author} {\bibinfo {author} {\bibfnamefont {C.}~\bibnamefont
  {Wang}}, \bibinfo {author} {\bibfnamefont {X.-W.}\ \bibnamefont {Zhang}},
  \bibinfo {author} {\bibfnamefont {X.}~\bibnamefont {Liu}}, \bibinfo {author}
  {\bibfnamefont {Y.}~\bibnamefont {He}}, \bibinfo {author} {\bibfnamefont
  {X.}~\bibnamefont {Xu}}, \bibinfo {author} {\bibfnamefont {Y.}~\bibnamefont
  {Ran}}, \bibinfo {author} {\bibfnamefont {T.}~\bibnamefont {Cao}},\ and\
  \bibinfo {author} {\bibfnamefont {D.}~\bibnamefont {Xiao}},\ }\bibfield
  {title} {\bibinfo {title} {{Fractional Chern Insulator in Twisted Bilayer
  MoTe$_2$}},\ }\href {https://doi.org/10.1103/physrevlett.132.036501}
  {\bibfield  {journal} {\bibinfo  {journal} {Phys. Rev. Lett.}\ }\textbf
  {\bibinfo {volume} {132}},\ \bibinfo {pages} {036501} (\bibinfo {year}
  {2024}{\natexlab{a}})}\BibitemShut {NoStop}%
\bibitem [{\citenamefont {Morales-Durán}\ \emph {et~al.}(2024)\citenamefont
  {Morales-Durán}, \citenamefont {Wei}, \citenamefont {Shi},\ and\
  \citenamefont {MacDonald}}]{MoralesDuran2024}%
  \BibitemOpen
  \bibfield  {author} {\bibinfo {author} {\bibfnamefont {N.}~\bibnamefont
  {Morales-Durán}}, \bibinfo {author} {\bibfnamefont {N.}~\bibnamefont {Wei}},
  \bibinfo {author} {\bibfnamefont {J.}~\bibnamefont {Shi}},\ and\ \bibinfo
  {author} {\bibfnamefont {A.~H.}\ \bibnamefont {MacDonald}},\ }\bibfield
  {title} {\bibinfo {title} {Magic angles and fractional chern insulators in
  twisted homobilayer transition metal dichalcogenides},\ }\href
  {https://doi.org/10.1103/physrevlett.132.096602} {\bibfield  {journal}
  {\bibinfo  {journal} {Phys. Rev. Lett.}\ }\textbf {\bibinfo {volume} {132}},\
  \bibinfo {pages} {096602} (\bibinfo {year} {2024})}\BibitemShut {NoStop}%
\bibitem [{\citenamefont {Li}\ and\ \citenamefont {Wu}(2024)}]{Wu2024}%
  \BibitemOpen
  \bibfield  {author} {\bibinfo {author} {\bibfnamefont {B.}~\bibnamefont
  {Li}}\ and\ \bibinfo {author} {\bibfnamefont {F.}~\bibnamefont {Wu}},\ }\href
  {https://doi.org/10.48550/ARXIV.2405.20307} {\bibinfo {title} {{Variational
  Mapping of Chern Bands to Landau Levels: Application to Fractional Chern
  Insulators in Twisted MoTe$_2$}}} (\bibinfo {year} {2024}),\ \Eprint
  {https://arxiv.org/abs/2405.20307} {arXiv:2405.20307 [cond-mat.mes-hall]}
  \BibitemShut {NoStop}%
\bibitem [{\citenamefont {Dong}\ \emph
  {et~al.}(2023{\natexlab{b}})\citenamefont {Dong}, \citenamefont {Wang},
  \citenamefont {Wang}, \citenamefont {Soejima}, \citenamefont {Zaletel},
  \citenamefont {Vishwanath},\ and\ \citenamefont {Parker}}]{Dong2023}%
  \BibitemOpen
  \bibfield  {author} {\bibinfo {author} {\bibfnamefont {J.}~\bibnamefont
  {Dong}}, \bibinfo {author} {\bibfnamefont {T.}~\bibnamefont {Wang}}, \bibinfo
  {author} {\bibfnamefont {T.}~\bibnamefont {Wang}}, \bibinfo {author}
  {\bibfnamefont {T.}~\bibnamefont {Soejima}}, \bibinfo {author} {\bibfnamefont
  {M.~P.}\ \bibnamefont {Zaletel}}, \bibinfo {author} {\bibfnamefont
  {A.}~\bibnamefont {Vishwanath}},\ and\ \bibinfo {author} {\bibfnamefont
  {D.~E.}\ \bibnamefont {Parker}},\ }\href
  {https://doi.org/10.48550/ARXIV.2311.05568} {\bibinfo {title} {{Anomalous
  Hall Crystals in Rhombohedral Multilayer Graphene I: Interaction-Driven Chern
  Bands and Fractional Quantum Hall States at Zero Magnetic Field}}} (\bibinfo
  {year} {2023}{\natexlab{b}}),\ \Eprint {https://arxiv.org/abs/2311.05568}
  {arXiv:2311.05568 [cond-mat.str-el]} \BibitemShut {NoStop}%
\bibitem [{\citenamefont {Dong}\ \emph
  {et~al.}(2023{\natexlab{c}})\citenamefont {Dong}, \citenamefont {Patri},\
  and\ \citenamefont {Senthil}}]{Dong2023a}%
  \BibitemOpen
  \bibfield  {author} {\bibinfo {author} {\bibfnamefont {Z.}~\bibnamefont
  {Dong}}, \bibinfo {author} {\bibfnamefont {A.~S.}\ \bibnamefont {Patri}},\
  and\ \bibinfo {author} {\bibfnamefont {T.}~\bibnamefont {Senthil}},\ }\href
  {https://doi.org/10.48550/ARXIV.2311.03445} {\bibinfo {title} {{Theory of
  fractional quantum anomalous Hall phases in pentalayer rhombohedral graphene
  moir\'{e} structures}}} (\bibinfo {year} {2023}{\natexlab{c}}),\ \Eprint
  {https://arxiv.org/abs/2311.03445} {arXiv:2311.03445 [cond-mat.str-el]}
  \BibitemShut {NoStop}%
\bibitem [{\citenamefont {Zhou}\ \emph {et~al.}(2023)\citenamefont {Zhou},
  \citenamefont {Yang},\ and\ \citenamefont {Zhang}}]{Zhou2023}%
  \BibitemOpen
  \bibfield  {author} {\bibinfo {author} {\bibfnamefont {B.}~\bibnamefont
  {Zhou}}, \bibinfo {author} {\bibfnamefont {H.}~\bibnamefont {Yang}},\ and\
  \bibinfo {author} {\bibfnamefont {Y.-H.}\ \bibnamefont {Zhang}},\ }\href
  {https://doi.org/10.48550/ARXIV.2311.04217} {\bibinfo {title} {{Fractional
  quantum anomalous Hall effects in rhombohedral multilayer graphene in the
  moir\'{e}less limit and in Coulomb imprinted superlattice}}} (\bibinfo {year}
  {2023}),\ \Eprint {https://arxiv.org/abs/2311.04217} {arXiv:2311.04217
  [cond-mat.str-el]} \BibitemShut {NoStop}%
\bibitem [{\citenamefont {Guo}\ \emph {et~al.}(2023)\citenamefont {Guo},
  \citenamefont {Lu}, \citenamefont {Xie},\ and\ \citenamefont
  {Liu}}]{Guo2023}%
  \BibitemOpen
  \bibfield  {author} {\bibinfo {author} {\bibfnamefont {Z.}~\bibnamefont
  {Guo}}, \bibinfo {author} {\bibfnamefont {X.}~\bibnamefont {Lu}}, \bibinfo
  {author} {\bibfnamefont {B.}~\bibnamefont {Xie}},\ and\ \bibinfo {author}
  {\bibfnamefont {J.}~\bibnamefont {Liu}},\ }\href
  {https://doi.org/10.48550/ARXIV.2311.14368} {\bibinfo {title} {{Theory of
  fractional Chern insulator states in pentalayer graphene moir\'{e}
  superlattice}}} (\bibinfo {year} {2023}),\ \Eprint
  {https://arxiv.org/abs/2311.14368} {arXiv:2311.14368 [cond-mat.str-el]}
  \BibitemShut {NoStop}%
\bibitem [{\citenamefont {Kwan}\ \emph {et~al.}(2023)\citenamefont {Kwan},
  \citenamefont {Yu}, \citenamefont {Herzog-Arbeitman}, \citenamefont {Efetov},
  \citenamefont {Regnault},\ and\ \citenamefont {Bernevig}}]{Kwan2023}%
  \BibitemOpen
  \bibfield  {author} {\bibinfo {author} {\bibfnamefont {Y.~H.}\ \bibnamefont
  {Kwan}}, \bibinfo {author} {\bibfnamefont {J.}~\bibnamefont {Yu}}, \bibinfo
  {author} {\bibfnamefont {J.}~\bibnamefont {Herzog-Arbeitman}}, \bibinfo
  {author} {\bibfnamefont {D.~K.}\ \bibnamefont {Efetov}}, \bibinfo {author}
  {\bibfnamefont {N.}~\bibnamefont {Regnault}},\ and\ \bibinfo {author}
  {\bibfnamefont {B.~A.}\ \bibnamefont {Bernevig}},\ }\href
  {https://doi.org/10.48550/ARXIV.2312.11617} {\bibinfo {title} {{Moir\'{e}
  Fractional Chern Insulators III: Hartree-Fock Phase Diagram, Magic Angle
  Regime for Chern Insulator States, the Role of the Moir\'{e} Potential and
  Goldstone Gaps in Rhombohedral Graphene Superlattices}}} (\bibinfo {year}
  {2023}),\ \Eprint {https://arxiv.org/abs/2312.11617} {arXiv:2312.11617
  [cond-mat.str-el]} \BibitemShut {NoStop}%
\bibitem [{\citenamefont {Christos}\ \emph {et~al.}(2022)\citenamefont
  {Christos}, \citenamefont {Sachdev},\ and\ \citenamefont
  {Scheurer}}]{Christos2022}%
  \BibitemOpen
  \bibfield  {author} {\bibinfo {author} {\bibfnamefont {M.}~\bibnamefont
  {Christos}}, \bibinfo {author} {\bibfnamefont {S.}~\bibnamefont {Sachdev}},\
  and\ \bibinfo {author} {\bibfnamefont {M.~S.}\ \bibnamefont {Scheurer}},\
  }\bibfield  {title} {\bibinfo {title} {{Correlated Insulators, Semimetals,
  and Superconductivity in Twisted Trilayer Graphene}},\ }\href
  {https://doi.org/10.1103/physrevx.12.021018} {\bibfield  {journal} {\bibinfo
  {journal} {Phys. Rev. X}\ }\textbf {\bibinfo {volume} {12}},\ \bibinfo
  {pages} {021018} (\bibinfo {year} {2022})}\BibitemShut {NoStop}%
\bibitem [{\citenamefont {Parker}\ \emph {et~al.}(2021)\citenamefont {Parker},
  \citenamefont {Ledwith}, \citenamefont {Khalaf}, \citenamefont {Soejima},
  \citenamefont {Hauschild}, \citenamefont {Xie}, \citenamefont {Pierce},
  \citenamefont {Zaletel}, \citenamefont {Yacoby},\ and\ \citenamefont
  {Vishwanath}}]{Parker2021}%
  \BibitemOpen
  \bibfield  {author} {\bibinfo {author} {\bibfnamefont {D.}~\bibnamefont
  {Parker}}, \bibinfo {author} {\bibfnamefont {P.}~\bibnamefont {Ledwith}},
  \bibinfo {author} {\bibfnamefont {E.}~\bibnamefont {Khalaf}}, \bibinfo
  {author} {\bibfnamefont {T.}~\bibnamefont {Soejima}}, \bibinfo {author}
  {\bibfnamefont {J.}~\bibnamefont {Hauschild}}, \bibinfo {author}
  {\bibfnamefont {Y.}~\bibnamefont {Xie}}, \bibinfo {author} {\bibfnamefont
  {A.}~\bibnamefont {Pierce}}, \bibinfo {author} {\bibfnamefont {M.~P.}\
  \bibnamefont {Zaletel}}, \bibinfo {author} {\bibfnamefont {A.}~\bibnamefont
  {Yacoby}},\ and\ \bibinfo {author} {\bibfnamefont {A.}~\bibnamefont
  {Vishwanath}},\ }\href {https://doi.org/10.48550/ARXIV.2112.13837} {\bibinfo
  {title} {{Field-tuned and zero-field fractional Chern insulators in magic
  angle graphene}}} (\bibinfo {year} {2021}),\ \Eprint
  {https://arxiv.org/abs/2112.13837} {arXiv:2112.13837 [cond-mat.str-el]}
  \BibitemShut {NoStop}%
\bibitem [{\citenamefont {Bultinck}\ \emph
  {et~al.}(2020{\natexlab{a}})\citenamefont {Bultinck}, \citenamefont {Khalaf},
  \citenamefont {Liu}, \citenamefont {Chatterjee}, \citenamefont {Vishwanath},\
  and\ \citenamefont {Zaletel}}]{Bultinck2020a}%
  \BibitemOpen
  \bibfield  {author} {\bibinfo {author} {\bibfnamefont {N.}~\bibnamefont
  {Bultinck}}, \bibinfo {author} {\bibfnamefont {E.}~\bibnamefont {Khalaf}},
  \bibinfo {author} {\bibfnamefont {S.}~\bibnamefont {Liu}}, \bibinfo {author}
  {\bibfnamefont {S.}~\bibnamefont {Chatterjee}}, \bibinfo {author}
  {\bibfnamefont {A.}~\bibnamefont {Vishwanath}},\ and\ \bibinfo {author}
  {\bibfnamefont {M.~P.}\ \bibnamefont {Zaletel}},\ }\bibfield  {title}
  {\bibinfo {title} {Ground state and hidden symmetry of magic-angle graphene
  at even integer filling},\ }\href
  {https://doi.org/10.1103/physrevx.10.031034} {\bibfield  {journal} {\bibinfo
  {journal} {Phys. Rev. X}\ }\textbf {\bibinfo {volume} {10}},\ \bibinfo
  {pages} {031034} (\bibinfo {year} {2020}{\natexlab{a}})}\BibitemShut
  {NoStop}%
\bibitem [{\citenamefont {Xie}\ and\ \citenamefont
  {MacDonald}(2020)}]{Xie2020}%
  \BibitemOpen
  \bibfield  {author} {\bibinfo {author} {\bibfnamefont {M.}~\bibnamefont
  {Xie}}\ and\ \bibinfo {author} {\bibfnamefont {A.}~\bibnamefont
  {MacDonald}},\ }\bibfield  {title} {\bibinfo {title} {{Nature of the
  Correlated Insulator States in Twisted Bilayer Graphene}},\ }\href
  {https://doi.org/10.1103/physrevlett.124.097601} {\bibfield  {journal}
  {\bibinfo  {journal} {Phys. Rev. Lett.}\ }\textbf {\bibinfo {volume} {124}},\
  \bibinfo {pages} {097601} (\bibinfo {year} {2020})}\BibitemShut {NoStop}%
\bibitem [{\citenamefont {Moon}\ and\ \citenamefont
  {Koshino}(2014)}]{Moon2014}%
  \BibitemOpen
  \bibfield  {author} {\bibinfo {author} {\bibfnamefont {P.}~\bibnamefont
  {Moon}}\ and\ \bibinfo {author} {\bibfnamefont {M.}~\bibnamefont {Koshino}},\
  }\bibfield  {title} {\bibinfo {title} {{Electronic properties of
  graphene/hexagonal-boron-nitride moir\'{e} superlattice}},\ }\href
  {https://doi.org/10.1103/physrevb.90.155406} {\bibfield  {journal} {\bibinfo
  {journal} {Phys. Rev. B}\ }\textbf {\bibinfo {volume} {90}},\ \bibinfo
  {pages} {155406} (\bibinfo {year} {2014})}\BibitemShut {NoStop}%
\bibitem [{Note1()}]{Note1}%
  \BibitemOpen
  \bibinfo {note} {This is akin to ED calculations for continuum Landau level
  based strong field FQH physics where the ED is necessarily limited to one or
  two Landau levels.}\BibitemShut {Stop}%
\bibitem [{\citenamefont {Wu}\ \emph {et~al.}(2012{\natexlab{b}})\citenamefont
  {Wu}, \citenamefont {Bernevig},\ and\ \citenamefont {Regnault}}]{Wu2012}%
  \BibitemOpen
  \bibfield  {author} {\bibinfo {author} {\bibfnamefont {Y.-L.}\ \bibnamefont
  {Wu}}, \bibinfo {author} {\bibfnamefont {B.~A.}\ \bibnamefont {Bernevig}},\
  and\ \bibinfo {author} {\bibfnamefont {N.}~\bibnamefont {Regnault}},\
  }\bibfield  {title} {\bibinfo {title} {{Zoology of fractional Chern
  insulators}},\ }\href {https://doi.org/10.1103/physrevb.85.075116} {\bibfield
   {journal} {\bibinfo  {journal} {Phys. Rev. B}\ }\textbf {\bibinfo {volume}
  {85}},\ \bibinfo {pages} {075116} (\bibinfo {year}
  {2012}{\natexlab{b}})}\BibitemShut {NoStop}%
\bibitem [{\citenamefont {Liu}\ \emph {et~al.}(2012)\citenamefont {Liu},
  \citenamefont {Bergholtz}, \citenamefont {Fan},\ and\ \citenamefont
  {L\"{a}uchli}}]{Liu2012}%
  \BibitemOpen
  \bibfield  {author} {\bibinfo {author} {\bibfnamefont {Z.}~\bibnamefont
  {Liu}}, \bibinfo {author} {\bibfnamefont {E.~J.}\ \bibnamefont {Bergholtz}},
  \bibinfo {author} {\bibfnamefont {H.}~\bibnamefont {Fan}},\ and\ \bibinfo
  {author} {\bibfnamefont {A.~M.}\ \bibnamefont {L\"{a}uchli}},\ }\bibfield
  {title} {\bibinfo {title} {{Fractional Chern Insulators in Topological Flat
  Bands with Higher Chern Number}},\ }\href
  {https://doi.org/10.1103/physrevlett.109.186805} {\bibfield  {journal}
  {\bibinfo  {journal} {Phys. Rev. Lett.}\ }\textbf {\bibinfo {volume} {109}},\
  \bibinfo {pages} {186805} (\bibinfo {year} {2012})}\BibitemShut {NoStop}%
\bibitem [{\citenamefont {Repellin}\ and\ \citenamefont
  {Senthil}(2020)}]{Repellin2020}%
  \BibitemOpen
  \bibfield  {author} {\bibinfo {author} {\bibfnamefont {C.}~\bibnamefont
  {Repellin}}\ and\ \bibinfo {author} {\bibfnamefont {T.}~\bibnamefont
  {Senthil}},\ }\bibfield  {title} {\bibinfo {title} {{Chern bands of twisted
  bilayer graphene: Fractional Chern insulators and spin phase transition}},\
  }\href {https://doi.org/10.1103/physrevresearch.2.023238} {\bibfield
  {journal} {\bibinfo  {journal} {Phys. Rev. Res.}\ }\textbf {\bibinfo {volume}
  {2}},\ \bibinfo {pages} {023238} (\bibinfo {year} {2020})}\BibitemShut
  {NoStop}%
\bibitem [{\citenamefont {Liu}\ \emph {et~al.}(2021)\citenamefont {Liu},
  \citenamefont {Abouelkomsan},\ and\ \citenamefont {Bergholtz}}]{Liu2021}%
  \BibitemOpen
  \bibfield  {author} {\bibinfo {author} {\bibfnamefont {Z.}~\bibnamefont
  {Liu}}, \bibinfo {author} {\bibfnamefont {A.}~\bibnamefont {Abouelkomsan}},\
  and\ \bibinfo {author} {\bibfnamefont {E.~J.}\ \bibnamefont {Bergholtz}},\
  }\bibfield  {title} {\bibinfo {title} {{Gate-Tunable Fractional Chern
  Insulators in Twisted Double Bilayer Graphene}},\ }\href
  {https://doi.org/10.1103/physrevlett.126.026801} {\bibfield  {journal}
  {\bibinfo  {journal} {Phys. Rev. Lett.}\ }\textbf {\bibinfo {volume} {126}},\
  \bibinfo {pages} {026801} (\bibinfo {year} {2021})}\BibitemShut {NoStop}%
\bibitem [{\citenamefont {Li}\ \emph {et~al.}(2021)\citenamefont {Li},
  \citenamefont {Kumar}, \citenamefont {Sun},\ and\ \citenamefont
  {Lin}}]{Li2021}%
  \BibitemOpen
  \bibfield  {author} {\bibinfo {author} {\bibfnamefont {H.}~\bibnamefont
  {Li}}, \bibinfo {author} {\bibfnamefont {U.}~\bibnamefont {Kumar}}, \bibinfo
  {author} {\bibfnamefont {K.}~\bibnamefont {Sun}},\ and\ \bibinfo {author}
  {\bibfnamefont {S.-Z.}\ \bibnamefont {Lin}},\ }\bibfield  {title} {\bibinfo
  {title} {{Spontaneous fractional Chern insulators in transition metal
  dichalcogenide moir\'e superlattices}},\ }\href
  {https://doi.org/10.1103/physrevresearch.3.l032070} {\bibfield  {journal}
  {\bibinfo  {journal} {Phys. Rev. Res.}\ }\textbf {\bibinfo {volume} {3}},\
  \bibinfo {pages} {L032070} (\bibinfo {year} {2021})}\BibitemShut {NoStop}%
\bibitem [{\citenamefont {Abouelkomsan}\ \emph {et~al.}(2020)\citenamefont
  {Abouelkomsan}, \citenamefont {Liu},\ and\ \citenamefont
  {Bergholtz}}]{Abouelkomsan2020}%
  \BibitemOpen
  \bibfield  {author} {\bibinfo {author} {\bibfnamefont {A.}~\bibnamefont
  {Abouelkomsan}}, \bibinfo {author} {\bibfnamefont {Z.}~\bibnamefont {Liu}},\
  and\ \bibinfo {author} {\bibfnamefont {E.~J.}\ \bibnamefont {Bergholtz}},\
  }\bibfield  {title} {\bibinfo {title} {{Particle-Hole Duality, Emergent Fermi
  Liquids, and Fractional Chern Insulators in Moir\'e Flatbands}},\ }\href
  {https://doi.org/10.1103/physrevlett.124.106803} {\bibfield  {journal}
  {\bibinfo  {journal} {Phys. Rev. Lett.}\ }\textbf {\bibinfo {volume} {124}},\
  \bibinfo {pages} {106803} (\bibinfo {year} {2020})}\BibitemShut {NoStop}%
\bibitem [{\citenamefont {Nomura}\ and\ \citenamefont
  {MacDonald}(2006)}]{Nomura2006}%
  \BibitemOpen
  \bibfield  {author} {\bibinfo {author} {\bibfnamefont {K.}~\bibnamefont
  {Nomura}}\ and\ \bibinfo {author} {\bibfnamefont {A.~H.}\ \bibnamefont
  {MacDonald}},\ }\bibfield  {title} {\bibinfo {title} {Quantum hall
  ferromagnetism in graphene},\ }\href
  {https://doi.org/10.1103/physrevlett.96.256602} {\bibfield  {journal}
  {\bibinfo  {journal} {Phys. Rev. Lett.}\ }\textbf {\bibinfo {volume} {96}},\
  \bibinfo {pages} {256602} (\bibinfo {year} {2006})}\BibitemShut {NoStop}%
\bibitem [{\citenamefont {Young}\ \emph {et~al.}(2012)\citenamefont {Young},
  \citenamefont {Dean}, \citenamefont {Wang}, \citenamefont {Ren},
  \citenamefont {Cadden-Zimansky}, \citenamefont {Watanabe}, \citenamefont
  {Taniguchi}, \citenamefont {Hone}, \citenamefont {Shepard},\ and\
  \citenamefont {Kim}}]{Young2012}%
  \BibitemOpen
  \bibfield  {author} {\bibinfo {author} {\bibfnamefont {A.~F.}\ \bibnamefont
  {Young}}, \bibinfo {author} {\bibfnamefont {C.~R.}\ \bibnamefont {Dean}},
  \bibinfo {author} {\bibfnamefont {L.}~\bibnamefont {Wang}}, \bibinfo {author}
  {\bibfnamefont {H.}~\bibnamefont {Ren}}, \bibinfo {author} {\bibfnamefont
  {P.}~\bibnamefont {Cadden-Zimansky}}, \bibinfo {author} {\bibfnamefont
  {K.}~\bibnamefont {Watanabe}}, \bibinfo {author} {\bibfnamefont
  {T.}~\bibnamefont {Taniguchi}}, \bibinfo {author} {\bibfnamefont
  {J.}~\bibnamefont {Hone}}, \bibinfo {author} {\bibfnamefont {K.~L.}\
  \bibnamefont {Shepard}},\ and\ \bibinfo {author} {\bibfnamefont
  {P.}~\bibnamefont {Kim}},\ }\bibfield  {title} {\bibinfo {title} {{Spin and
  valley quantum Hall ferromagnetism in graphene}},\ }\href
  {https://doi.org/10.1038/nphys2307} {\bibfield  {journal} {\bibinfo
  {journal} {Nat. Phys.}\ }\textbf {\bibinfo {volume} {8}},\ \bibinfo {pages}
  {550} (\bibinfo {year} {2012})}\BibitemShut {NoStop}%
\bibitem [{\citenamefont {Li}\ \emph {et~al.}(2016)\citenamefont {Li},
  \citenamefont {Zhang},\ and\ \citenamefont {MacDonald}}]{Li2016}%
  \BibitemOpen
  \bibfield  {author} {\bibinfo {author} {\bibfnamefont {X.}~\bibnamefont
  {Li}}, \bibinfo {author} {\bibfnamefont {F.}~\bibnamefont {Zhang}},\ and\
  \bibinfo {author} {\bibfnamefont {A.}~\bibnamefont {MacDonald}},\ }\bibfield
  {title} {\bibinfo {title} {{SU(3) Quantum Hall Ferromagnetism in SnTe}},\
  }\href {https://doi.org/10.1103/physrevlett.116.026803} {\bibfield  {journal}
  {\bibinfo  {journal} {Phys. Rev. Lett.}\ }\textbf {\bibinfo {volume} {116}},\
  \bibinfo {pages} {026803} (\bibinfo {year} {2016})}\BibitemShut {NoStop}%
\bibitem [{\citenamefont {Zou}\ \emph {et~al.}(2018)\citenamefont {Zou},
  \citenamefont {Po}, \citenamefont {Vishwanath},\ and\ \citenamefont
  {Senthil}}]{Zou2018}%
  \BibitemOpen
  \bibfield  {author} {\bibinfo {author} {\bibfnamefont {L.}~\bibnamefont
  {Zou}}, \bibinfo {author} {\bibfnamefont {H.~C.}\ \bibnamefont {Po}},
  \bibinfo {author} {\bibfnamefont {A.}~\bibnamefont {Vishwanath}},\ and\
  \bibinfo {author} {\bibfnamefont {T.}~\bibnamefont {Senthil}},\ }\bibfield
  {title} {\bibinfo {title} {{Band structure of twisted bilayer graphene:
  Emergent symmetries, commensurate approximants, and Wannier obstructions}},\
  }\href {https://doi.org/10.1103/physrevb.98.085435} {\bibfield  {journal}
  {\bibinfo  {journal} {Phys. Rev. B}\ }\textbf {\bibinfo {volume} {98}},\
  \bibinfo {pages} {085435} (\bibinfo {year} {2018})}\BibitemShut {NoStop}%
\bibitem [{\citenamefont {Zhang}\ \emph {et~al.}(2019)\citenamefont {Zhang},
  \citenamefont {Mao},\ and\ \citenamefont {Senthil}}]{Zhang2019}%
  \BibitemOpen
  \bibfield  {author} {\bibinfo {author} {\bibfnamefont {Y.-H.}\ \bibnamefont
  {Zhang}}, \bibinfo {author} {\bibfnamefont {D.}~\bibnamefont {Mao}},\ and\
  \bibinfo {author} {\bibfnamefont {T.}~\bibnamefont {Senthil}},\ }\bibfield
  {title} {\bibinfo {title} {{Twisted bilayer graphene aligned with hexagonal
  boron nitride: Anomalous Hall effect and a lattice model}},\ }\href
  {https://doi.org/10.1103/physrevresearch.1.033126} {\bibfield  {journal}
  {\bibinfo  {journal} {Phys. Rev. Research}\ }\textbf {\bibinfo {volume}
  {1}},\ \bibinfo {pages} {033126} (\bibinfo {year} {2019})}\BibitemShut
  {NoStop}%
\bibitem [{\citenamefont {Wu}\ \emph {et~al.}(2019)\citenamefont {Wu},
  \citenamefont {Lovorn}, \citenamefont {Tutuc}, \citenamefont {Martin},\ and\
  \citenamefont {MacDonald}}]{Wu2019}%
  \BibitemOpen
  \bibfield  {author} {\bibinfo {author} {\bibfnamefont {F.}~\bibnamefont
  {Wu}}, \bibinfo {author} {\bibfnamefont {T.}~\bibnamefont {Lovorn}}, \bibinfo
  {author} {\bibfnamefont {E.}~\bibnamefont {Tutuc}}, \bibinfo {author}
  {\bibfnamefont {I.}~\bibnamefont {Martin}},\ and\ \bibinfo {author}
  {\bibfnamefont {A.}~\bibnamefont {MacDonald}},\ }\bibfield  {title} {\bibinfo
  {title} {{Topological Insulators in Twisted Transition Metal Dichalcogenide
  Homobilayers}},\ }\href {https://doi.org/10.1103/physrevlett.122.086402}
  {\bibfield  {journal} {\bibinfo  {journal} {Phys. Rev. Lett.}\ }\textbf
  {\bibinfo {volume} {122}},\ \bibinfo {pages} {086402} (\bibinfo {year}
  {2019})}\BibitemShut {NoStop}%
\bibitem [{\citenamefont {Bultinck}\ \emph
  {et~al.}(2020{\natexlab{b}})\citenamefont {Bultinck}, \citenamefont
  {Chatterjee},\ and\ \citenamefont {Zaletel}}]{Bultinck2020}%
  \BibitemOpen
  \bibfield  {author} {\bibinfo {author} {\bibfnamefont {N.}~\bibnamefont
  {Bultinck}}, \bibinfo {author} {\bibfnamefont {S.}~\bibnamefont
  {Chatterjee}},\ and\ \bibinfo {author} {\bibfnamefont {M.~P.}\ \bibnamefont
  {Zaletel}},\ }\bibfield  {title} {\bibinfo {title} {{Mechanism for Anomalous
  Hall Ferromagnetism in Twisted Bilayer Graphene}},\ }\href
  {https://doi.org/10.1103/physrevlett.124.166601} {\bibfield  {journal}
  {\bibinfo  {journal} {Phys. Rev. Lett.}\ }\textbf {\bibinfo {volume} {124}},\
  \bibinfo {pages} {166601} (\bibinfo {year} {2020}{\natexlab{b}})}\BibitemShut
  {NoStop}%
\bibitem [{\citenamefont {Xie}\ and\ \citenamefont
  {Das~Sarma}(2024)}]{XieMing2024}%
  \BibitemOpen
  \bibfield  {author} {\bibinfo {author} {\bibfnamefont {M.}~\bibnamefont
  {Xie}}\ and\ \bibinfo {author} {\bibfnamefont {S.}~\bibnamefont
  {Das~Sarma}},\ }\bibfield  {title} {\bibinfo {title} {{Integer and fractional
  quantum anomalous Hall effects in pentalayer graphene}},\ }\href
  {https://doi.org/10.1103/physrevb.109.l241115} {\bibfield  {journal}
  {\bibinfo  {journal} {Physical Review B}\ }\textbf {\bibinfo {volume}
  {109}},\ \bibinfo {pages} {L241115} (\bibinfo {year} {2024})}\BibitemShut
  {NoStop}%
\bibitem [{\citenamefont {Zhang}\ \emph {et~al.}(2010)\citenamefont {Zhang},
  \citenamefont {Sahu}, \citenamefont {Min},\ and\ \citenamefont
  {MacDonald}}]{Zhang2010}%
  \BibitemOpen
  \bibfield  {author} {\bibinfo {author} {\bibfnamefont {F.}~\bibnamefont
  {Zhang}}, \bibinfo {author} {\bibfnamefont {B.}~\bibnamefont {Sahu}},
  \bibinfo {author} {\bibfnamefont {H.}~\bibnamefont {Min}},\ and\ \bibinfo
  {author} {\bibfnamefont {A.~H.}\ \bibnamefont {MacDonald}},\ }\bibfield
  {title} {\bibinfo {title} {{Band structure of $ABC$-stacked graphene
  trilayers}},\ }\href {https://doi.org/10.1103/physrevb.82.035409} {\bibfield
  {journal} {\bibinfo  {journal} {Phys. Rev. B}\ }\textbf {\bibinfo {volume}
  {82}},\ \bibinfo {pages} {035409} (\bibinfo {year} {2010})}\BibitemShut
  {NoStop}%
\bibitem [{\citenamefont {Jung}\ and\ \citenamefont
  {MacDonald}(2013)}]{Jung2013}%
  \BibitemOpen
  \bibfield  {author} {\bibinfo {author} {\bibfnamefont {J.}~\bibnamefont
  {Jung}}\ and\ \bibinfo {author} {\bibfnamefont {A.~H.}\ \bibnamefont
  {MacDonald}},\ }\bibfield  {title} {\bibinfo {title} {{Gapped broken symmetry
  states in ABC-stacked trilayer graphene}},\ }\href
  {https://doi.org/10.1103/physrevb.88.075408} {\bibfield  {journal} {\bibinfo
  {journal} {Phys. Rev. B}\ }\textbf {\bibinfo {volume} {88}},\ \bibinfo
  {pages} {075408} (\bibinfo {year} {2013})}\BibitemShut {NoStop}%
\bibitem [{\citenamefont {Wang}\ \emph
  {et~al.}(2024{\natexlab{b}})\citenamefont {Wang}, \citenamefont {Vila},
  \citenamefont {Zaletel},\ and\ \citenamefont {Chatterjee}}]{Wang2024a}%
  \BibitemOpen
  \bibfield  {author} {\bibinfo {author} {\bibfnamefont {T.}~\bibnamefont
  {Wang}}, \bibinfo {author} {\bibfnamefont {M.}~\bibnamefont {Vila}}, \bibinfo
  {author} {\bibfnamefont {M.~P.}\ \bibnamefont {Zaletel}},\ and\ \bibinfo
  {author} {\bibfnamefont {S.}~\bibnamefont {Chatterjee}},\ }\bibfield  {title}
  {\bibinfo {title} {Electrical control of spin and valley in spin-orbit
  coupled graphene multilayers},\ }\href
  {https://doi.org/10.1103/physrevlett.132.116504} {\bibfield  {journal}
  {\bibinfo  {journal} {Phys. Rev. Lett.}\ }\textbf {\bibinfo {volume} {132}},\
  \bibinfo {pages} {116504} (\bibinfo {year} {2024}{\natexlab{b}})}\BibitemShut
  {NoStop}%
\bibitem [{\citenamefont {Kudin}\ \emph {et~al.}(2002)\citenamefont {Kudin},
  \citenamefont {Scuseria},\ and\ \citenamefont {Can\`{e}s}}]{Kudin2002}%
  \BibitemOpen
  \bibfield  {author} {\bibinfo {author} {\bibfnamefont {K.~N.}\ \bibnamefont
  {Kudin}}, \bibinfo {author} {\bibfnamefont {G.~E.}\ \bibnamefont
  {Scuseria}},\ and\ \bibinfo {author} {\bibfnamefont {E.}~\bibnamefont
  {Can\`{e}s}},\ }\bibfield  {title} {\bibinfo {title} {{A black-box
  self-consistent field convergence algorithm: One step closer}},\ }\href
  {https://doi.org/10.1063/1.1470195} {\bibfield  {journal} {\bibinfo
  {journal} {J. Chem. Phys.}\ }\textbf {\bibinfo {volume} {116}},\ \bibinfo
  {pages} {8255} (\bibinfo {year} {2002})}\BibitemShut {NoStop}%
\end{thebibliography}%

\end{document}